\documentclass[12pt]{article}
\usepackage{amsmath}
\usepackage{graphicx,psfrag,epsf}
\usepackage{enumerate}
\usepackage{natbib}
\usepackage{url} 

\newcommand{\blind}{1}

\long\def\ignore#1{} 
\usepackage[inline,shortlabels]{enumitem}
\usepackage{float}
\usepackage{mathtools}
\usepackage{statmath}
\usepackage{subdepth}
\usepackage[colorlinks,citecolor=blue,urlcolor=blue]{hyperref}
\usepackage{tikz}
\usetikzlibrary{arrows.meta}
\usetikzlibrary{decorations.markings}

\tikzset{+ /.tip = {Bar[sep=-3pt 2,width=3pt 4]_[sep=0]}}
\usepackage[english]{babel}
\usepackage{longtable}
\usepackage{color}
\usepackage{mathtools}
\usepackage{amssymb,amsmath,amsthm}
\usepackage{multirow}
\usepackage{dsfont}
\newcommand{\gs}{{\gamma_{1:n}}}
\newcommand{\prns}[1]{\left(#1\right)}

\newcommand{\IF}{\mathbb{IF}}

\renewcommand{\P}{\mathsf{P}}

\newcommand{\I}{\mathds{1}}

\newcommand{\Xb}{\text{X}}
\newcommand{\Vb}{\text{V}}
\newcommand{\Zb}{\text{Z}}

\newcommand{\Xt}{\widetilde{\Xb}}
\newcommand{\xt}{\widetilde{\xb}}

\newcommand{\xb}{\text{x}}
\newcommand{\vb}{\text{v}}

\newcommand{\Ex}{\mathsf{E}}

\newcommand{\one}{\mathds{1}}
\renewcommand{\P}{\mathsf{P}}

\newtheorem{assumption}{Assumption}[section]
\DeclarePairedDelimiterX{\norm}[1]{\lVert}{\rVert}{#1}

\pgfdeclarelayer{background}
\pgfsetlayers{background,main}
\usetikzlibrary{arrows,positioning}
\tikzset{
>=stealth',
punkt/.style={
rectangle,
rounded corners,
draw=black, very thick,
text width=6.5em,
minimum height=2em,
text centered},
pil/.style={
->,
thick,
shorten <=2pt,
shorten >=2pt,}
}
\newcommand{\Vertex}[3]
{\node[minimum width=0.6cm,inner sep=0.05cm] (#2) at (#1) {#3};
}
\newcommand{\Vertexr}[3]
{\node[rectangle, draw, minimum width=0.6cm,inner sep=0.05cm] (#2) at (#1) {#2};
}
\newcommand{\ArrowR}[3]%
{ \begin{pgfonlayer}{background}
\draw[->,#3] (#1) to[bend right=30] (#2);
\end{pgfonlayer}
}
\newcommand{\ArrowLW}[3]%
{ \begin{pgfonlayer}{background}
\draw[->,#3] (#1) to[bend left=30] (#2);
\end{pgfonlayer}
}

\newcommand{\ArrowL}[3]%
{ \begin{pgfonlayer}{background}
    \draw[->,#3] (#1) to[bend left=45] (#2);
  \end{pgfonlayer}
}

\newcommand{\EdgeL}[3]%
{ \begin{pgfonlayer}{background}
\draw[dashed,#3] (#1) to[bend right=-45] (#2);
\end{pgfonlayer}
}

\newcommand{\Arrow}[3]%
{ \begin{pgfonlayer}{background}
\draw[->,#3] (#1) -- +(#2);
\end{pgfonlayer}
}

\allowdisplaybreaks
\pgfarrowsdeclare{arcs}{arcs}{...}
{
  \pgfsetdash{}{0pt} 
  \pgfsetroundjoin   
  \pgfsetroundcap    
  \pgfpathmoveto{\pgfpoint{-5pt}{5pt}}
  \pgfpatharc{180}{270}{5pt}
  \pgfpatharc{90}{180}{5pt}
  \pgfusepathqstroke
}
\newcommand{\ArrowB}[3]%
{ \begin{pgfonlayer}{background}
    \draw[|-arcs,line width=0.4mm,shorten <= 0.3cm,shorten >= 0.3cm,#3] (#1) -- +(#2);
  \end{pgfonlayer}
}

\newcommand{\satett}{\mathsf{ATE}}
\newcommand{\csatett}{\mathsf{CATE}}

\begin{document}

\def\spacingset#1{\renewcommand{\baselinestretch}%
{#1}\small\normalsize} \spacingset{1}


\if1\blind
{
  \title{\bf Modern Causal Inference Approaches to Improve Power for Subgroup Analysis in Randomized Controlled Trials}
   \author{Antonio D'Alessandro, Jiyu Kim, Samrachana Adhikari\hspace{.2cm}\\
    Division of Biostatistics, Department of Population Health, \\ New York University School of Medicine, \\ 
    New York, NY, 10016\\
     \\
    Donald Goff\\
    Department of Psychiatry, \\
    New York University School of Medicine, \\ 
    New York, NY, 10016 \\
    \\
    Falco J. Bargagli-Stoffi\thanks{This article is based upon work supported the Amazon Web Services (AWS) grant on ``AI/ML for Identifying Social Determinants of Health''. Corresponding author. E-mail: falco@g.ucla.edu} \\
    Department of Biostatistics, University of California, \\
    Los Angeles, CA, 90095\\
     \\
    Michele Santacatterina\thanks{
    This article is based upon work supported by the National Science Foundation under Grant No 2306556,  National Institute of Health Grant No 1R01AI197146-01.} \\
    Division of Biostatistics, Department of Population Health, \\ New York University School of Medicine, \\ 
    New York, NY, 10016}
    \date{}
  \maketitle

} \fi

\if0\blind
{
  \bigskip
  \bigskip
  \bigskip
  \begin{center}
    {\LARGE\bf Modern Causal Inference Approaches to Improve Power for Subgroup Analysis in Randomized Controlled Trials}
\end{center}
  \medskip
} \fi

\thispagestyle{empty}

\newpage

\thispagestyle{empty}

\newpage

\bigskip
\newpage
\begin{abstract}
Randomized controlled trials (RCTs) often include subgroup analyses to assess whether treatment effects vary across pre-specified patient populations. However, these analyses frequently suffer from small sample sizes which limit the power to detect heterogeneous effects. Power can be improved by leveraging predictors of the outcome---i.e., through covariate adjustment---as well as by borrowing external data from similar RCTs or observational studies. The benefits of covariate adjustment may be limited when the trial sample is small. Borrowing external data can increase the effective sample size and improve power, but it introduces two key challenges: (i) integrating data across sources can lead to model misspecification, and (ii) practical violations of the positivity assumption---where the probability of receiving the target treatment is near-zero for some covariate profiles in the external data---can lead to extreme inverse-probability weights and unstable inferences, ultimately negating potential power gains. To account for these shortcomings, we present an approach to improving power in pre-planned subgroup analyses of small RCTs that leverages both baseline predictors and external data. We propose debiased estimators that accommodate parametric, machine learning, and nonparametric Bayesian methods. To address practical positivity violations, we introduce three estimators: a covariate-balancing approach, an automated debiased machine learning (DML) estimator, and a calibrated DML estimator. We show improved power in various simulations and offer practical recommendations for the application of the proposed methods. Finally, we apply them to evaluate the effectiveness of citalopram for negative symptoms in first-episode schizophrenia patients across subgroups defined by duration of untreated psychosis, using data from two small RCTs.

\end{abstract}

\noindent%
\vfill

\newpage

\pagenumbering{arabic}
\setcounter{page}{1}

\spacingset{1.45} 
\section{Introduction}
\label{sec:intro}
In randomized controlled trials (RCTs), subgroup analyses are often conducted to evaluate treatment effect heterogeneity within pre-specified subgroups \citep{lipkovich2024modern, lipkovich2011subgroup}. However, these analyses frequently suffer from small sample sizes, reducing the power to detect meaningful differences \citep{alosh2015statistical}. For example, the DECIFER trial \citep{goff2019citalopram}, which evaluated the effect of add-on citalopram in patients with first-episode schizophrenia, found a non-significant indication of a negative effect on the \textit{Calgary Depression Scale for Schizophrenia} (CDSS) score at 52 weeks among participants with a duration of untreated psychosis (DUP) less than 18 weeks. The point estimate was –0.75, with a 95\% confidence interval of (–2.55, 1.05). In contrast, among those with DUP greater than 18 weeks, there was a non-significant indication of a positive effect, with a point estimate of 1.15 and a 95\% confidence interval of (–0.66, 2.95). While these findings suggest potential clinical implications, they are not statistically significant, highlighting the need for methods that increase precision---i.e., narrow confidence intervals, thus improving power---while maintaining valid inference.

There are two primary approaches to increasing precision in RCTs. The first is \textit{covariate adjustment}, which leverages baseline predictors of the outcome (often referred to as \textit{prognostic variables}) to reduce variance and improve the precision of treatment effect estimates \citep{van2024covariate,lin2013agnostic}. Covariate adjustment achieves this by accounting for outcome variation explained by baseline characteristics, allowing the treatment effect to be estimated from the residual variation not attributable to those covariates. The second approach involves \textit{incorporating external data} from comparable randomized or observational studies \citep{brantner2024comparison}, thereby augmenting the effective sample size. We provide a more detailed literature review on these approaches in the next section.

In the case of the DECIFER trial, we have access to baseline predictors, such as age, gender, and race, that can be used for covariate adjustment. Additionally, we have access to a source of external data: the \textit{Recovery After an Initial Schizophrenia Episode} (RAISE) trial, which assessed the efficacy of specialized coordinated care for individuals with first-episode schizophrenia (FES) \citep{kane2015raise}.

While these approaches help improve precision, as shown in our simulation results, each comes with limitations. Covariate adjustment relies solely on data from the randomized trial itself, which may be limited in size, as in the DECIFER trial ($N = 95$, $52$ completers). Incorporating external data can mitigate this limitation by effectively increasing the sample size, but it requires an additional identification assumption and correct models specification---external data are typically not randomized with respect to the treatment arms of the target trial, and the probability of treatment assignment conditional on covariates must be estimated correctly. In addition, outcome models must also be estimated correctly, as differences in covariate–outcome relationships across sources can lead to bias if not properly accounted for.

Furthermore, while treatment assignment is guaranteed to be positive within the trial (positivity is formally introduced in Section \ref{sec:identification}), the same may not hold in the external population, where the probability of receiving a particular treatment could be small or even near zero. For instance, in the RAISE dataset, the probability of receiving add-on citalopram, conditional on age, race, and gender, was less than $0.05$ for $29$ patients, indicating a potential practical positivity violation. Such violations lead to large inverse probability weights and unreliable inferences, negating the potential efficiency gains from using external data.

To mitigate model misspecification, debiased estimators that combine outcome and treatment models, while allowing for flexible, data-adaptive methods, have been proposed \citep{kang2007demystifying,Kennedy2022}. To address the effects of practical positivity violations, covariate balancing estimators \citep{kallus2021more, hirshberg2021augmented}, as well as automated \citep{Chern2024} and calibrated \citep{Lars2024} debiased machine learning (DML) approaches, have been developed.

In this article, we present a structured approach to improving precision in pre-planned subgroup analyses of small RCTs, leveraging baseline predictors and external data. We propose debiased estimators that accommodate parametric, machine learning, and nonparametric Bayesian methods, while also incorporating modern strategies to address practical positivity violations. Our work extends the contributions of \citet{wang2024efficient} and \citet{brantner2024comparison}. We demonstrate the proposed methods using a combined dataset from the DECIFER and RAISE studies. 

\section{Related work}

The bulk of statistical techniques discussed in this work relies on incorporating external data to improve precision. The practice of combining external and randomized trial data can be traced back to \cite{pocock1976}. This foundational work has since led to the generation of a substantial body of literature on improving precision in RCTs by incorporating external controls or so called real world data (RWD). A significant portion of the work in this field has focused on using Bayesian techniques to construct prior distributions \citep{ibrahim2000} or leveraging Bayesian hierarchical models \citep{neu2010} to remove the bias introduced by using external controls. A useful overview, particularly of early Bayesian methods, can be found in \cite{viele2014}. More recently, frequentist approaches that focus on study design, as in \cite{yuan2019}, or integrating machine learning methods have emerged. The use of machine learning is varied and can range from (i) using machine learning to flexibly choose between data sources for controls by optimizing the bias-variance trade-off like in \cite{laan2024}, (ii) integrating predictions from machine learning models trained on large observational datasets to boost precision without introducing bias \citep{gagnon2023}, or (iii) using data-adaptive methods to estimate subgroup specific effects by combining data from multiple RCTs \citep{brantner2024}.  

All the statistical methods covered in this paper make use of the causal inference framework. Within this field, a literature parallel to the one described above has emerged, which focuses on combining data from randomized experiments and observational studies, so-called ``data-fusion'' \citep{pearl2016}. Much work within this literature focuses on deriving the causal assumptions required \textit{transportability} \citep{bare2014, west2017, Dahabreh2024, josey2021transporting, josey2022calibration, josey2025real} and \textit{generalizability} \citep{stuart2015, stuart2018, Dahabreh2021}. Our work makes use of developments from these fields to make explicit necessary causal assumptions and to investigate the conditions under which they may or may not hold in the context of borrowing data to improve precision. A review of the generalizability and transportability literature can be found in \cite{colnet2020} and \cite{Degtiar2023}.  

When baseline predictors are strongly correlated with the study outcome, their inclusion can reduce variance in estimates and substantially improve the statistical power of the analysis. The literature surrounding the use and development of covariate adjustment has grown in recent years, especially as it relates to clinical trials, as more extensive baseline information is collected on patients \citep{van2022,lin2013agnostic}. Current work investigates using covariate adjustment in the presence of missing or incomplete data \citep{chang2023}, time-to-event and binary outcomes \citep{benk2020, li2023}, and implementing machine learning based estimators \citep{zhang2019, williams2024}. The article by \cite{Kahan2014} provides a helpful overview of the risks and rewards of using covariate adjustment specifically in the setting of a randomized trial.  

The literature at the intersection of machine learning and causal inference, especially for estimating Conditional Average Treatment Effects (CATE), has expanded dramatically in the past decade. CATE is the primary estimand targeted in subgroup analyses, as it captures treatment effect heterogeneity across covariate-defined subpopulations.
Significant work has been done to adapt widely used machine learning algorithms to the task of estimating CATE. Examples of this approach can be found in \cite{hill2011}, \cite{athey2016}, \cite{wager2018} and \cite{tian2018}.  Alternatively, a class of more general algorithms or ``meta-learners'' has emerged, which allow any data-adaptive method to be used to estimate CATE \citep{Kunzel2019}. 

\section{Notation and setup}
\label{sec:setup}

\subsection{Notation} 
\label{sec:notation}

Suppose we observe an independent and identically distributed (iid) sample $(\Zb_1, \ldots, \Zb_n)$ of size $n$ drawn from a distribution $\P$. For each subject $i$ in the sample, the vector $\Zb$ can be decomposed as $\Zb = (\Xb, A, S, Y)$, where we dropped the subscript $i$ for readability. Let $\Xb$ denote the vector of baseline covariates, which may include both discrete and continuous random variables, and let $\mathcal{X}$ denote the support of $\Xb$. Additionally, suppose $\Vb$ is a vector of discrete covariates such that $\Vb \subset \Xb$, representing the pre-planned subgroups of interest. Let $\widetilde{\Xb} = \Xb \setminus\Vb$, that is $\widetilde{\Xb}$ is the vector of random variables composed of those elements of $\Xb$ not in $\Vb$. Let $A$ denote the randomized treatment indicator such that $A \in \mathcal{A} = \{0,1\}$, where $1$ indicates the reception of the active treatment and $0$ the reception of the control. Let $S$ represent the population indicator variable with support $\mathcal{S} = \{0,1\}$, describing membership in the original target trial ($S=1$) where the subgroup analysis is conducted, or external dataset ($S=0$), while the random variable $Y$ represents a real-valued outcome. In this manuscript we consider the ``target'' or ``target-trial'', the original trial (S=1). Throughout this paper, we will use counterfactual notation \citep{pearl2010introduction}, where $Y(a)$ denotes the counterfactual outcome that would have been observed in a hypothetical world where treatment $A=a$ had been administered. 

In this paper, our focus is on the subgroup average treatment effect in the target trial ($\satett$), defined as
\begin{align}
\csatett(\xt, \vb, 1) &= \Ex[\, Y(1) - Y(0) \mid \Xt=\xt, \Vb=\vb,  S=1], \\
\satett(\vb, 1) &= \Ex[\, \csatett(\vb,\Xt,1) \mid \Vb=\vb, S=1].
\end{align}
\noindent
In words, we are targeting the mean difference in outcome between the two treatment arms in a specific subgroup in the original trial population. For instance, in our case study, this is the mean difference in change CDSS at 52 weeks between add-on citalopram and placebo among those participants with a duration of untreated psychosis of less than or more than 18 weeks in the DECIFER trial population. This is a special case of the target parameter considered by \cite{wang2024efficient}.

\subsection{Nonparametric identification}
\label{sec:identification}

Non-parametric identification expresses the causal target, which is based on counterfactual (unobservable quantities), in terms of the observed data distribution, without assuming a specific functional form \citep{pearl1995causal}. We now summarize the identification results presented in \cite{wang2024efficient}, starting with the identification assumptions. 

\begin{assumption}[Weak ignorability]  \label{ass:wexch}  
 $$\Ex[Y(a) \mid \Xt=\xt, \Vb=\vb,   S=s] = \Ex[Y(a) \mid A=a,\Xt=\xt, \Vb=\vb, S=s]$$ for each $a \in \mathcal{A}$ and $s\in\mathcal{S}.$ \\
\end{assumption}
\vspace{-0.75cm}
Assumption~\ref{ass:wexch}, also known as exchangeability, states that once we control for $\Xt$ in $\Vb=v, S=s$, the counterfactual outcome under $a$ is independent from the treatment assignment (on expectation). We expect this to hold by design because of randomization when $S=1$--the original trial--while we need to assume it when $S=0$, the external data. It is worth noticing that, to increase precision in the original trial, we are willing to assume untestable assumption--\textit{i.e.,} it is a function of counterfactuals which are unobservable; this could lead to biased estimates, and a covariate adjustment strategy may be preferred.

\begin{assumption}[Weak exchangeability over source] \label{ass:sexch} 
$$\Ex[Y(a) \mid  A=a, \Xt=\xt, \Vb=\vb,  S=s] = \Ex[Y(a) \mid A=a, \Xt=\xt, \Vb=\vb]$$ for each $a \in \mathcal{A}$ and $s\in\mathcal{S}$. \\
\end{assumption}
\vspace{-0.75cm}
Assumption~\ref{ass:sexch} states that once we control for $\Xb$, the counterfactual outcome under $a$ is independent from the source (on expectation) under both treatment arms.  Similar to Assumption~\ref{ass:wexch}, this is another untestable assumption that we are willing to make to improve precision, \textit{i.e.,} we can combine/pool data from different sources thus increasing the sample size to learn conditional expectations theoretically improving precision. Note that Assumption \ref{ass:wexch} is not strictly required for S=0, since Assumption \ref{ass:sexch} is transporting to the target population under the same variable set X. 

\begin{assumption}[Consistency] \label{ass:cons}
If $A_i = a$ then $Y_i(a) = Y_i$ for all individuals $i$ and treatments $a \in \mathcal{A}$. \\
\end{assumption}
\vspace{-0.75cm}
Assumption~\ref{ass:cons} is a standard causal inference assumption that states that the observed outcome conditioned on $A=a$ is the same as the counterfactual outcome $Y(a)$ for all $a$ in $\{0,1\}$. In the randomized trial (S=1) we assume the trial protocol strictly controls all necessary features related to the administered treatment. We expect by design, this assumption to hold when S=1 while we need to assume it in the external data (S=0). 

\begin{assumption}[Positivity of the probability of treatment] \label{ass:posa}
 $$\P(A=a \, | \, \Xt=\xt, \Vb=\vb,  S=s) > 0$$ for all $x=(\xt,v) \in \mathcal{X}$, $s\in\mathcal{S}$, and for each $a \in \mathcal{A}.$\\
\end{assumption}
\vspace{-0.75cm}
Assumption~\ref{ass:posa} states that all covariate profiles $\xt$ have a positive probability of receiving such treatment in both populations $S=1$ and $S=0$ for $\Vb=\vb$. Again, this assumption holds by design in the original trial, while we need to assume it for $S=0$. In addition, this assumption while it can hold theoretically, can be violated in practice, i.e. when the probability of treatment is close to zero for some subjects in a given sample. 

\begin{assumption}[Positivity of probability of participation] \label{ass:poss}
For all $x=(\xt,v)  \in \mathcal{X}$,  $\P(S=s \, | \, \Xt=\xt, \Vb=\vb) > 0$ with probability $1$, for each $s \in \mathcal{S}$.\\
\end{assumption}
\vspace{-0.75cm}
Assumption~\ref{ass:poss} states that the probability of belonging to each source, original trial and external, is positive across all covariate patterns $\xb$.
Under Assumptions~\ref{ass:wexch}-\ref{ass:poss}, \cite{wang2024efficient} showed that $\csatett(\xt, \vb, 1)$ is identified as $$\Ex[ Y \, | \, A=1,\Xt=\xt, \Vb=\vb] - \Ex[ Y \, | \,  A=0, \Xt=\xt, \Vb=\vb]$$ and $\satett(\vb,1)$ is consequently identified by taking the average of the above expression for $\csatett(\xt, \vb, 1)$ over the distribution of $\Xt$. We provide an alternative derivation in the Appendix for convenience. In addition, it is worth saying that when only using data from the original trial (not leveraging external data) as for example when using a covariate adjustment estimator, Assumption~\ref{ass:sexch} and Assumption~\ref{ass:poss}, are not required, with the other assumptions holding by design under proper randomization.

\section{Estimators that improve power for subgroup analysis when targeting $\satett(\vb,1)$} \label{sec:methods}


\subsection{Debiased estimators}

Based on the identification results, we start by proposing a debiased covariate-adjustment estimator for $\satett(\vb,1)$ (Section \ref{drcovadj}). As aforementioned, this estimator leverages baseline predictors of the outcome to increase power instead of requiring external data. While this estimator allows for the use of flexible data-driven techniques such as machine learning, in this paper, we use parametric regression techniques to learn the nuisance functions since the original trial is small. We then review a debiased estimator that leverages external data and propose to use parametric frequentist and Bayesian techniques in addition to machine learning approaches, such as random forests, and non-parametric Bayesian methods, like Bayesian additive regression trees (BART), to learn the nuisance functions (Section \ref{drext}). Finally, we propose three new debiased estimators that leverage external data while dealing with practical positivity violations, namely a covariate balancing, an auto-DML and a calibrated-DML estimator (Section \ref{sec:auto-dml}). 

In this paper, we propose debiased estimators as an alternative to estimators based solely on outcome regression or inverse probability weighting. These estimators provide consistent estimates by debiasing the outcome model with weighted residuals, which leads to more robust inferences under model misspecification. They possess desirable properties, including asymptotic normality and fast convergence rates. Moreover, they allow for the incorporation of machine learning algorithms while maintaining valid statistical inferences. To derive these estimators, we follow standard practice of constructing them based on efficient influence functions (EIF)s \citep{bickel1993efficient,fisher2021visually,hines2022demystifying,kennedy2021semiparametric}.

\subsection{A debiased covariate adjustment estimator that only leverages baseline predictors in the original trial} \label{drcovadj}

As previously discussed, the idea of covariate adjustment is to leverage baseline predictors of the outcome to reduce variance and improve precision \citep{van2024covariate}. We derive the EIF of $\satett(\vb,1)$ under Assumption~\ref{ass:wexch}, \ref{ass:cons}, and \ref{ass:posa} as:

\begin{align*}
&\frac{1}{\P(\Vb=\vb,S=1)} \Big(  \left[ \frac{\I(A=1,\Vb=\vb,S=1)}{ \P(A=1 \mid \Xt, \Vb=\vb,S=1)} \right] \lbrace Y - \Ex[Y \mid A=1, \Xt, \Vb=\vb, S=1]\rbrace \\
                    &-   \left[ \frac{\I(A=0,\Vb=\vb, S=1)}{1- \P(A=1 \mid \Xt, \Vb=\vb,S=1)} \right]  \lbrace Y - \Ex[Y \mid A=0,  \Xt, \Vb=\vb,S=1]\rbrace \\
                    &+    \I(\Vb=\vb, S=1) \lbrace \Ex[Y \mid A=1,  \Xt, \Vb=\vb,S=1] - \Ex[Y \mid A=0,  \Xt, \Vb=\vb,S=1] - \satett(\vb,1) \rbrace  \Big),
\end{align*} 

\noindent
The technical details of the derivation can be found in the Appendix. The EIF suggests the following debiased covariate adjustment estimator

\begin{align*}
    \widehat{\satett}_{cov}(\vb,1) &= \frac{\hat{\alpha}^{-1}}{n} \sum_{i=1}^{n} \Big[ \I(A_i =1, \Vb_i = \vb, S_i =1)\frac{1}{\hat{\pi}(\xt_i, \vb_i,1)}\{y_i - \hat{m}(1,\xt_i,\vb_i,1)\}\\
    &- \I(A_i=0, \Vb_i = \vb, S_i=1)\frac{1}{(1-\hat{\pi}(\xt_i, \vb_i,1))}\{ y_i - \hat{m}(0,\xt_i, \vb_i,1) \} \\
    &+ \I(\Vb_i = \vb, S_i=1)\{ \hat{m}(1,\xt_i, \vb_i,1) - \hat{m}(0, \xt_i, \vb_i,1) \} \Big],
\end{align*}
\noindent
 where $\hat{\alpha}$ is the sample proportion of $\I(\Vb=\vb, S=1)$, $\hat{m}(a,\xt_i, \vb_i,1)$ and $\hat{\pi}(\xt_i, \vb_i,1)$ are estimators for the conditional expectation $\Ex[Y \, | \, A=a, \Xt, \Vb=\vb,S=1] $ and $\P(A=1  \mid \Xt, \Vb=\vb,S=1)$, respectively. These functions may be estimated by using standard parametric generalized linear models.

\subsection{A debiased estimator that leverages external data and uses flexible data-driven techniques} \label{drext}

When leveraging external data, following \citep{wang2024efficient}, the derived EIF of $\satett(\vb,1)$ under Assumption~\ref{ass:wexch}-\ref{ass:poss} is given by:

\begin{align}
    \frac{1}{\P(\Vb=\vb, S=1)} 
    \Big[  
    &  \I(A=1, \Vb=\vb) \frac{\P(S=1 \mid \Xt, \Vb=\vb)}{\P(A=1 \mid \Xt, \Vb=\vb)} \{Y - \Ex[Y \mid A=1, \Xt, \Vb=\vb]\} \nonumber \\
    & -\I(A=0, \Vb=\vb) \frac{\P(S=1 \mid \Xt, \Vb=\vb)}{\P(A=0 \mid \Xt, \Vb=\vb)} \{Y - \Ex[Y \mid A=0, \Xt, \Vb=\vb]\}  \nonumber \\ 
    & +\I(\Vb=\vb,S=1) \{ \E[Y \mid A=1, \Xt, \Vb=\vb] - \Ex[Y \mid A=0, \Xt, \Vb=\vb] - \satett(\vb,1) \}\Big].\label{if_ext}
\end{align}

\noindent
Its derivations can be found in the Appendix. The following debiased estimator can then be used
\begin{align} \label{eq:3}
    \widehat{\satett}_{D}(\vb) &= \frac{\hat{\alpha}^{-1}}{n} \sum_{i=1}^{n} \Big[ \I(A_i =1, \Vb_i = \vb)\frac{\hat{\eta}(\xt_i,\vb_i)}{\hat{\pi}(\xt_i,\vb_i)}\{y_i - \hat{m}(1,\xt_i,\vb_i)\} \nonumber\\
    &- \I(A_i = 0,\Vb_i = \vb)\frac{\hat{\eta}(\xt_i,\vb_i)}{(1-\hat{\pi}(\xt_i,\vb_i))}\{ y_i - \hat{m}(0,\xt_i,\vb_i) \} \nonumber\\
    &+ \I(\Vb_i=\vb, S_i =1)\{ \hat{m}(1,\xt_i,\vb_i) - \hat{m}(0, \xt_i,\vb_i) \} \Big] 
\end{align}
\newline
\noindent
where $\hat{\alpha}$ is obtained as above, $\hat{m}(a,\xt_i,\vb_i)$ is an estimator for the conditional expectation of the outcome $\Ex[Y \mid A=a, \Xt, \Vb=\vb]$, and $\hat{\eta}(\xt_i,\vb_i)$ and  $\hat{\pi}(\xt_i,\vb_i)$ are estimators for $\P(S=1 \mid \Xt,\Vb=\vb)$ and $\P(A=a \mid \Xt,\Vb=\vb)$, respectively. 

In this article, we propose to use parametric frequentist and Bayesian regression models, a random forest, and BART to estimate these quantities \citep{Chipman2010, Breiman2001}. We discuss how to obtain estimators' variances, construct confidence intervals and hypothesis tests in our practical guidelines in Section \ref{sec:practical_guidelines}.

\subsection{Debiased estimators that leverage external data, uses flexible data-driven techniques, and deals with practical positivity violation} \label{sec:auto-dml}

In the previous section, we introduced EIFs, forming the foundation for constructing debiased estimators such as $\widehat{\satett}_{D}(v)$. 
As shown above, these estimators first learn a model for the observed outcomes, such as $\Ex(Y\mid A=a, \Xt=\xt, \Vb=\vb)$, and then employ a weighted sum of residuals to \textit{debias} it. We showed that these weights take the form of $\one{(A_i=1, V_i=v)} \frac{\hat{\eta}(\xt_i,\vb_i)}{\hat{\pi}(\xt_i,\vb_i)} + \one{(A_i=0, V_i=v)} \frac{\hat{\eta}(\xt_i,\vb_i)}{(1-\hat{\pi}(\xt_i,\vb_i))}$. We refer to these as inverse weights. While this choice is justified by asymptotic arguments, $\hat{\pi}(\xt_i,\vb_i)$ can become very small, resulting in excessively large weights, a phenomenon also known as practical positivity violation \citep{petersen2012diagnosing} or lack of overlap \citep{crump2009dealing}. This can happen in scenarios where there is large covariate shift in the distribution of baseline covariates across populations. 

We follow recently proposed techniques designed to address this issue, including methods based on covariate balancing \citep[among others]{kallus2021optimal, kallus2022optimal, kallus2021more, hirshberg2019minimax, hirshberg2021augmented}, automatic learning \citep{Chern2024}, and calibration \citep{Lars2024}.

\subsubsection{A covariate balancing estimator} 
\label{sec:covbal}

Rather than plugging in inverse probability weights directly, the covariate balancing literature proposes learning approximate weights by minimizing a measure of covariate imbalance while penalizing for the complexity of the weights. These weights are obtained by solving an optimization problem that trades off balance and precision. To provide intuition, we begin by considering a single treatment arm $A=a$ based on our results regarding the identification of the causal parameter $\csatett$ in the Appendix, and consider $Z = \{A, \Xt, \Vb, S\}$ for clarity. This leads to focusing on the following statistical functional, 

\begin{align*}
    \Psi^a(m) &= \Ex[ \ \Ex[\ Y \ | \ A=a, \Xt=\xt, \Vb=\vb] \ | \ \Vb=\vb, S=1]\\ 
    &= \Ex[\ m(a,\Xt=\xt, \Vb=\vb) \ | \ \Vb=\vb, S=1] \\ 
    &= \frac{ \I(\Vb=\vb, S=1) }{\P(\Vb=\vb, S=1)}\Ex[m(Z)], \\ 
\end{align*}

\noindent 
where $m(Z) =  m(a,\Xt=\xt, \Vb=\vb) = \Ex[\ Y \ | \ A=a, \Xt=\xt, \Vb=\vb]$ is a function of the entire observations but does not use information on $S$ (as a result of Assumption~\ref{ass:wexch}). To show how our proposed balancing weights are obtained, we start by defining $Y_i = m(Z_i) + \epsilon_{i}$, 
where $\Ex[\epsilon_{a,i} \ | \ Z_i] = 0$, and $\sigma^2_a = \Ex[\epsilon_{a,i}^2 | Z_i] = Var[Y_i | Z_i]$ where $A_i=a$ for $a=0,1$ and all $i$. Let $\delta_{m}(Z_i) = \hat{m}(Z_i) - m(Z_i)$ (the regression  error) and recall the augmented estimator for treatment $a$, ie., setting $A_i=a$ for all $i$, we introduced in section \ref{drext}, 

\begin{align*}
    \hat{\Psi}^{a}_{D} &= \frac{ \alpha^{-1} }{n} \sum^{n}_{i=1} \big[ \I(V_i=v, S_i =1)\hat m(Z_i)    \big] \\ 
    &- \frac{ \alpha^{-1} }{n}\sum^{n}_{i=1} \big[ \gamma(Z_i)(\hat m(Z_i) - Y_i)\big], \\ 
\end{align*}

\noindent
where $\gamma(Z_i) = \I[A_i=a, \Vb_i=\vb_i] \gamma(\Xt_i,\Vb,S)$ and where $\gamma(\Xt_i,\Vb,S)$ could be chosen to be set to $\frac{\eta(\Xt_i,\Vb)}{\pi(\Xt_i,\Vb)}$ for $A_i=1$ or $\frac{\eta(\Xt_i,\Vb)}{(1-\pi(\Xt_i,\Vb))}$ for $A_i=0$, as done in section \ref{drext}. We can decompose this estimator's error as 

\begin{align*}
    \hat{\Psi}^{a}_{D} - \Psi^a(m) &= \frac{ \alpha^{-1} }{n} \sum^{n}_{i=1} \big[ \I(V_i=v, S_i =1)(\delta_{m}(Z_i) + m(Z_i) ) \big] \\ 
    &- \frac{ \alpha^{-1} }{n}\sum^{n}_{i=1}\big[ \gamma(Z_i)\delta_{m}(Z_i)   \big] \\ 
    &+ \frac{ \alpha^{-1} }{n}\sum^{n}_{i=1}\big[\gamma(Z_i)\epsilon_i \big] - \Psi^a(m)\\
    &= \underbrace{ \frac{ \alpha^{-1} }{n} \Big[\sum^{n}_{i=1} \I(V_i=v , S_i =1)\delta_{m}(Z_i) - \sum^{n}_{i=1}\gamma(Z_i)\delta_{m}(Z_i) \Big]}_{\text{imbalance in } \delta_{m}}\\
    &+ \underbrace{ \frac{ \alpha^{-1} }{n} \sum^{n}_{i=1}\gamma(Z_i)\epsilon_i}_{\text{noise - mean zero}} + \underbrace{ \frac{ \alpha^{-1} }{n} \sum^{n}_{i=1} \I(V_i=v , S_i =1)m(Z_i)  - \Psi^a(m)}_{\text{sampling variation - mean zero}}.
\end{align*}\\

We assume that $\delta_{m}(Z_i)$ is contained in an absolutely convex set of functions $\mathcal{M}$ (i.e. a Hilbert Space) and define the following worst-case imbalance in $\mathcal{M}$, 

\begin{align*}
\mathbb{I}_{\mathcal{M}}(\gamma) = \sup_{\delta_{m} \in \mathcal{M}} | \frac{ \alpha^{-1} }{n}  \big(\sum^{n}_{i=1}\I(V_i = v, S_i=1)\delta_{m}(Z_i) - \sum^{n}_{i=1} \gamma(Z_i)\delta_{m}(Z_i)\big)  |    .
\end{align*}
\noindent

We then propose to obtain $\gamma(Z_i)$  by minimizing the worst-case imbalance while accounting for some measure of the complexity of the weights, ie., $\frac{\sigma^2_a}{n^2}\sum_i \I[S_i=1] \gamma(Z_i)^2$ (which controls complexity specifically for the trial population, $S=1$). This leads to the following optimization problem,
\begin{align*}
    \hat{\gamma} &=  \underset{\gamma}{\arg\min}\Big[\mathbb{I}^{2}_{\mathcal{M}}(\gamma) + \lambda \frac{\sigma^2_a}{n^2}\sum_i \I[S_i=1]\gamma(Z_i)^2\Big], 
\end{align*}
\noindent
where $\lambda$ is an arbitrary penalization parameter. Following standard practice \citep{ben2021balancing,hirshberg2021augmented,hirshberg2019minimax,kallus2020generalized,kallus2021optimal,kallus2022optimal,pham2023stable}, we choose as a model $\mathcal{M}$ the unit ball $\mathcal{B}_{\mathcal{H}} = \{ m(Z) \in \mathcal{H} : \| m \|_{\mathcal{H}} \leq 1 \}$ of a Reproducing Kernel Hilbert Space (RKHS) $\mathcal{H}$. 
Define the matrix $K_a\in\mathbb R^{n\times n}$ as $K_{aij}=\mathcal K_a(Z_{i},Z_{j})$ where $A_i=a$ for all $i$, and setting $\gamma_i = \gamma(Z_i)$. By the representer theorem, we have that
\begin{align*}
\mathbb{I}_{\mathcal{M}}(\gamma) &=  \sup_{\|m\|^2_{\mathcal{H}} \leq 1} \prns{  \sum_{i = 1}^n  \prns{\I[V_i=v, S_i=1] -\I[A_i=a,V_i=v]\gamma_i} m(Z_i)}^2 \\
&= \sup_{\sum_{i,j=1}^n\alpha_i\alpha_j\mathcal K_a(Z_i,Z_j)\leq1} \prns{  \sum_{i = 1}^n  \prns{\I[V_i=v, S_i=1] -\I[A_i=a,V_i=v]\gamma_i} \sum_{j=1}^n\alpha_j\mathcal K_a(Z_i,Z_j)}^2\\
&= \sup_{\alpha^TK_a\alpha\leq1} \prns{\alpha^TK_a(e_{vs}-I_{av}\gs)}^2\\
&=(I_{av}\gs-e_{vs})^TK_a(I_{av}\gs-e_{vs}) \\
&= \prns{\gs^TI_{av}K_aI_{av}\gs-2 e_{vs}^TK_aI_{av}\gs+e_{vs}^T K_a e_{vs}},
\end{align*}

\noindent
where $e_{vs}$ is the length-$n$ vector with observations with $V=v$ and $S=1$, and $I_{av}$  is the $n$-by-$n$ diagonal matrix with $\I[A_i=a,V_i=v]$ in the $i^{th}$ diagonal entry. Note that, since the true regression error $\delta_m(Z_i)$ is unknown, we replaced it with a generic function $m(Z_i) \in \mathcal{M}$. This allows us to define a worst-case imbalance that bounds the conditional bias uniformly over all functions in $\mathcal{M}$, rather than relying on knowledge of a specific $\delta_m$.

Based on these results, we choose the weights $\gamma_{1:n}$ to solve the optimization problem

\begin{equation}
   \label{eq:optimization_problem}
    \underset{\gamma_{1:n} \ge 0}{\min} \gs^T \prns{I_{1v}K_1I_{1v} + I_{0v}K_0I_{0v} + \Sigma_{\lambda} }\gs-2 e_{vs}^T \prns{K_1I_{1v} + K_0I_{0v} }\gs ,
\end{equation}

\noindent
where $\Sigma_{\lambda}$ is the $n$-by-$n$ diagonal matrix with $\I[A_i=1]\sigma_1^2 + \I[A_i=0]\sigma_0^2$ in its $i^{th}$ diagonal entry multiplied by the penalization parameter $\lambda$. This optimization problem depends on a choice of kernel, $K_a(z,z')$, the conditional variance $\sigma_a^2$ for $a \in \lbrace 0,1 \rbrace$ and the penalization parameter $\lambda$. We provide some details on these choices in section \ref{sec:practical_guidelines}.

Finally, the obtained weights are then plugged into the debiased estimator previously considered, 

\begin{align} 
\label{satt_cb}
    \widehat{\satett}_{CB}(\vb) &= \frac{\hat{\alpha}^{-1}}{n} \sum_{i=1}^{n} \Big[ \I(A_i =1, \Vb_i = \vb) \hat \gamma\{y_i - \hat{m}(1,\xt_i,\vb_i)\} \nonumber\\
    &- \I(A_i = 0,\Vb_i = \vb)\hat \gamma\{ y_i - \hat{m}(0,\xt_i,\vb_i) \} \nonumber\\
    &+ \I(\Vb_i=\vb, S_i =1)\{ \hat{m}(1,\xt_i,\vb_i) - \hat{m}(0, \xt_i,\vb_i) \} \Big],
\end{align}
\noindent
thus avoiding the issues associated with directly plugging in estimated conditional probabilities into the inverse weights consequently dealing with practical positivity violation as shown in our simulation results. More details are provided in the practical guidelines. 

\subsubsection{Automatic Debiased Machine Learning (auto-DML)} \label{sec:autodml}

In the previous section, we introduced a covariate balancing method to obtain more stable weights that can handle practical positivity violations and approximate the inverse weights from Section \ref{drext}. In this section, we propose an alternative estimator based on auto-DML, which learns the weights by solving a different optimization problem than the one used for covariate balancing. 

Specifically, following the notation introduced in the previous section and considering \( Z = \{A, \Xt, \Vb, S\} \) for clarity, the statistical estimand of interest is given by
\begin{align*}
\frac{ \I(\Vb = \vb, S = 1) }{\P(\Vb = \vb, S = 1)} \Ex[q(Z, m)],     
\end{align*}
where \( q(Z, m) = m(1, \Xt, \Vb) - m(0, \Xt, \Vb) \). This is a continuous linear functional of \( m \). By the Riesz representation theorem \citep{chernozhukov2021automatic}, there exists a random variable \( \gamma(Z) \) such that, for all functions \( m(Z) \) with \( \Ex[m(Z)^2] < \infty \),
\begin{align*}
    \Ex[q(Z; m)] = \Ex[\gamma_0(Z) m(Z)].
\end{align*}

In our setting, the true representer corresponds to the set of inverse weights previously derived, namely,
\begin{align*}
    \gamma_0(Z) = \I(A = 1, \Vb = \vb) \frac{\P(S = 1 \mid \Xt, \Vb = \vb)}{\P(A = 1 \mid \Xt, \Vb = \vb)} + \I(A = 0, \Vb = \vb) \frac{\P(S = 1 \mid \Xt, \Vb = \vb)}{\P(A = 0 \mid \Xt, \Vb = \vb)}.
\end{align*}

The work of \cite{chernozhukov2021automatic} shows that the Riesz representer can be viewed as the minimizer of the following loss function:
\begin{align*}
\gamma_0 &= \arg\min_{\gamma} \Ex\big[(\gamma(Z) - \gamma_0(Z))^2\big] \\
&= \arg\min_{\gamma} \Ex\big[\gamma(Z)^2 - 2\gamma_0(Z)\gamma(Z) + \gamma_0(Z)^2\big] \\
&= \arg\min_{\gamma} \Ex\big[\gamma(Z)^2 - 2q(Z; \gamma)\big],
\end{align*}
where the second equality comes from expanding the square, and the third follows from the Riesz representation theorem (\( \Ex[q(Z; \gamma)] = \Ex[\gamma_0(Z) \gamma(Z)] \)) and the fact that \( \Ex[\gamma_0(Z)^2] \) is constant with respect to the minimizer \citep{Chern2022}. 

Rather than directly plugging in estimated conditional probabilities to estimate \( \gamma_0 \), auto-DML learns the representer by solving the following optimization problem within some space \( \mathcal{A} \):
\begin{align*}\label{autodmlmin}
\hat{\gamma}^\ast = \arg\min_{\gamma \in \mathcal{A}} \frac{1}{n} \sum_{i=1}^n\big[\gamma(Z_i)^2 - 2q(Z_i; \gamma)\big].
\end{align*}

Similar to covariate balancing weights, this approach avoids the issues associated with directly plugging in estimated conditional probabilities into the inverse weights consequently dealing with practical positivity violations, as shown in our simulation results. We then consider,

\begin{align} 
    \widehat{\satett}_{AM}(\vb) &= \frac{\hat{\alpha}^{-1}}{n} \sum_{i=1}^{n} \Big[ \I(A_i =1, \Vb_i = \vb) \hat \gamma^\ast\{y_i - \hat{m}(1,\xt_i,\vb_i)\} \nonumber\\
    &- \I(A_i = 0,\Vb_i = \vb)\hat \gamma^\ast\{ y_i - \hat{m}(0,\xt_i,\vb_i) \} \nonumber\\
    &+ \I(\Vb_i=\vb, S_i =1)\{ \hat{m}(1,\xt_i,\vb_i) - \hat{m}(0, \xt_i,\vb_i) \} \Big] 
\end{align}

More details on the implementation are provided in the practical guidelines. Note that, the use of the \textit{representer theorem} in the kernel formulation ensures that the solution to the optimization problem lies in the span of the kernel evaluations at the observed points. This allows us to rewrite the infinite-dimensional problem over functions \( m(Z) \) in the RKHS as a finite-dimensional problem over the coefficients \( \alpha \). In contrast, auto-DML leverages the \textit{Riesz representation theorem}, which establishes that for any continuous linear functional on a Hilbert space (e.g., the functional \( q(Z; m) \mapsto \Ex[q(Z; m)] \)), there exists a unique representer function \( \gamma_0(Z) \) such that the functional can be written as an inner product: $
\Ex[q(Z; m)] = \Ex[\gamma_0(Z) m(Z)]$.
While the Riesz theorem guarantees the existence of such a representer, the representer theorem in the kernel case explicitly constructs it in terms of the kernel evaluations at the observed points, reducing the problem to finite-dimensional operations involving the kernel matrix \( K_a \). Thus, both theorems serve to simplify the infinite-dimensional functional problem into a tractable, finite-dimensional representation, albeit through different mathematical mechanisms tailored to their respective settings.

\subsubsection{Calibrated Debiased Machine Learning (CDML)} \label{cdml}

In addition to learning the weights in a data-adaptive way that avoids directly plugging in estimated conditional probabilities, such as in covariate balancing and auto-DML approaches, calibration can also help address practical positivity violations. Calibration, a machine learning technique commonly used in prediction and classification tasks \citep{zadrozny2001obtaining,gupta2020distribution}, can be employed to smooth extreme values, thus stabilizing the inverse weights and mitigating issues caused by small propensities. Specifically, as introduced in \cite{van2024stabilized} for inverse probability weighting and extended in \cite{Lars2024} for DML estimators, we propose to calibrate the conditional probabilities $\pi(\Xt,\Vb)$ and $\eta(\Xt,\Vb)$ and outcome models $m(a,\Xt,\Vb)$ using isotonic regression. Isotonic regression fits a non-decreasing function to these conditional expectations, effectively ``flattening'' large fluctuations that arise from practical positivity violations, stabilizing the inverse weights. We propose the following calibrated DML estimator for $\satett(v)$:

\begin{align} \label{eq:cdml}
    \widehat{\satett}_{Cal}(\vb) &= \frac{\hat{\alpha}^{-1}}{n} \sum_{i=1}^{n} \Big[ \I(A_i =1, \Vb_i = \vb)\frac{\eta^\ast(\xt_i,\vb_i)}{\pi^\ast(\xt_i,\vb_i)}\{y_i - m^\ast(1,\xt_i,\vb_i)\} \nonumber\\
    &- \I(A_i = 0,\Vb_i = \vb)\frac{\eta^\ast(\xt_i,\vb_i)}{(1-\pi^\ast(\xt_i,\vb_i))}\{ y_i - m^\ast(0,\xt_i,\vb_i) \} \nonumber\\
    &+ \I(\Vb_i=\vb, S_i =1)\{ m^\ast(1,\xt_i,\vb_i) - m^\ast(0, \xt_i,\vb_i) \} \Big],
\end{align}
\noindent
where $\pi^\ast$, $\eta^\ast$, and $\mu^\ast$ are obtained using isotonic regression. Following \cite{Lars2024}, we implement isotonic regression using XGBoost with a monotonicity constraint, ensuring that the predicted values increase monotonically. In addition to its stability, \cite{Lars2024} show that CDML enjoys desirable properties like asymptotic linearity and double robustness, which are outside the scope of this paper but worth noting. More detail on its implementation is provided in the practical guidelines.

\section{Simulations}\label{sec:simulation}

\subsection{Simulation setup}


In this section, we evaluate the performance of the estimators listed in Table~\ref{table_methods} across three simulation scenarios with respect to power, absolute bias, variance, and 95\% confidence interval coverage under different scenarios: (1) increasing the amount of external data relative to the original trial, (2) practical violations of the positivity assumption, and (3) model misspecification.

\paragraph*{Data-generating process for Scenario 1} We fixed the number of patients in the target trial $n_{S=1}$ at $100$ and considered the external data source of size $n_{S=0}$ taking values of $\{100,200,\ldots,900\}$. The total number of patients is $n$ such that $n_{S=1} + n_{S=0} = n$.  Data for each subject $i \in \{1,\ldots, n\}$ were generated according to the below steps. This process was repeated $1000$ times for each pair $(n_{S=1}=100,n_{S=0}=100), (n_{S=1}=100,n_{S=0}=200), \ldots, (n_{S=1}=100,n_{S=0}=900)$. \label{sec:scen1}  

\begin{itemize}
    \item[] \textbf{Step 1.} For each subject $i$ randomly draw covariate $\Xt_i$ such that $\Xt_i \sim \text{N}(0,1)$.
    \item[] \textbf{Step 2.} For each subject $i$ randomly draw covariate $V_i$ such that $V_i \sim \text{Bernoulli}(0.5)$.
    \item[] \textbf{Step 3.}  Let $\eta_i$ be the probability subject $i$ is enrolled in the target trial, 
    and randomly draw population indicator $S_i$ such that $S_i \sim \text{Bernoulli}(\eta_i)$ where $\eta_i = (1+\text{exp}(-C + 0.5*\Xt_i + 1.2*V_i))^{-1}$ . Here, \( C \) is determined by solving an optimization problem such that  
$\eta$ equals the target proportion $100/n_{S=0}$, thereby maintaining the RCT sample size fixed at 100 while allowing the external data size to vary.
    \item[] \textbf{Step 4.} If subject $i$ is in the target population ($S_i = 1$) randomly draw treatment indicator $A_i$ from $A_i \sim \text{Bernoulli}(0.5)$. 
    \item[] \textbf{Step 5.} If subject $i$ is in the external population ($S_i = 0$) let $\pi_i$ be the probability subject $i$ obtains treatment. Compute this quantity as $\pi_i= (1+\text{exp}(-0.045 + 0.09*\Xt_i + 0.09*V_i))^{-1}$ and randomly draw treatment indicator $A_i$ from $A_i \sim \text{Bernoulli}(\pi_i)$.
    \item[] \textbf{Step 6.} For subject $i$ compute their potential outcome under control ($A_i=0$) as $Y_i(0) = 1.5*\Xt_i + 0.5*V_i + \epsilon_i$ such that $\epsilon_i \sim N(0,1)$. 
    \item[] \textbf{Step 7.} Next compute the potential outcome for subject $i$ under treatment ($A_i=1$) as $Y_i(1) = Y_i(0) + V_i - 0.5$.
    \item[] \textbf{Step 8.} Finally generate the observed outcome for subject $i$ as $Y_i = A_i*Y_i(1) + (1-A_i)*Y_i(0)$. 
    
\end{itemize}

\paragraph*{Estimands} The estimand of interest is $\satett(v,1)$, which was set to $0.5$ in the subgroup defined by $V=1$ and $-0.5$ in the subgroup $V=0$. 

\paragraph*{Methods} For each dataset generated within each scenario we used the methods summarized in Table \ref{table_methods}.

\begin{table}
\small
\caption{Methods used in the estimation of $\satett$. \label{table_methods}}
\begin{tabular}{lcc}
\multicolumn{1}{c}{\textbf{Method}} & \multicolumn{1}{c}{\textbf{Acronym}} & \multicolumn{1}{r}{\textbf{Sec}} \\
\hline
Covariate adjustment with generalized linear models & \texttt{cov-adj} & ~\ref{drcovadj}\\ \hline
Debiased estimator from (\ref{eq:3}) using linear/logistic regression for $\pi, \eta$ and $m$ & \texttt{D-glm} & ~\ref{drext} \\ \hline
Debiased estimator from (\ref{eq:3}) using Bayesian glms for $\pi, \eta$ and $m$ & \texttt{D-bayglm} & ~\ref{drext}\\ \hline
Debiased estimator from (\ref{eq:3}) using random forest for $\pi, \eta$ and $m$ & \texttt{D-ranger} & ~\ref{drext}\\ \hline
Debiased estimator from (\ref{eq:3}) using BART for $\pi, \eta$ and $m$ & \texttt{D-bart} & ~\ref{drext} \\ \hline
Covariate balancing penalty 0.01 & \texttt{covbal0} & ~\ref{sec:covbal} \\ \hline
Automatic  debiased machine learning  & \texttt{riesz} & ~\ref{sec:autodml} \\ \hline
Calibrated debiased machine learning & \texttt{cdml} & ~\ref{cdml}\\ \hline
Difference in sample means & \texttt{naive} \\ \hline
\end{tabular}
\end{table}

\paragraph*{Performance metrics} We report absolute bias, variance,  coverage of the 95\% confidence interval (and it's Monte Carlo SE as discussed in Table 6 of \cite{morris2019}), and power. Power was calculated for each estimator in Table~\ref{table_methods} using a Wald test statistic, testing the null hypothesis $\satett(v,1) = 0$ against the two-sided alternative. The proportion of simulations in which the resulting p-value was less than 0.05 is reported as the estimated power. We provide the data generating processes for scenario (2) and (3) in the appendix.

\subsection{Results} \label{sec:results}


\subsubsection{Scenario 1 - Power}

\ignore{
\begin{itemize}
Expected plots:
    \item Scenario 1: power, bias, variance, coverage (plots) plus tables in the appendix under $V=1$, $V=0$ in the appendix or vice-versa.
    \item Scenario 2: bias, variance, coverage (plots) plus tables in the appendix, plus current table 2 and figure 9 (distribution of estimate)
\end{itemize}
}

\begin{figure}
\begin{center}
\includegraphics[scale=0.4]{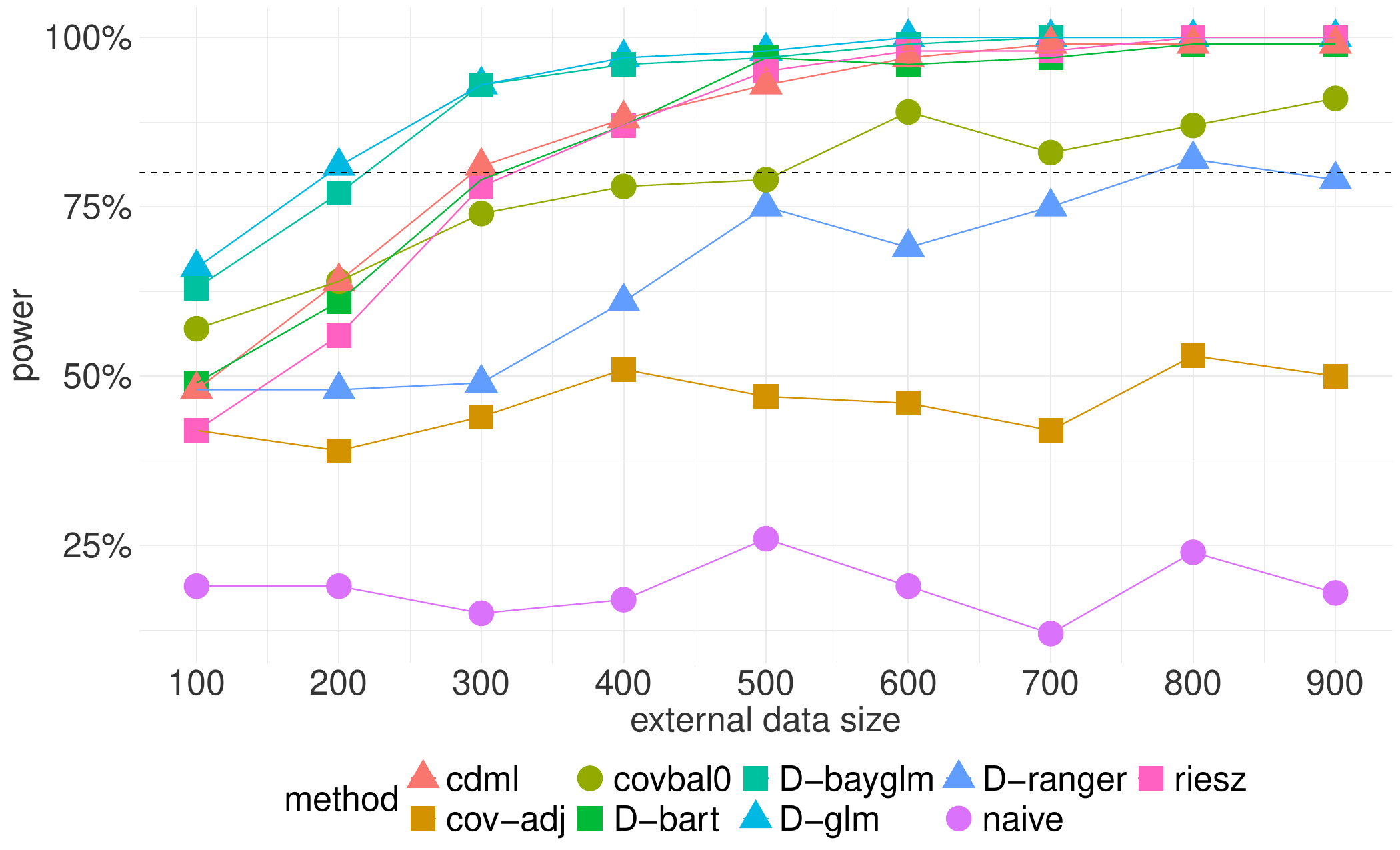}
\end{center}
\caption{Power of the estimators in Table \ref{table_methods} as a function of external data size to detect a subgroup specific effect in the subgroup defined by $V=1$. The dashed line indicates the $80\%$ power threshold. \label{fig:powv1} }
\end{figure}

 Figure \ref{fig:powv1} shows the power to detect $\satett(v=1,1) = 0.5$ as a function of the external data size $n_{S=0}$ for each estimator. For estimators that incorporate external data, power increases as the size of the external source grows. As expected, estimators that rely solely on the target trial data (e.g., \texttt{cov-bal} and \texttt{naive}) maintain a consistent level of power throughout, with only minor fluctuations due to sampling variability. Estimators utilizing external data reach and exceed the $80\%$ power threshold at different rates. In particular, \texttt{D-bayglm} and \texttt{D-glm} appear to gain power at a faster rate compared to the ML based estimators. This could be due to correct model specification, i.e. both \texttt{D-bayglm} and \texttt{D-glm} use parametric models to estimate the underlying nuisance functions which correctly match the form used in the DGP. 
 
\begin{figure}[]
\caption{Mean absolute bias of the estimators in Table \ref{table_methods} as a function of external data size (top). The sampling variability of the same estimators
as a function of external data size (bottom). These plots correspond to the subgroup defined by $V=1$.\label{fig:biasvarv1}}
\vspace{-7.5mm}
\begin{center}
\includegraphics[scale=0.4]{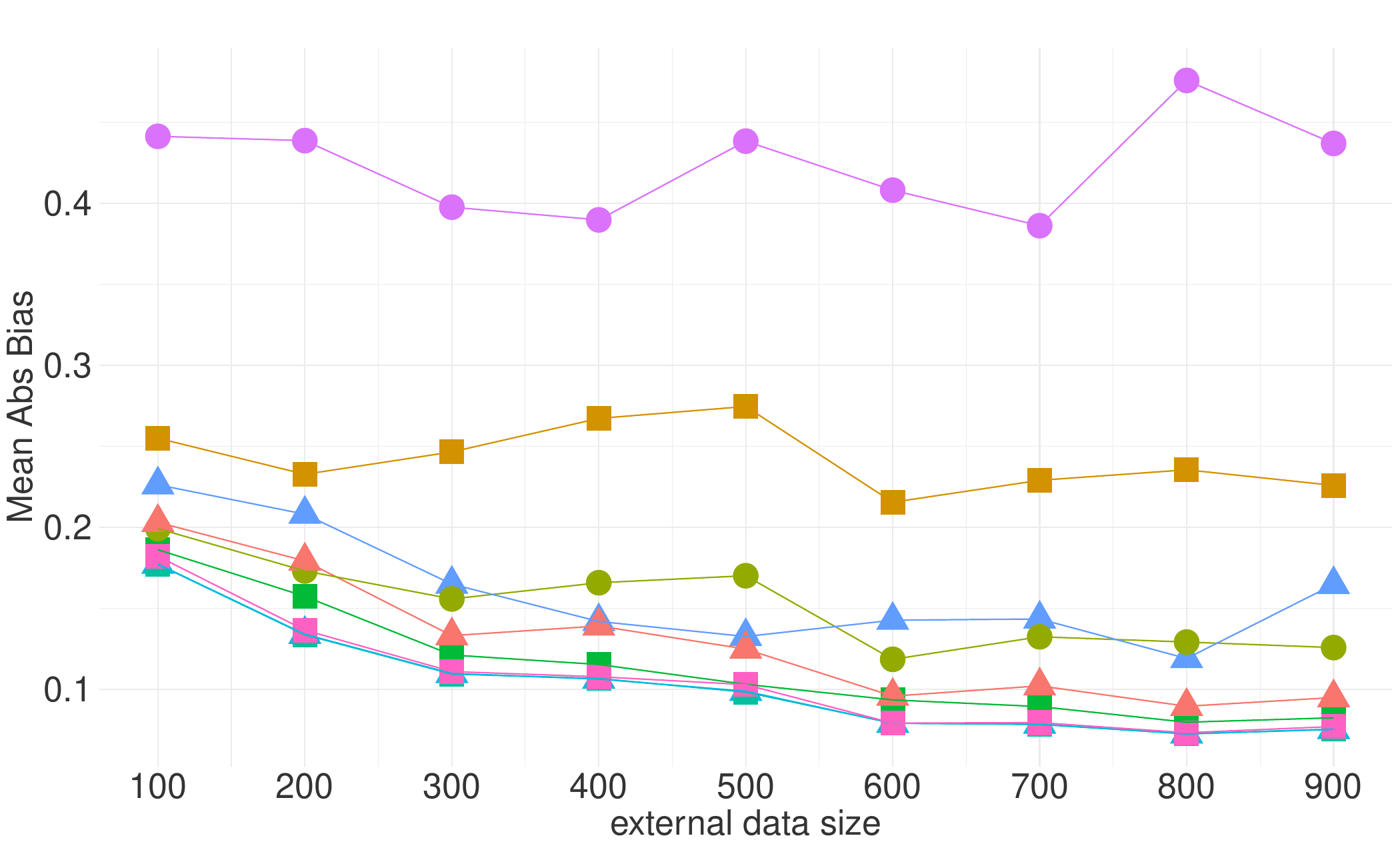}
\includegraphics[scale=0.4]{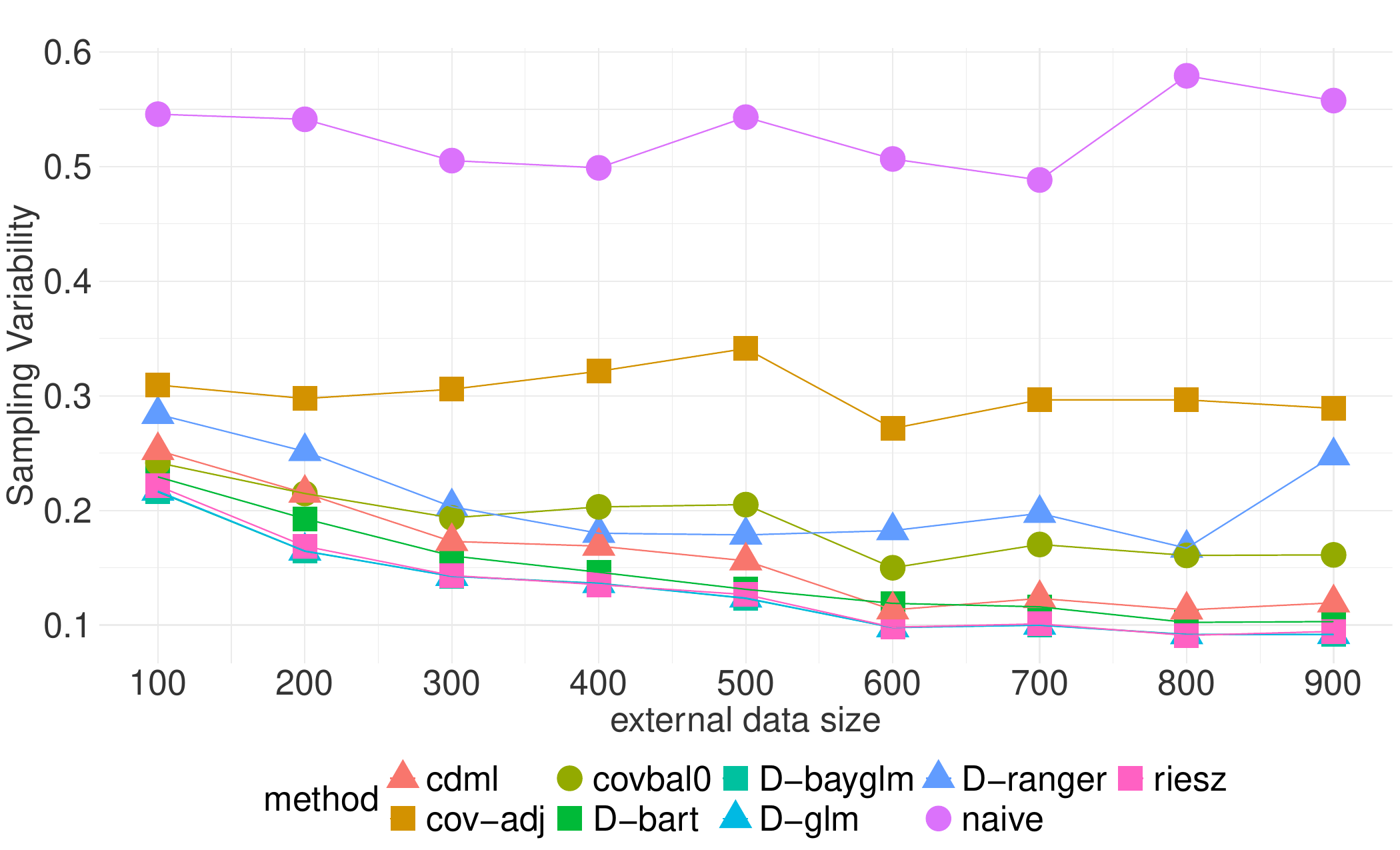}
\end{center}

\end{figure}

 Figure \ref{fig:biasvarv1} shows the mean absolute bias and sampling variability for $\satett(v=1,1)$ as a function of the external data size. All estimators that incorporate external data exhibit decreases in both bias and variance. The observed reduction in finite-sample bias is expected, as such bias can occur even for consistent estimators but typically diminishes with increasing sample size. All of the estimators appear to maintain nominal coverage of the $95\%$ confidence interval across varying sizes of the external data set, which can be seen in either Figure \ref{fig:covv1} or Table \ref{table_cov1}. Monte Carlo SE results are presented in the appendix (Table \ref{table_mc0} and Table \ref{table_mc1}).

\clearpage
\clearpage

\begin{figure}[]
\begin{center}
\includegraphics[scale=0.4]{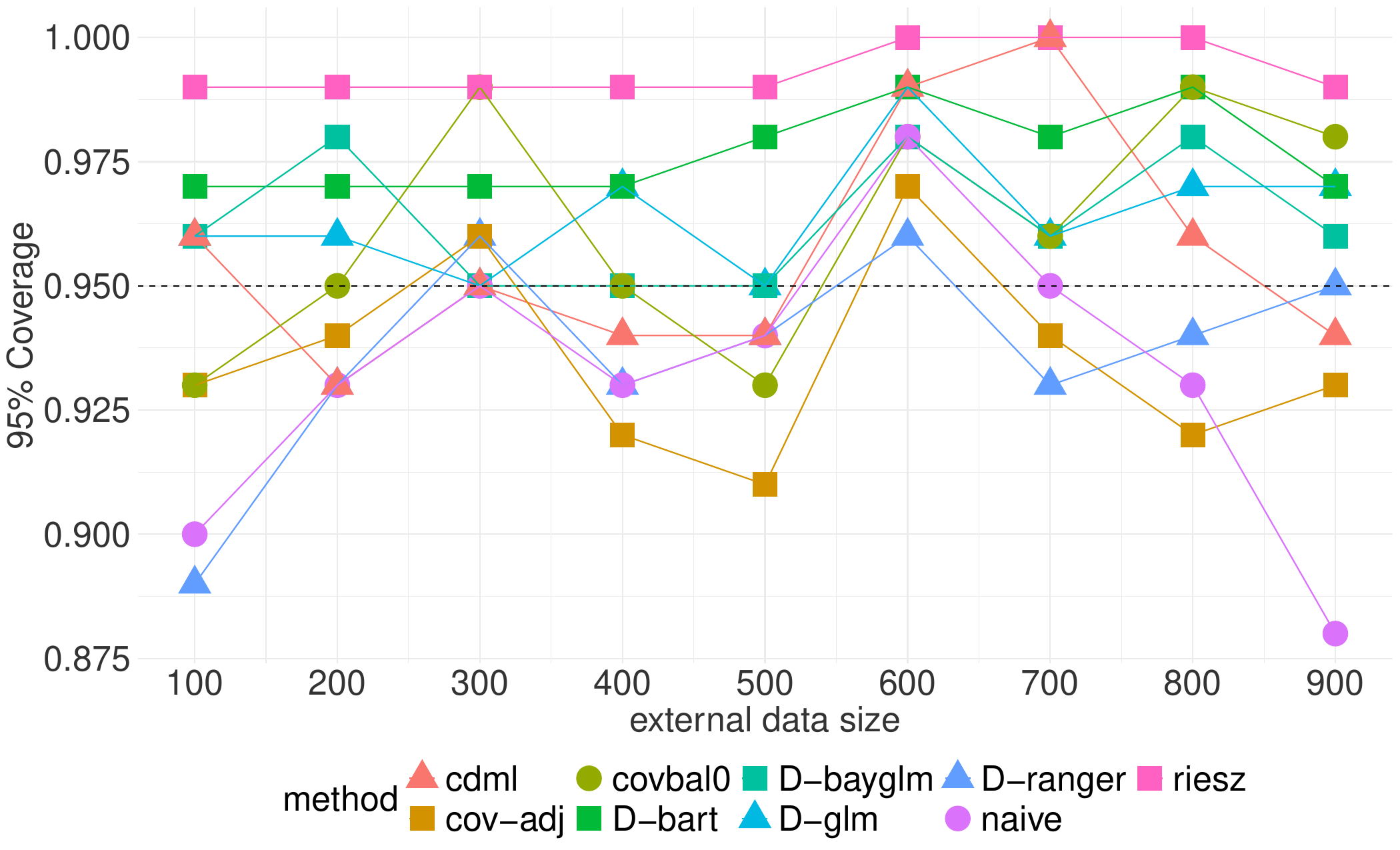}
\end{center}
\caption{Coverage of the $95\%$ confidence interval of the estimators in Table \ref{table_methods} as a function of external data size for the subgroup $V=1$.\label{fig:covv1}}
\end{figure}

\begin{table}[]
\centering
\begin{tabular}{lccccccccc}
 
                                             \multicolumn{10}{c}{\textbf{External data size}} \\  
\hline
 \textbf{Method} & \textbf{100} & \textbf{200} & \textbf{300} & \textbf{400} & \textbf{500} & \textbf{600} & \textbf{700} & \textbf{800} & \textbf{900} \\
\hline
\texttt{cov-adj}    & 0.92 & 0.96 & 0.97 & 0.93 & 0.95 & 0.97 & 0.95 & 0.92 & 0.96 \\
\texttt{D-glm}     & 0.95 & 0.94 & 0.99 & 0.94 & 0.94 & 1.00 & 0.96 & 0.96 & 0.97\\ 
\texttt{D-bayglm}  & 0.97 & 0.95 & 0.99 & 0.96 & 0.99 & 0.99 & 0.98 & 0.96 & 0.98 \\
\texttt{D-ranger}  & 0.91 & 0.95 & 0.98 & 0.92 & 0.88 & 0.98 & 0.91 & 0.93 & 0.90  \\
\texttt{D-bart}    & 0.98 & 0.97 & 1.00 & 0.97 & 1.00 & 0.99 & 1.00 & 0.99 & 1.00  \\
\texttt{covbal0}    & 0.95 & 0.96 & 0.97 & 0.93 & 0.97 & 1.00 & 0.97 & 0.97 & 0.98 \\
\texttt{riesz}      & 1.00 & 0.98 & 1.00 & 0.98 & 1.00 & 1.00 & 1.00 & 0.99 & 1.00 \\
\texttt{cdml}       & 0.96 & 0.94 & 0.95 & 0.89 & 0.96 & 0.98 & 0.95 & 0.93 & 0.97 \\
\texttt{naive}      & 0.95 & 0.91 & 0.98 & 0.92 & 0.93 & 0.96 & 0.92 & 0.95 & 0.92 \\
\hline
\end{tabular}
\caption{Coverage of the $95\%$ confidence interval of the estimators in Table \ref{table_methods} as a function of external data size for the subgroup $V=1$. \label{table_cov1}}
\end{table}
\clearpage

\subsubsection{Scenario 2 - Positivity Violations}

Table~\ref{table_ppv} reports the mean absolute bias, sampling variability and coverage across estimators under practical positivity violations (PPV). The datasets used in this simulation were chosen such that the maximum estimated IPW weight was at least $50$. Estimators that directly plug in inverse probability weights, whether using parametric models (\texttt{D-glm} and \texttt{D-bayglm}) or data-adaptive approaches (\texttt{D-ranger} and \texttt{D-bart}), suffer from large weights and exhibit greater sampling variability compared to the naive estimator. This suggests that under PPV, even with correctly specified models, the potential gains in precision are offset by the instability introduced by extreme weights. The covariate adjustment estimator (\texttt{cov-adj}), which uses only target trial data, performs similarly to the naive estimator, indicating that the covariates may not explain substantial outcome variation, though in our DGP they were set to have a moderate effect (i.e., $1.5 \cdot \Xt$). 

\begin{figure}[]
\begin{center}
\includegraphics[scale=0.45]{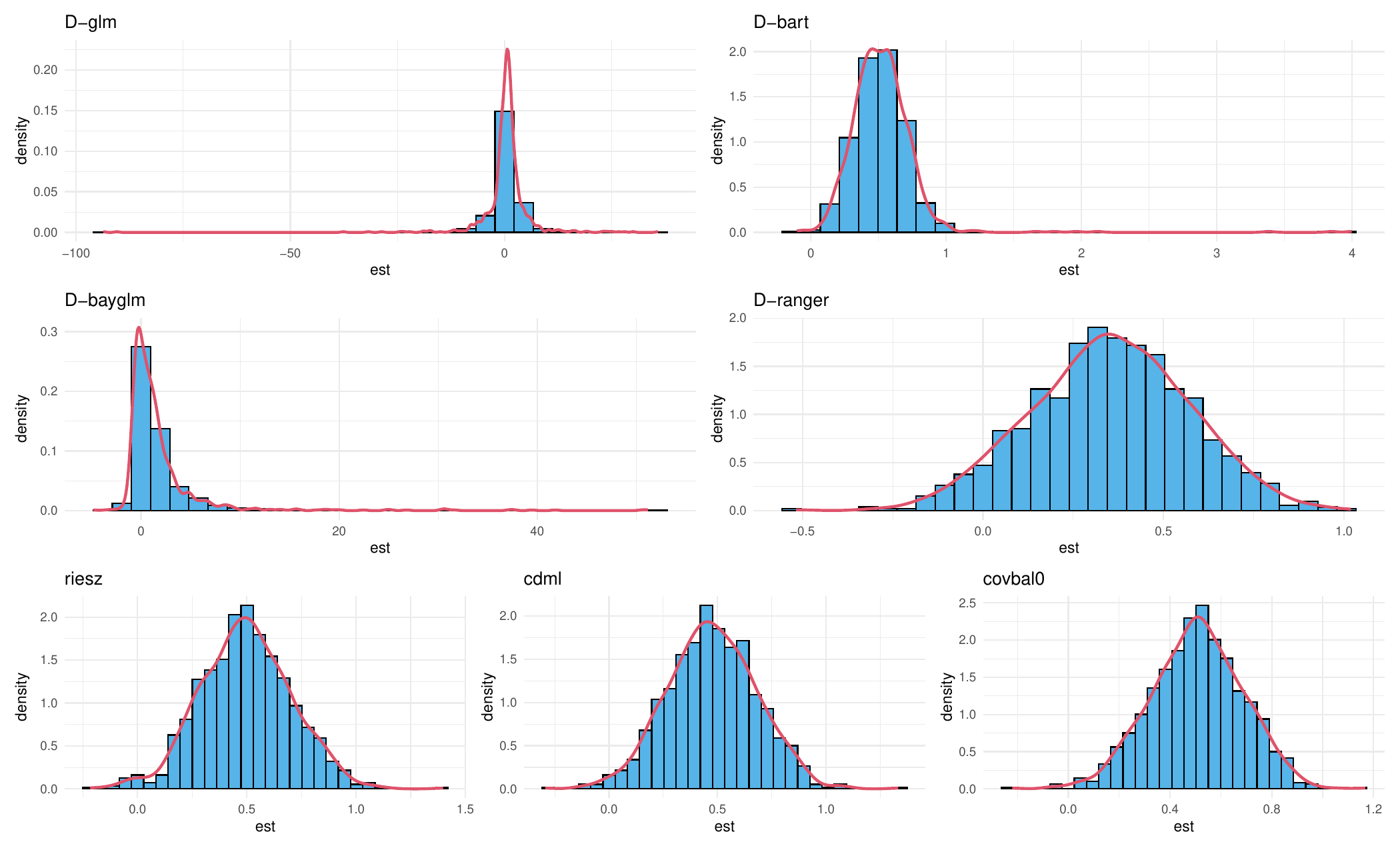}
\end{center}
\caption{Estimated sampling distribution of subgroup specific treatment effect estimates for subgroup $V=1$ by method. Here we exclude the naive and covariate adjustment estimator since they are unaffected by practical positivity violations in the external data source. \label{fig:ppvdens}}
\end{figure}

As expected, methods that adaptively learn or calibrate weights—such as \texttt{covbal0}, \texttt{riesz}, and \texttt{cdml}, demonstrate improved precision. The increase in bias observed in both the parametric (\texttt{D-glm}, \texttt{D-bayglm}) and machine learning-based (\texttt{D-ranger}, \texttt{D-bart}) debiased estimators is attributable to high variance, which in turn leads to unstable point estimates of $\satett(v,1)$. Figure \ref{fig:ppvdens}  summarizes the sampling distribution of each estimator which incorporates inverse probability weights. Those techniques which target the PPV setting (\texttt{riesz}, \texttt{cdml}, \texttt{covbal0}) have sampling distributions which appear to be approaching a normal distribution while those which do not (\texttt{D-ranger}, \texttt{D-bart}, \texttt{D-glm}, \texttt{D-bayglm}) are highly skewed.

\begin{table*}[]
\caption{Mean absolute bias (MAB), variance and coverage of the $95\%$ confidence interval for each method in subgroup $V=1$ under practical positivity violations. \label{table_ppv}}
\vspace{5mm}
\centering
\begin{tabular}{@{}lccc@{}}
\hline
\multicolumn{1}{c}{\textbf{Method}} & \multicolumn{1}{c}{\textbf{MAB}} & \multicolumn{1}{c}{\textbf{Variance}} & \multicolumn{1}{c}{\textbf{Coverage}}  \\
\hline
\texttt{cov-adj}   & 0.36  & 0.43   &  0.93 \\
\texttt{D-glm}    & 21.11 & 60.13  &  1.00\\
\texttt{D-bayglm} & 13.54 & 39.78  &  1.00 \\
\texttt{D-ranger} & 0.45  &  0.75  &  0.79 \\
\texttt{D-bart}   & 0.94  & 6.45   &  1.00 \\
\texttt{covbal0}   & 0.27  & 0.32   &  0.95 \\
\texttt{riesz}     & 0.24  & 0.29   &  1.00 \\
\texttt{cdml}      & 0.32  & 0.380  &  0.83 \\
\texttt{naive}     & 0.39  &  0.46  &  0.94\\
\hline
\end{tabular}
\end{table*}
\clearpage

\subsubsection{Scenario 3 - Model Misspecification}

\begin{figure}[]
\caption{Mean absolute bias of the estimators in Table \ref{table_methods} under different model misspecification scenarios. The sampling variability of the same estimators
under the same misspecification scenarios (bottom) for the subgroup $V=1$.\label{fig:missbiasvarv1}}
\begin{center}
\includegraphics[scale=0.4]{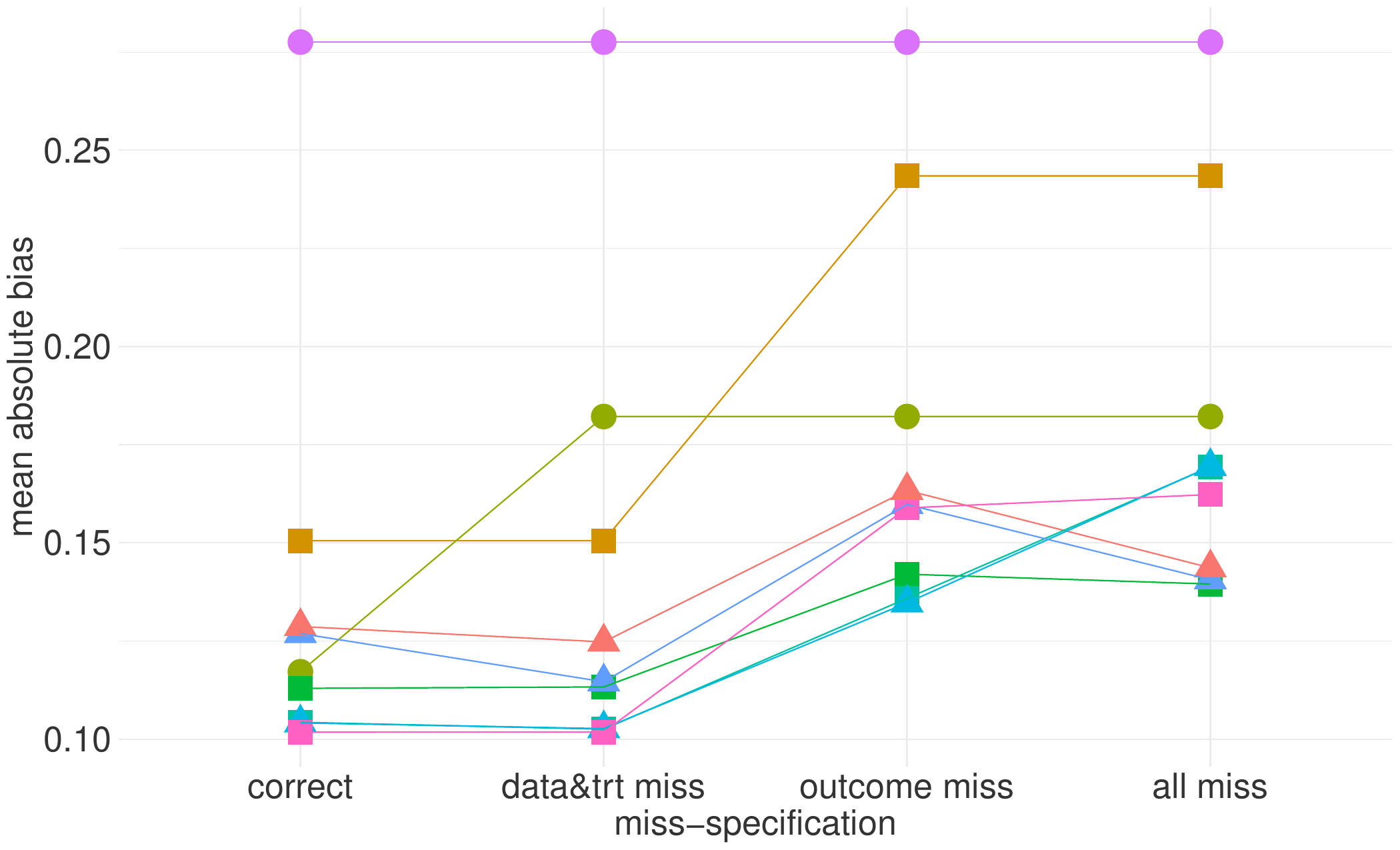}
\includegraphics[scale=0.4]{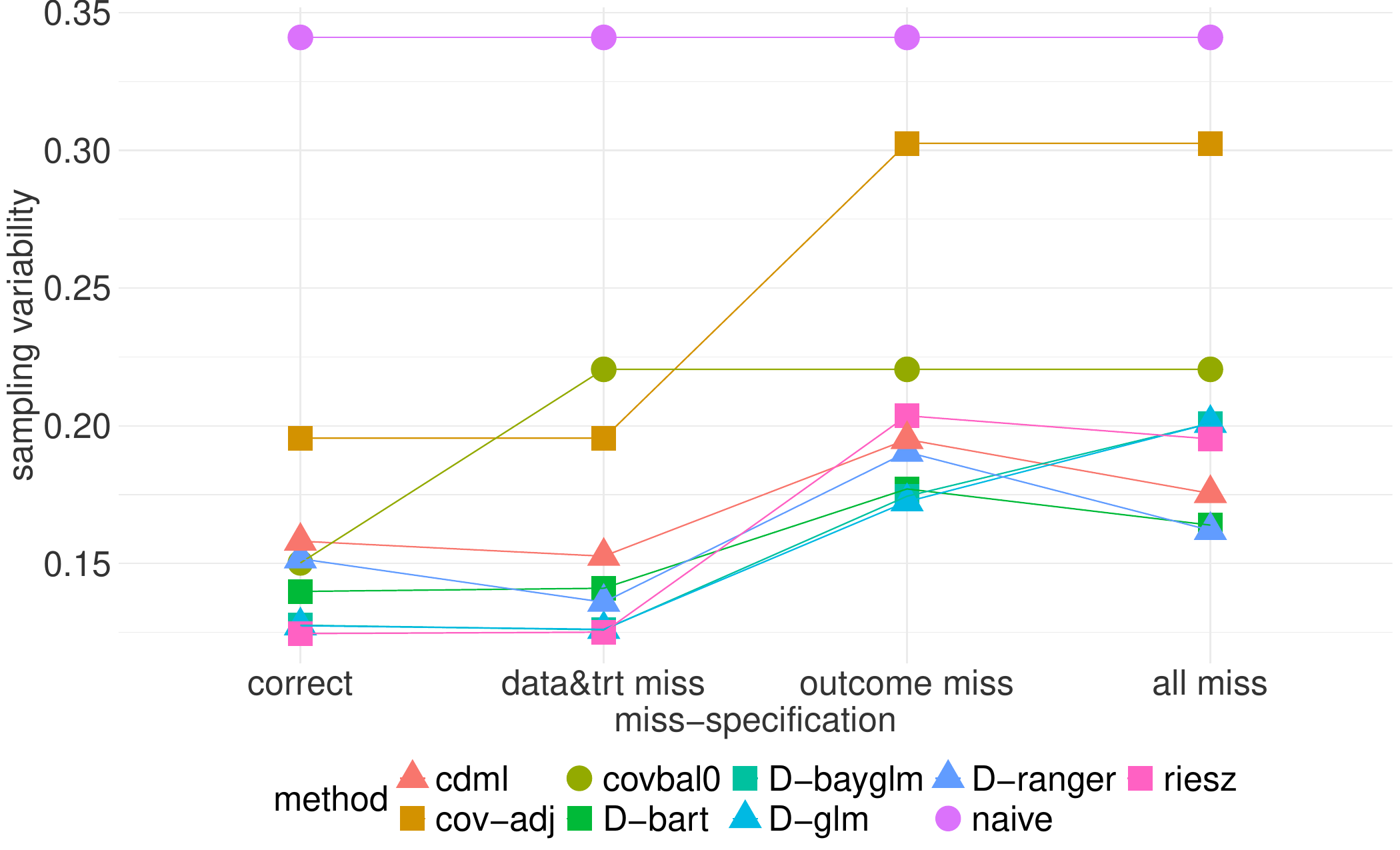}
\end{center}
\end{figure}

Figure ~\ref{fig:missbiasvarv1} displays mean absolute bias and sampling variability of each estimator in Table \ref{table_methods} across different model misspecification settings. Relative to the naive estimator, debiased estimators perform better with respect to bias and variance regardless of which combination of nuisance models are misspecified. The covariate balancing estimator does experience an increase in bias and variance once misspecification is introduced. When the outcome model is misspecified the covariate adjustment estimator shows notable loss in performance across all three metrics. This drop in performance could be due to the limitation in overall sample size compared to the other methods. For the covariate adjustment estimator the effective sample size available in estimation is half that of the other methods (excluding the naive). The coverage of the $95\%$ confidence interval for each method is provided in Table \ref{table_missv1}. Additional performance results for subgroup $V=0$ can be found in the Appendix.

\subsubsection{Summary of results} 

Methods which borrow data improve efficiency as the size of the external source grows. However, using these methods requires a trade-off in assumptions, i.e. researchers must make additional causal assumptions which are untestable. Debiased estimators provide consistent estimates with relatively small increases in bias and variance. 
In addition, estimators that deal with practical positivity violations 
provide more stable subgroup specific treatment effect estimates, compared to standard plug-in approaches.     

\clearpage

\begin{table*}[]
\caption{Coverage of the $95\%$ confidence interval for each method by misspecification scenario in subgroup $V=1$. In the first column all models are correctly specified, in the second column the data and treatment models are misspecified, in the third column only the outcome model is misspecified, in the forth column all models are misspecified. \label{table_missv1}}
\vspace{5mm}
\centering
\begin{tabular}{@{}ccccc@{}}
\hline
\multicolumn{1}{c}{\textbf{Method}} & \multicolumn{1}{c}{\textbf{all correct}} & \multicolumn{1}{c}{\textbf{data \& treatment miss}} & \multicolumn{1}{c}{\textbf{outcome miss}} & \multicolumn{1}{c}{\textbf{all miss}} \\
\hline
\texttt{cov-adj}      & 0.95 & 0.95 & 0.97 & 0.97 \\
\texttt{D-glm}       & 0.96 & 0.98 & 0.99 & 0.96 \\
\texttt{D-bayglm}    & 0.99 & 0.96 & 0.99 & 0.95 \\
\texttt{D-ranger}    & 0.96 & 0.97 & 0.98 & 0.97 \\
\texttt{D-bart}      & 0.96 & 0.98 & 1.00 & 0.97 \\
\texttt{covbal0}      & 0.94 & 0.96 & 0.96 & 0.96 \\
\texttt{riesz}        & 1.00 & 1.00 & 0.98 & 1.00 \\
\texttt{cdml}         & 0.91 & 0.93 & 0.94 & 0.93 \\
\texttt{naive}        & 0.96 & 0.96 & 0.96 & 0.96 \\
\hline
\end{tabular}
\end{table*}

\clearpage
\section{Practical considerations} 
\label{sec:practical_guidelines}

In this section, we discuss some practical considerations for the implementation of the proposed estimators.

\textbf{Standard errors, confidence intervals and hypothesis testing.}  To estimate the variance (and consequently the standard error) of the proposed estimators, we recommend relying on the existing literature for each method (\citep{Kennedy2022, hirshberg2017augmented, Chern2022, Antonelli2022, Lars2024}). Specifically, for \texttt{cov-adj}, \texttt{D-glm}, \texttt{D-ranger}, \texttt{covbal0}, and \texttt{riesz}, we consider  estimating the variance by computing the empirical variance of the estimated EIF, evaluated using the respective set of weights (see \citep{Kennedy2022} for \texttt{cov-adj}, \texttt{D-glm}, and \texttt{D-ranger}; \citep{hirshberg2017augmented} for \texttt{covbal0}; and \citep{Chern2022} for \texttt{riesz}. For \texttt{D-bayglm} and \texttt{D-bart}, we consider following the algorithm proposed by \cite{Antonelli2022}, which accommodates the use of Bayesian methods for estimating nuisance functions while maintaining desirable frequentist properties. We note that this estimator has been shown to be theoretically conservative. For confidence intervals, researchers may consider constructing Wald-type intervals based on asymptotic normality for all methods except \texttt{cdml}. For \texttt{cdml}, we suggest using the automated bootstrap-assisted procedure described in \cite{Lars2024}, as mentioned above. Finally, we propose conducting hypothesis testing for all methods using Wald tests based on the respective estimated standard errors. \\

\textbf{Which learners to use for debiased estimators?} The debiased estimators proposed in this paper depend in practice on the choice of learners. For non-Bayesian techniques, we suggest either canonical parametric generalized linear models or random forest, as a flexible nonparametric alternative. We suggest using parametric models when there is strong prior knowledge supporting correct model specification; otherwise, more flexible approaches such as random forest may be preferred. Alternatively, ensemble learning methods such as the super learner \citep{van2007super} may be employed.

When the total sample size is relatively small, it may be difficult to leverage flexible, data-driven methods like random forest, and parametric models may be preferred for greater stability. In settings without practical positivity violations, we suggest considering Bayesian methods. When prior knowledge regarding the tree structure (e.g., favoring shallow trees), the distribution of leaf node values (e.g., centered at zero with small variance), and the hyperparameters governing tree depth and leaf regularization is available, BART can be appropriately applied, leveraging its Bayesian framework to balance flexibility and regularization. Similarly, when prior knowledge regarding the regression model, the distribution of regression coefficients (e.g., centered priors with specified variance), and the likelihood model (e.g., Gaussian for continuous outcomes or binomial for binary outcomes) is available, Bayesian generalized linear models can be appropriately applied. In our simulations we selected a t-distribution with scale and degrees of freedom both equal to $100$ for the regression coefficients. 

For \texttt{covbal0}, we suggest using Gaussian process regression as described in detail in the section below. For \texttt{riesz}, we suggest learning the nuisance functions using generalized linear models and setting the algorithm space denoted $\mathcal{A}$ to be the space of neural networks. To compute estimates given this choice for $\mathcal{A}$ we suggest using either R package \texttt{SuperRiesz} or \texttt{RieszNet}. If enough sample data is available to allow for cross-fitting it is possible to consider setting $\mathcal{A}$ to be the set of random forests, and using the \texttt{ForestRiesz} package in R. This alternative which was shown in \cite{chernozhukov2021automatic} to have performance gains, but we note that the \texttt{ForestRiesz} package specifically was not uded in our simulation. \\


\ignore{
Learning the Riesz representer as described in \ref{sec:autodml} required minimizing the loss function \ref{autodmlmin} over an algorithm space denoted $\mathcal{A}$ which needs to be specified by the researcher. In this work we used the R package \texttt{SuperRiesz} which sets $\mathcal{A}$ as the space of neural networks. Alternatively one may have chosen the algorithm space to be the set of random forests and estimated the Riesz representer using the R package \texttt{ForestRiesz}. In \cite{chernozhukov2021automatic} the authors explore using both neural networks and random forest in their simulated experiments and real world application, obtaining comparable results. The authors do note that the forest based approach did see performance gains from cross fitting, a technique which we did not employ in this work for reasons we will explain in a subsequent section. \\
}

\textbf{Details on the choice of the kernel and the tuning of the kernel's hyperparameters.} The covariate balancing estimator depends on the choice of the kernel and the tuning of its hyperparameters. As proposed previously \citep{kallus2021more,kallus2021optimal,kallus2022optimal}, we generally propose to use a polynomial kernel:
\begin{equation}\label{polykernel}
\mathcal{K}_a(z, z') = C \cdot \left( (z^\top z')^{d_1} \right) + \sigma_a^2 \, \delta^\ast(z, z')
\end{equation}
\noindent 
where $C$ is the constant term, $d_1$ is the degree of the kernel, $\sigma_a^2$ is the noise variance, and $\delta^\ast(z, z')$ is the Kronecker delta function. In this paper, we set $d_1 = 1$ and treat $C$ and $\sigma_a^2$ as hyperparameters.  We model the data using Gaussian Processes (GPs), which are non-parametric Bayesian models that define a distribution over functions \citep{rasmussen2003gaussian}. A GP is fully specified by its mean function and covariance function (kernel), where the latter encodes assumptions about the function's smoothness and structure. In our case, the kernel is defined by Equation~\eqref{polykernel}. Specifically, we suppose $m(1,\Xt, \Vb)$ and $m(0,\Xt, \Vb)$ come from a GP with kernels $\mathcal K_1,\mathcal K_0$ and that each $Y_i$ was observed from $m(Z_i)$ with Gaussian noise of variance $\sigma^2_a$.
The hyperparameters $C$ and $\sigma_a^2$ are learned through maximum likelihood estimation (MLE) during model fitting. Specifically, we optimize the log marginal likelihood of the observed data, which measures how well the model explains the data after integrating over all possible functions that the GP could represent. This optimization balances model fit and complexity by trading off data fidelity against model smoothness. In other words, it seeks hyperparameters that allow the model to capture the underlying patterns in the data (avoiding underfitting) while preventing the model from fitting noise (avoiding overfitting). The term $C$ controls the overall variance explained by the polynomial component of the kernel, while $\sigma_a^2$ captures observation noise. This balance is an inherent property of the GP framework, as the log marginal likelihood naturally penalizes overly complex models while rewarding good data fit.   This method is implemented in the \textsf{Python} library \textsf{sklearn.gaussianprocess}. It may be helpful to set the penalization parameter $\lambda$ to a small value, such as 0.01 (primarily to control bias), but not larger than the logarithm of the number of covariates \citep{hirshberg2017augmented}. However, when the number of subjects in the subgroup of interest is small, increasing the value of the penalty term may help to improve efficinecy gains. For example, in our simulation scenario 1 the subgroup defined by V=0, on any given iteration on has around 30 members across both arms within the target trial. In this case covariate balancing has difficulty gaining power even as the external data source grows in size. Increasing the penalty term can help to resolve this problem (at the expense of bias), see Figure \ref{fig:powv0} and Figure \ref{fig:powv0cov1} in the Appendix. \\

\textbf{Which calibration techniques to use for CDML?} CDML depends on the choice of learners for both the nuisance function estimation and the calibration step. For the nuisance functions, we recommend following the same guidelines discussed previously. For calibration, we suggest following the approach originally outlined in \cite{Lars2024}. Specifically, we recommend using gradient-boosted decision trees (GBDT), implemented via the \texttt{xgboost} package in R, to learn the isotonic regression model used for calibration. The use of machine learning requires selecting hyperparameters that control the complexity of the resulting model. In the case of GBDT, these include parameters controlling the depth of the tree (\texttt{max\_depth}) and the minimum number of observations required in each node (\texttt{min\_child\_weight}), among others. In our simulations the hyper-parameters \texttt{max\_depth} and \texttt{min\_child\_weight} were fixed at $15$ and $20$ respectively. \\

\textbf{Cross-fitting and Donsker conditions.} All of the estimators considered in this work, with the exception of the \texttt{naive} method, require learning nuisance functions from the data. The same data used to estimate the requisite nuisance functions is also used to compute the final subgroup-specific treatment effect estimates. Employing the same data in this way introduces the risk of overfitting, which can bias the second-stage estimation. There are two primary strategies to address this issue: cross-fitting or invoking so-called Donsker conditions. In brief, cross-fitting is a sample-splitting technique in which the collected data are divided into subsamples; nuisance functions are estimated on one subsample and used to compute the target quantity on a different subsample, and the results are then averaged. For a detailed discussion, we refer the reader to Section 4.2 of \cite{Kennedy2022}. Alternatively, one can assume that the class of estimated nuisance functions satisfies a Donsker property,  that is, the set of functions has limited complexity, ensuring sufficiently fast empirical process convergence. Under such conditions, it is possible to use the same data for both nuisance estimation and target estimation without introducing overfitting bias. For standard GLMs, the Donsker property holds. However, for more flexible, data-driven methods, it may not hold, and cross-fitting is therefore recommended. When the sample size is sufficiently large, we recommend using cross-fitting to mitigate overfitting bias. In contrast, when the sample size is relatively small, such that even using two folds would result in very limited data in each fold, potentially leading to practical positivity violations, researchers may consider foregoing cross-fitting and instead relying on the Donsker property to justify estimation. If the Donsker property does not hold and cross-fitting is not employed, the usual sufficient conditions for asymptotic linearity are not guaranteed, the resulting estimators may suffer from bias, and standard errors or confidence intervals may fail to achieve nominal coverage, as for example using random forest for which Donsker does not hold. Importantly, the practical impact of violating Donsker-type conditions depends on the complexity of the underlying nuisance functions: when the true nuisance models are relatively simple (as in our simulation design), flexible learners may still approximate them well, and thus we would not expect substantial bias; consistent with this, in Section 5.2.1 we do not see much empirical bias. This should not be interpreted as evidence that Donsker conditions (or cross-fitting) are unnecessary in general. \\



\textbf{Recommendations for estimator selection.} If baseline covariates are collected in the target RCT, which are known to be prognostic or strong predictors of the outcome,it may be beneficial to use a covariate adjustment based estimator like the one described in Section \ref{drcovadj}. Although this estimator does not leverage external data, it also does not require invoking the additional assumptions \ref{ass:wexch} and \ref{ass:sexch}. Both of these assumptions are untestable in practice and if they are violated may introduce bias into resulting treatment effect estimates. 

Alternatively, if the baseline predictors in the target RCT are weak and researchers believe assumptions \ref{ass:wexch} and \ref{ass:sexch} are reasonable based on subject specific knowledge, then using an estimator which borrows data may be worth considering.  
In this case we suggest one in which practical positivity violation is controlled, like calibrated de-biased machine learning. The \texttt{cdml} estimator performs well in simulations where extreme positivity violations are present and is fairly straight forward to implement using standard R packages. \\


\textbf{Discussion of the effect of random seeds on estimation.} Using machine learning algorithms for effect estimation can induce a dependence on random number seeds, i.e. the results one obtains may change based on what seed is chosen. Under scenario 1 of our simulation study we investigated the possible effects of varying the random seed prior to estimation on each dataset,  following proposed solution number two in \cite{schader2024}.  Specifically, we considered two settings where a) the random number seed was fixed at the start and remained the same across all 1000 datasets;
b) the random seed was changed with each dataset 1,...,1000, before the estimation step, i.e. the RNG seed was changed 1000 times and then averaged the final estimate.  
In summary, the confidence intervals (Figures \ref{fig:civ1},\ref{fig:civ0}, in the Appendix) and sampling distributions (Figures \ref{fig:densv1} and \ref{fig:densv0} in the Appendix) between a) and b) above seem to overlap, suggesting that under our simulation settings results did not depend heavily on the RNG seed. Similar conclusions were obtained for our case study. Applied researchers may consider varying the random seed at pre-specified increments to avoid possible issues with random seed dependence, as shown in our additional analyses of the case study in the appendix. Further helpful discussion may be found in \cite{zivich2025} and \cite{naimi2024}.

\section{Case Study}\label{sec:casestudy}

\subsection{The DECIFER trial}

Small randomized clinical trials, which are common in psychiatry, often lack sufficient power for subgroup analyses aimed at detecting treatment effect heterogeneity. Our case study uses data from the DECIFER trial \citep{goff2019citalopram}, a NIMH-funded randomized clinical trial (RCT) that evaluated the effectiveness of citalopram over 12 months on negative symptoms and subsyndromal depressive symptoms in patients with first-episode schizophrenia (FES). The primary outcome of interest was the change in the Calgary Depression Scale for Schizophrenia (CDSS) score between baseline and 52 weeks. The DECIFER trial enrolled 95 participants, with 52 completing the 12-month assessment. In this trial, researchers observed differences in the magnitude and direction of citalopram’s effect on depressive symptoms across subgroups defined by duration of untreated psychosis (DUP), although these effects did not reach statistical significance, likely due to the small sample size and limited power. DUP is a clinically important prognostic factor in early psychosis, with longer durations associated with worse outcomes, motivating its use as a stratifying variable. Among the 52 participants who completed the 12-month assessments, 28 had a DUP below 18 weeks, while 23 had a DUP of 18 weeks or longer. As an external data source, we used the Recovery After an Initial Schizophrenia Episode (RAISE) trial \citep{kane2015raise}, which evaluated the effectiveness of coordinated specialty care for individuals with FES. The RAISE trial included 159 complete-case patients. Demographic information, including age, sex, and race, collected in both trials was used as covariates in our analysis. DUP, the use of add-on citalopram and CDSS at 52 week were also recorded in the RAISE trial. More details on the demographics of these two populations are provided in the Appendix.

\subsection{Intervention/treatment and endpoint}

We evaluate the effect of add-on citalopram versus not add-on citalopram on the change in CDSS score between baseline and 52 weeks. 

\subsection{Estimators}

We deployed the proposed estimators, \texttt{naive}, \texttt{cov-adj}, \texttt{D-glm},
\texttt{D-bayglm}, \texttt{D-ranger}, \texttt{D-bart}, \texttt{covbal0}, \texttt{riesz}, and \texttt{cdml} following the recommendations discussed in section \ref{sec:practical_guidelines}.

\subsection{Results}



\begin{table*}
\caption{Difference in mean CDSS within the patient subgroup defined by DUP. The column \textbf{SEn/SE} is the ratio of the standard error of the naive estimator over the standard error of each method, a ratio greater than 1 indicates an increase in precision. \label{table_applied}}
\label{postable}
\vspace{5mm}
\begin{tabular}{@{}lcccc|cccc@{}}
\hline
\multicolumn{5}{c|}{ \textbf{DUP $\geq 18$}} & \multicolumn{4}{c}{ \textbf{DUP $< 18$}} \\
\hline
\textbf{Method} & \textbf{Est} & \textbf{SE} & \textbf{CI} & \textbf{SEn/SE} & \textbf{Est} & \textbf{SE} & \textbf{CI} & \textbf{SEn/SE} \\
\hline
\texttt{cov-adj}   & 1.25 & 0.80  &  (-0.32 , 2.82)   &  1.01  & -0.88 & 0.89  & (-2.63 , 0.88)  &  1.07 \\
\texttt{D-glm}    & 1.96 & 0.65  &  (0.65 , 3.26)    &  1.25  & -0.80 & 0.67  & (-2.11 , 0.50)  &  1.43 \\
\texttt{D-bayglm} & 1.96 & 0.75  &  (0.49 , 3.42)    &  1.08  & -0.75 & 0.72  & (-2.16 , 0.67) &  1.33 \\
\texttt{D-ranger} & 1.46 & 0.58  &  (0.33 , 2.59)    &  1.39  & -0.70 & 0.55  & (-1.77 , 0.37) &  1.74 \\
\texttt{D-bart}   & 1.70 & 0.97  & (-0.19 , 3.60)    &  .83  & -0.60 & 0.77  & (-2.11 , 0.92) &  1.25 \\
\texttt{covbal0}   & 1.12 & 0.58  & (-0.02 , 2.26)    &  1.39  & -0.29 & 0.61  & (-1.49 ,  0.91) &  1.58 \\
\texttt{riesz}     & 1.68 & 0.89  & (-0.08  , 3.43)   &  .91  & -0.63 & 0.90  & (-2.40 , 1.14) &  1.07 \\
\texttt{cdml}      & 1.69 & 0.70  &  (0.32 , 3.06)    &  1.15  & -0.55 & 0.80  & (-2.12 , 1.02) &  1.2 \\
\texttt{naive}     & 1.15 & 0.81  & (-0.43  , 2.73)   &  1  & -0.75 & 0.96  & (-2.63 , 1.13) &  1 \\
\hline
\end{tabular}
\end{table*}

Table \ref{table_applied} shows the point estimates, standard errors, and 95\% confidence interval bounds for the difference in outcomes between DUP subgroups across methods. The \texttt{naive} estimator yielded point estimates of 1.15 for DUP $\geq$ 18 and $-0.75$ for DUP $<$ 18, with standard errors of 0.81 and 0.96, respectively, leading to confidence intervals that contained zero (i.e., non-significant results). The \texttt{cov-adj} estimator did not substantially improve the standard errors compared to the \texttt{naive} estimator (0.81 vs. 0.80 for DUP $\geq$ 18, and 0.96 vs. 0.89 for DUP $<$ 18), suggesting that the available covariates were not strong predictors of the outcome. Among the methods that borrow external data from the RAISE trial, most approaches improved precision for both subgroups, with the exception of \texttt{D-bart} and \texttt{riesz} for the DUP $\geq$ 18 subgroup (left panel of Table \ref{table_applied}). It is possible we see these results because methods like \texttt{D-bart} and \texttt{riesz} are more sensitive to small sample sizes and weak covariate-outcome associations, which can limit their ability to leverage external information effectively. For both subgroups, the point estimates from the data-borrowing methods were generally close to that of the \texttt{naive} estimator, at least within its confidence intervals. Although most estimates remained non-significant (with the exception of \texttt{D-glm}, \texttt{D-bayglm}, \texttt{D-ranger} and \texttt{cdml} for DUP $\geq$ 18 which may be due to higher variability), the resulting confidence intervals were notably narrower, suggesting that the use of larger external datasets or stronger prognostic covariates could potentially lead to statistically significant results in future analyses.

\section{Conclusion}
\label{sec:conclusion}

Researchers conducting RCTs are often interested in assessing treatment effect heterogeneity in pre-specified subgroups. However, RCTs frequently struggle to enroll a sufficient number of participants, which directly impacts the reliability of statistical analyses. Trials with low enrollment suffer from reduced statistical power and limited precision in effect estimates, typically reflected in wide confidence intervals. In this work, we draw on the causal inference literature and modern statistical techniques to propose a suite of methods designed to address these challenges.

We first propose a covariate-adjusted estimator that relies only on data from the RCT and requires minimal assumptions. By leveraging baseline covariates that are predictive of the outcome, this approach has the potential to improve precision. While straightforward to implement, its effectiveness is constrained by the sample size of the original trial and the availability of strongly prognostic baseline variables—information that may not always be accessible at the time of study design or analysis.

To address these limitations, we propose multiple alternative estimators that incorporate information from external datasets. These methods have desirable large-sample properties, allow flexible estimation of nuisance functions, and are empirically validated through simulation. However, like the covariate-adjusted estimator, they require tradeoffs. In particular, borrowing external data necessitates stronger assumptions for identification and estimation. The identification assumptions (\ref{ass:sexch}, \ref{ass:wexch}) are untestable, as they concern counterfactual distributions that are inherently unobservable.  We caution that while borrowing external data can improve precision, it also introduces the potential for bias. In our case study, we found that point estimates from data-borrowing methods were broadly consistent with those of the naive estimator, suggesting that external information may be used without substantially introducing bias.  Additionally, the positivity assumption (\ref{ass:posa}) may be violated in practice, especially when the external data source is observational. To address PPV, we introduced three novel estimators specifically designed to mitigate these issues. The techniques we propose for addressing practical positivity violations target the same parameter and the same population as the true inverse probability weights. They do so by approximating the true set of inverse weights in finite samples, and by converging to the true set asymptotically (see \cite{hirshberg2021augmented}, Theorem 1 for covariate balancing estimators; Section 2.3.1 of \cite{chern2021} for Auto-DML; and \cite{Lars2024} Theorem 1 for CDML). This contrasts with overlap weights or truncated weights, which change the target parameter (e.g., truncated weights apply truncation even with infinite data; see Table 1 of \cite{li2018balancing}). 

Our proposed covariate balancing estimator circumvents the plug-in step by learning the weights $\gamma(Z)$ directly. It does so via a worst-case strategy: choosing weights that minimize imbalance under the setting where the regression error $\delta(Z)$ is most unfavorable (if we had a model such that $\delta(Z)=0$, debiasing would not be needed). To control the complexity of the weights, we add a penalization parameter $\lambda$. Setting $\lambda=0$ minimizes bias without regard for weight complexity, whereas larger values of $\lambda$ improve precision at the cost of introducing bias. In the setting of targeting the ATT \citep{kallus2022optimal, kallus2020generalized, santacatterina2019optimal}, this trade-off corresponds to minimizing the conditional mean squared error (CMSE). Thus, the estimator minimizes worst-case CMSE. Importantly, these weights are nonetheless shown to converge to the true inverse weights. In our simulations and case study, we used small values of $\lambda$ (e.g., 0.001), primarily to control bias, and never larger than the logarithm of the number of covariates (see Section \ref{sec:practical_guidelines}, details on kernel choice and hyperparameter tuning). Since we consider using small values of $\lambda$, even in finite samples the method primarily targets bias and finds weights that minimize imbalance.

There is work on external controls that seeks to avoid relying exclusively on the exchangeability and positivity assumptions (identification assumptions \ref{ass:wexch} and \ref{ass:posa}), in particular,  the articles by \cite{dang2022}, \cite{yang2023} and \cite{karl2025}. Methods discussed by these authors are worth exploring as they may generalize to our problem setting while offering an alternative approach with different assumptions. Our work also assumes that the set of effect modifiers denoted $V$ is discrete and low dimensional. We suggest following \cite{kennedy2023} when $V$ is non-categorical or high dimensional. 

Our proposed framework assumes that both treatment assignment and the endpoint of interest are available in the external data. This may not always hold. For instance, outcomes may be recorded at different time points across datasets, or proxy outcomes may be observed instead. Treatment arms may also differ in definition. In our case study, we defined treatment as add-on citalopram use, which could be reasonably aligned across datasets. However, differences in treatment distributions between the target and external populations could still introduce heterogeneity, which must be considered in practice. 

Several questions remain open and warrant further investigation. Many of the data-adaptive methods we employ rely on hyperparameters that govern model complexity and influence estimation. In our analysis, we used default values for convenience. However, examining the effect of tuning these parameters—particularly via cross-validation—in settings with large external datasets could yield important insights. Additionally, while our work focuses on traditional two-arm parallel-group trials, future research could explore how these methods perform in more complex designs such as cluster-randomized or factorial trials or with different types of data, like repeated measurements, missing data and time-to-event.

\section*{Financial disclosure}

This article is based upon work supported by the National Science Foundation under Grant No 2306556,  National Institute of Health Grant No 1R01AI197146-01 and the Amazon Web Services (AWS) grant on ``AI/ML for Identifying Social Determinants of Health''.

\section*{Conflict of interest}

The authors declare no potential conflict of interests.

\section*{Supporting information}

Additional supporting information may be found in the online version of the article at the publisher’s website.

\bibliographystyle{agsm}

\bibliography{Bibliography-MM-MC, refs, refs_balancing}

\newpage
\setcounter{page}{1}
\setcounter{assumption}{0}
\begin{center}
{\large\bf SUPPLEMENTARY MATERIAL \\}
{\large\bf  Modern causal inference approaches to improve power for subgroup analysis in randomized controlled trials\\}
\vspace{5mm}
  \author{Antonio D'Alessandro, 
  Jiyu Kim, 
  Samrachana Adhikari, 
  Donald Goff, \\
  Falco J. Bargagli-Stoffi, Michele Santacatterina
    
    }
\end{center}

\section*{Identification of the causal parameter $\csatett$}



\begin{align*}
    &\Ex(Y(a) \mid V=v, S=s) \\ &= \Ex( \Ex(Y(a) | X=x, S=s) \mid V=v, S=s)  &&\text{$(\because$ iterated expectation)} \\
    &= \Ex( \Ex(Y(a) | A=a, X=x, S=s) \mid V=v, S=s)  &&\text{($\because$ Assumption~\ref{ass:wexch})} \\
    &= \Ex( \Ex(Y(a) | A=a, X=x) \mid V=v, S=s)  &&\text{($\because$ Assumption~\ref{ass:sexch})} \\
    &= \Ex( \Ex(Y | A=a, X=x) \mid V=v, S=s)   &&\text{($\because$ Assumptions~\ref{ass:cons},~\ref{ass:posa},~\ref{ass:poss})}
\end{align*}

\noindent
Note that in the third equality, we are still assuming Assumption~\ref{ass:wexch} in the form of $\E(Y(a) | A=a, X=x, S=s)$ for $S=0$, in addition to $S=1$. In other words, weak exchangiability needs to hold in the external population as well.

\section*{More details about the covariate balancing estimator}
\label{sec:cov_bal_derivation}
We here firstly provide more details on the error decomposition. 

\subsection*{Details about the error decomposition}

The error decomposition obtained in section \ref{sec:covbal} looks like that because we want to rewrite $\hat{\Psi}^{a}_{aipw}$ in terms of $\delta_{m_a}$, which leads
\begin{align*}
& \frac{\alpha^{-1}}{n} \sum^{n}_{i=1} 
 \I(V_i=v, S_i =1)( \delta_{m_{a}}(\Xt_i, \Vb_i) + m(Z_i) ) \\
 &= \frac{\alpha^{-1}}{n} \sum^{n}_{i=1}  \I(V_i=v, S_i =1) \hat m(Z_i)  \text{ and} \\
 &\frac{ \alpha^{-1} }{n}\sum^{n}_{i=1} \gamma(Z_i) \delta_{m(Z_i)} \\
 &= \frac{ \alpha^{-1} }{n}\sum^{n}_{i=1}\gamma(Z_i) (\hat m(Z_i) - m(Z_i) ) \\
 &=\frac{\alpha^{-1} }{n}\sum^{n}_{i=1} \gamma(Z_i)(\hat m(Z_i) - Y_i + \epsilon_i ) \\
 &=\frac{ \alpha^{-1} }{n}\sum^{n}_{i=1} \gamma(Z_i)(\hat m(Z_i) - Y_i  ) + \frac{ \alpha^{-1} }{n} \sum^{n}_{i=1} \gamma(Z_i)\epsilon_i 
\end{align*}
\noindent
The noise term, $\frac{\alpha^{-1}}{n} \sum_{i=1}^{n} \gamma(Z_i) \epsilon_i$, has mean zero due to $\E[\epsilon_{a,i} \mid Z_i] = 0$. Similarly, the sampling variation term $\frac{\alpha^{-1}}{n} \sum_{i=1}^{n} \I(V_i = v, S_i = 1)m(Z_i) - \Psi^a(m)$ averages out due to finite-sample randomness. Also note that the imbalance in $\delta_{m_a}$ 
\begin{align*}
    &\frac{ \alpha^{-1} }{n} \Big[\sum^{n}_{i=1} \I(V_i=v , S_i =1)\delta_{m}(Z_i) - \sum^{n}_{i=1}\gamma(Z_i)\delta_{m}(Z_i) \Big] \\
    &=
    \frac{ \alpha^{-1} }{n} \sum^{n}_{i=1} \Big[ \I(S_i =1,V_i =v) - \I(A_i =a,V_i =v) \gamma(Z_i)  \Big] \delta_{m}(Z_i),
\end{align*}
\noindent
can be interpreted as that we want to find weights that re-balance the regression errors among the $V=v$ observations in those treated with $A=a$ as those as in the trial population $V=v , S=1$. 

\section*{Derivation of the EIF for the estimator which leverages external data and uses flexible data-driven techniques}

We employ the $\mathbb{IF}$ operator and building block rules proposed by \cite{Kennedy2022} to derive the influence function. Note that in the following proof we use the following notational shorthand,  $m(a,x) = \Ex[ Y\mid A=a, X=x]$ and $\P(x) = \P(X=x)$. \\

Recall the identification result 
\begin{align*}
 \Psi = \Ex[\ Y(a) \mid \Vb=\vb, S=s] =  \Ex[ \Ex[Y \mid A=a, X=x] \mid V=v, S=s]
\end{align*}

We may find the influence function by applying the $\IF$ operator as follows, 
\begin{align*}
\IF(\Psi) &= \IF(\Ex[ \Ex[Y \mid A=a, X=x] \mid V=v, S=s])  \\
&=\IF\Big(\sum_x   \Ex[Y \mid A=a, X=x]\P(X=x\mid V=v, S=s)\Big) \\
&= \IF\Big(\sum_x m(a,x)p(x\mid v, s)\Big) \\
&= \sum_x \IF(m(a,x))p(x\mid v,s) + m(a,x)\IF(p(x\mid v,s))  &&\text{($\because$ $\IF$ linear \& product rule )}\\
\end{align*}
Now consider the first term, 
\begin{align*}
    \sum_x\IF(m(a,x)\P(x\mid v,s)= \\
    &= \sum_x \frac{\I(A=a, X=x) }{\P(a,x)}\{Y - m(a,x)\}\P(x\mid v, s) \\
    &= \sum_x \frac{\I(A=a, X=x) }{\P(a,x)}\{Y - m(a,x)\} \frac{\P(v,s \mid x)\P(x)}{\P(v,s)} \\
    &= \sum_x \frac{\I(A=a, X=x) }{\P(a,x)}\{Y - m(a,x)\}\frac{\I(V=v)\P(1\mid x)\P(x)}{\P(v,s)} \\
    &= \sum_x  \frac{\I(A=a, X=x)\I(V=v)}{\P(a \mid x) \P(x)\P(v,s)} \P(s\mid x)\P(x)\{Y - m(a,x)\} \\
    &= \sum_x \I(X=x) \frac{\I(A=a,V=v)}{\P(v,s)}\frac{\P(s\mid x)}{\P(a\mid x)}\{Y - m(a,x)\} \\
    &= \frac{\I(A=a,V=v)}{\P(v,s)}\frac{\P(s\mid X)}{\P(a\mid X)}\{Y - m(a,X)\}
\end{align*}
The second line on the right follows from an application of Bayes rule to $\P(x \mid v, s)$ while the third equality follows by using the substitution: 
\begin{align*}
\P(v, s \mid x) &= \\
&=\Ex[\I(v=v,S=s)\mid X=x] \\
&= \Ex[\I(v=v)\I(S=s)\mid X=x] \\
&= \I(V=v)\Ex[\I(S=s)\mid X=x] &&\text{($\because$ $V \subset X \implies \I(V=v)$ is known given $X=x$)}\\
& = \I(V=v)\P(s \mid x)
\end{align*}
Now consider the second term, 
\begin{align*}
    \sum_x m(a,x)\IF(\P(x\mid v, 1)) = \\
    &= \sum_x m(a,x)\IF (\Ex[\I(X=x) \mid V=v, S=s] ) \\
    &= \sum_x m(a,x) \Big( \frac{\I(V=v,S=s)}{\P(v,s)} \{\I(X=x) - \Ex[\I(X=x) \mid V=v, S=s] \} \Big) \\
    &= \sum_x \I(X=x)\frac{\I(V=v,S=s)}{\P(v,s)}m(a,x) - \P(x\mid v, s)\frac{\I(V=v,S=s)}{\P(v,s)}m(a,x) \\
    &= \frac{\I(V=v,S=s)}{\P(v,s)}m(a,X) - \frac{\I(V=v,S=s)}{\P(v,s)} \Psi \\
    & = \frac{\I(V=v,S=s)}{\P(v,s)} [m(a,X) - \Psi]
\end{align*}
Combining these two results we now obtain: 
\begin{align*}
\IF(\Psi) &= \frac{\I(A=a,V=v)}{\P(v,s)}\frac{\P(s\mid X)}{\P(a\mid X)}\{Y - m(a,X)\} +  \frac{\I(V=v,S=s)}{\P(V=v,S=s)} [m(a,X) - \Psi] \\
&= \frac{1}{\P(v,s)}\Big[ \I(A=a,V=v)\frac{\P(s\mid X)}{\P(a\mid X)}\{Y - m(a,X)\}  +\I(V=v,S=s) [m(a,X) - \Psi]                    \Big] 
\end{align*}

\section*{Derivation of the EIF for the covariate adjustment estimator}

Note the covariate adjustment technique relies on only data from the target trial ($S=1$). Under the standard causal inference assumptions of weak ignorability (A2.1), consistency (A2.3) and positivity (A2.4) the causal estimand $\Ex[\ Y(1) \mid \Vb=\vb]$ is identified by the following:

\begin{align*}
    \Psi = \Ex[\ Y(1) \mid \Vb=\vb, S=1] =\\ 
    &= \Ex[\ \Ex[\ Y(1) \mid \Vb=\vb, S=1, \Xt \ ] \mid \Vb=\vb, S=1] &&\text{($\because$ IE)} \\ 
    &= \Ex[\ \Ex[\ Y(1) \mid A=1, \Vb=\vb, S=1,  \Xt \ ] \mid \Vb=\vb, S=1] &&\text{($\because$ A2.1)} \\ 
    &= \Ex[\ \Ex[\ Y \mid A=1, \Vb=\vb, S=1, \Xt \ ] \mid \Vb=\vb, S=1] &&\text{($\because$ A2.3 + A2.4)} \\
    &= \Ex \left[ \ \frac{\mathds{1}{[\Vb=\vb, S=1]}}{\P(\Vb=\vb, S=1)}\Ex[\ Y \mid A=1, \Vb=\vb, S=1, \Xt \ ] \right] \\
\end{align*}

Let $m(1,\vb,1,\xt) = \Ex[\ Y \mid A=1, \Vb=\vb, S=1, \Xt=\xt \ ]$ and $\P(\xt)= \P(\Xt=\xt)$. We employ the $\mathbb{IF}$ operator and building block rules proposed by \cite{Kennedy2022} to derive the influence function. Using the above substitutions and pretending the data are discrete: 

\begin{align*}
    &\mathbb{IF} \left( \ \Ex[\ Y(1) \mid \Vb=\vb, S=1] ) \ \right) \\
    &= \mathbb{IF} \left(  \sum_{\xt} \frac{\mathds{1}{[\Vb=\vb, S=1]}m(1,\vb,1,\xt) \P(\xt)}{\P(\Vb=\vb, S=1)} \right) =\mathbb{IF}\left( \frac{n^1}{d} \right) \\
\end{align*}
Now  $\mathbb{IF}(n^1)=$
\begin{align*}
    &= \sum_{\xt} \mathds{1}{[\Vb=\vb, S=1]}\mathbb{IF} \left( \ m(1,\vb,\xt) \P(\xt) \ \right)  \\
    &= \mathds{1}{[\Vb=\vb, S=1]}\sum_{\xt}  \mathbb{IF}\left(m(1,\vb,\xt) \right)\P(\xt) + m(1,\vb,\xt)  \mathbb{IF}\left( \P(\xt) \right) \\
    &= \mathds{1}{[\Vb=\vb, S=1]} \sum_{\xt} \frac{\mathds{1}(A=1,\Xt=\xt,\Vb=\vb)}{\P(A=1\ | \ \Vb=\vb, \Xt=\xt  )\P( \Vb=\vb, \Xt=\xt  ) } \lbrace Y - m(1,\vb,\xt) \rbrace \P(\xt) \\
    &+ m(1,\vb,\xt)\lbrace \mathds{1}(\Xt=\xt) - \P(\Xt=\xt) \rbrace \\
    &= \mathds{1}{[\Vb=\vb, S=1]} \sum_{\xt} \frac{\mathds{1}(A=1,\Xt=\xt,\Vb=\vb)}{\P(A=1\ | \ \Vb=\vb, \Xt=\xt  )\P( \Vb=\vb \ | \ \Xt=\xt  ) } \lbrace Y - m(1,\vb,\xt) \rbrace \\
    &+ m(1,\vb,\xt)\mathds{1}(\Xt=\xt) - m(1,\vb,\xt)\P(\Xt=\xt) \\
    &= \mathds{1}{[\Vb=\vb, S=1]} \left[  \frac{A}{\P(A=1\ | \ \Vb=\vb, \Xt  )\P( \Vb=\vb \ | \ \Xt  ) }  \lbrace Y - m(1,\vb,\Xt) \rbrace + m(1,\vb,\Xt) \right] - n^1 
\end{align*}
\\

Under the indicator $\mathds{1}{[\Vb=\vb, S=1]}$  the term $\P(\Vb=\vb\ | \ \Xt  ) = 1 $ , it follows that  
\begin{center}
$\mathbb{IF}(n^1) = \mathds{1}{[\Vb=\vb, S=1]} \left[  \frac{A}{\P(A=a\ | \ \Vb=\vb, \Xt=\xt  )}  \lbrace Y - m(1,\vb,\Xt) \rbrace + m(1,\vb,\Xt) \right]  - n^1 $
\end{center}

\noindent
Next we compute $\mathbb{IF}(d) = \mathbb{IF}(\P(\Vb=\vb, S=1)) =  \mathds{1}(\Vb=\vb, S=1) - \P(\Vb=\vb, S=1) = \mathds{1}(\Vb=\vb, S=1) - d$. Using the above with the derivative quotient rule find the influence function of interest 

\vspace{4mm}

\noindent

$\mathbb{IF}\left( \frac{n^1}{d} \right)  =$
\begin{align*}
&=  \frac{\mathbb{IF}(n^1)}{d} - \frac{n^1}{d}\frac{\mathbb{IF}(d)}{d} \\
&= \frac{1}{d} \left[ \mathbb{IF}(n^1) - \frac{n^1}{d}\mathbb{IF}(d) \right] \\
&=\frac{1}{\P(\Vb=\vb, S=1)}\Big[ \mathds{1}{[\Vb=\vb, S=1]} \Big[  \frac{A}{\P(A=1\ | \ \Vb=\vb, \Xt ) }  \lbrace Y - m(1,\vb,\Xt) \rbrace + m(1,\vb,\Xt) \Big]\\ 
&-  n^1  -\frac{n^1}{d}\mathds{1}(\Vb=\vb, S=1) + \frac{n^1}{d}d \Big]\\
& = \frac{ \mathds{1}{[\Vb=\vb, S=1]}}{\P(\Vb=\vb, S=1)} \left[  \frac{A}{\P(A=1\ | \ \Vb=\vb, \Xt  ) }  \lbrace Y - m(1,\vb,\Xt) \rbrace + m(1,\vb,\Xt) - \frac{n^1}{d} \right]
\end{align*}

We can similarly compute $\mathbb{IF}\left(\frac{n^0}{d} \right)$ and find $\mathbb{IF}\left(  
\frac{n^1 - n^0}{d}\right) = \mathbb{IF}\left( \frac{n^1}{d}\right) - \mathbb{IF}\left( \frac{n^0}{d}\right) $

\section*{Details for running CDML}

As noted in section \ref{cdml} this technique benefits from using \textit{calibration}, a technique popular in machine learning. To calibrate predictions and estimate $\widehat{\satett}$ we used the R package \texttt{xgboost} to fit gradient-boosted decision tree (GBDT) models and the following procedure: 

\begin{enumerate}
\item With the $\texttt{ranger}$ package in R, and the complete dataset learn $\widehat{m}(1,\xt,v)$ and $\widehat{m}(0,\xt,v)$.
\item Estimate $\widehat{\pi}(\xt, v)$ and $\widehat{\eta}(\xt, v)$ using logistic regression and all available data.  
\item Generate predictions for all subjects using the models from steps 1 and 2. 
\item Calibrate the predicted conditional probability of treatment:
   \begin{enumerate}
       \item[i] Fit a GBDT with a monotonic constraint, regressing the treatment indicator $A$ on the predictions from $\widehat{\pi}(\xt,v)$.
       \item[ii] Input the predictions from $\widehat{\pi}(\xt,v)$ to the fit GBDT model to obtain $\pi^\ast(\xt,v)$. 
       \item[iii] Fit a GBDT with a monotonic constraint, regressing $1-A$ on $1-\widehat{\pi}(\xt,v)$.
       \item[iv] Input the predictions from $1-\widehat{\pi}(\xt,v)$ to the fit GBDT model to obtain $1-\pi^\ast(\xt,v)$.
   \end{enumerate}
\item Replace $A$ with $S$ and $\widehat{\pi}(\xt,v)$ with $\widehat{\eta}(\xt,v)$ and repeat step 4 to calibrate the predicted conditional probability of population membership. 
\item Calibrate predicted conditional outcomes within the treatment and control groups:  
   \begin{enumerate}
       \item[i] Fit a GBDT with a monotonic constraint, regressing the observed $Y$ values within the treatment group ($A=1$) on the predictions from $\widehat{m}(1,\xt,v)$.
       \item[ii] Input the predictions from $\widehat{m}(1,\xt,v)$ to the fit GBDT model to obtain $m^\ast(1,\xt,v)$.
       \item[iii] Repeat the above for $\widehat{m}(0,\xt,v)$ and the observed $Y$ within the control group ($A=0$) to obtain $m^\ast(0,\xt,v)$. 
   \end{enumerate}   
\item Compute $\widehat{\satett}_{Cal}$ using the calibrated predictions and the formula in \ref{cdml}.  
\end{enumerate}

Estimates of the standard error were obtained using a bootstrap algorithm described in \cite{Lars2024}: 

\begin{enumerate}
    \item Using the models from steps 1 and 2 in the calibration procedure, generate predictions  $\widehat{m}(1,\xt_i,v_i), \widehat{m}(0,\xt_i,v_i), \widehat{\eta}(\xt_i,v_i), \widehat{\pi}(\xt_i,v_i), 1-\widehat{\pi}(\xt_i,v_i)$ for all subjects $i$ and add to the original dataframe. 
    \item Sample with replacement from this new dataframe to create a bootstrapped dataset of size n. 
    \item Using the bootstrapped dataset, run steps 4-7 from the calibration procedure and store the estimated $\widehat{\satett}_{Cal}(v)$. 
    \item Repeat the above steps 2-3 until $B$ bootstrapped estimates of $\widehat{\satett}_{Cal}(v)$ are obtained. 
    \item Compute the sample standard error of the $B$ bootstrapped estimates. 
\end{enumerate}

\section*{Data generating processes for simulation scenarios 2 and 3}

\subsubsection*{The data-generating process for scenario 2} Here we explore the setting of a small two-arm trial ($n_{S=1} = 50$), with access to an external data source of moderate size ($n_{S=0}=500$), such that positivity violations exist in the external data. Similar to scenario 1 the total number of subjects is $n$ such that $n = n_{S=1} + n_{S=0}$. For each subject $i \in \{1,\ldots, n\}$ data were generated according to the below steps:   \label{sec:scen2}  

\begin{itemize}
    \item[] \textbf{Step 1.} For each subject $i$ randomly draw covariate $W_i$ such that $W_i \sim \text{N}(0,1)$.
    \item[] \textbf{Step 2.} For each subject $i$ randomly draw covariate $V_i$ such that $V_i \sim \text{Bernoulli}(0.5)$. 
    \item[] \textbf{Step 3.}  Let $\pi_i$ be the probability subject $i$ is enrolled in the target trial, compute this probability as $\eta_i$ = $.0909$. Then randomly draw population indicator $S_i$ such that $S_i \sim \text{Bernoulli}(\eta_i)$. 
    \item[] \textbf{Step 4.} If subject $i$ is in the target population ($S_i = 1$) randomly draw treatment indicator $A_i$ from $A_i \sim \text{Bernoulli}(0.5)$. 
    \item[] \textbf{Step 5.} If subject $i$ is in the external population ($S_i = 0$) let $\pi_i$ be the probability subject $i$ obtains treatment. Compute this quantity as $\pi_i= (1+\text{exp}(-0.045 + 9 * W_i + 9 * V_i))^{-1}$ and randomly draw treatment indicator $A_i$ from $A_i \sim \text{Bernoulli}(\pi_i)$.  
    \item[] \textbf{Step 6.} For subject $i$ compute their potential outcome under control ($A_i=0$) as $Y_i(0) = 1.5*W_i + 0.5*V_i + \epsilon_i$ such that $\epsilon_i \sim N(0,1)$. 
    \item[] \textbf{Step 7.} Next compute the potential outcome for subject $i$ under treatment ($A_i=1$) as $Y_i(1) = Y_i(0) + V_i - 0.5$
    \item[] \textbf{Step 8.} Generate the observed outcome for subject $i$ as $Y_i = A_i*Y_i(1) + (1-A_i)*Y_i(0)$. 
    \item[] \textbf{Step 9.} Model $\P(S=1\mid V,W)$ and $\P(A=1 \mid V,W)$ using logistic regression.
    \item[] \textbf{Step 10.} Using the resulting models, for each subject $i$ predict $\hat{\eta}_{i}=\P(S_i =1\mid V_i, W_i)$ and   $\hat{\pi}_{i}=\P(A_i =1 \mid V_i, W_i)$.
    \item[] \textbf{Step 11.} Compute $M^{*} = \max_i$   $ \Big[ \hat{\eta}_{i}  \, / \, \hat{\pi}_{i} \Big ]$. 
    If $M^*>50$ save the generated data and return to step 1, otherwise delete the data set and return to step 1. Repeat until $1000$ data sets are saved. 
\end{itemize}

Note the inclusion of steps $9$-$11$ is to ensure that we obtain $1000$ simulated datasets where in each one positivity violations exist in the external population. 

\subsubsection*{The data-generating process for scenario 3}  In this setting we explore estimator performance when different combinations of the required nuisance functions are misspecified. We consider a moderately sized two-arm randomized ($n_{S=1}=250$) trial where an external data source of comparable size ($n_{S=0}=250$) is available to borrow from. The total number of subjects is given by $n = n_{S=1} + n_{S=0}$ and data for each subject $i \in \{1,\ldots, n\}$ were generated according to the below steps. This process was repeated $1000$ times. \label{sec:scen3}  

\begin{itemize}
    \item[] \textbf{Step 1.} For each subject $i$ randomly draw covariate $W_i$ such that $W_i \sim \text{N}(0,1)$.
    \item[] \textbf{Step 2.} For each subject $i$ randomly draw covariate $V_i$ such that $V_i \sim \text{Bernoulli}(0.5)$. 
    \item[] \textbf{Step 3.} For each subject $i$ compute the transformed covariate $Z_i$ using the relationship $Z_i = \text{sin}\Big( \frac{W_i}{W_i +1}+2\Big)$ . 
    \item[] \textbf{Step 4.}  Let $\eta_i$ be the probability subject $i$ is enrolled in the target trial, compute this probability as $\eta_i$ = $(1+\text{exp}(-1 + 0.5*W_i + 1.2*V_i ))^{-1}$. Then randomly draw population indicator $S_i$ such that $S_i \sim \text{Bernoulli}(\eta_i)$. 
    \item[] \textbf{Step 5.} If subject $i$ is in the target population ($S_i = 1$) randomly draw treatment indicator $A_i$ from $A_i \sim \text{Bernoulli}(0.5)$. 
    \item[] \textbf{Step 6.} If subject $i$ is in the external population ($S_i = 0$) let $\pi_i$ be the probability subject $i$ obtains treatment. Compute this quantity as $\pi_i= (1+\text{exp}(-0.045 + 0.09*W_i + 0.09*V_i))^{-1}$ and randomly draw treatment indicator $A_i$ from $A_i \sim \text{Bernoulli}(\pi_i)$.
    \item[] \textbf{Step 7.} For subject $i$ compute their potential outcome under control ($A_i=0$) as $Y_i(0) = 1.5*W_i + 0.5*V_i + \epsilon_i$ such that $\epsilon_i \sim N(0,1)$. 
    \item[] \textbf{Step 8.} Next compute the potential outcome for subject $i$ under treatment ($A_i=1$) as $Y_i(1) = Y_i(0) + V_i - 0.5$
    \item[] \textbf{Step 9.} Finally generate the observed outcome for subject $i$ as $Y_i = A_i*Y_i(1) + (1-A_i)*Y_i(0)$. 
\end{itemize}

We explored the following four scenarios: all models are correctly specified, only the outcome is correct, only the data and treatment models are correct, and all models are incorrect. For each misspecification scenario 1000 data sets of size 500 were created with an equal split between external and target populations. The rationale behind this approach can be found in the second order remainder term described in section 3. Recall the second order remainder term for the estimator in equation 1 is such that $||\hat{m} - m||\left( ||\hat{\pi}-\pi || + ||\hat{\eta} - \eta || \right)= \text{o}( n^{-1/2} )$ where $\hat{m}$ , $\hat{\pi}$ and $\hat{\eta}$ are the outcome, treatment and data models respectively. One can see by inspection that this term is zero if and only if the outcome model is correct or both the treatment and data model are correctly specified. To simulate model misspecification, we replaced the covariate $W$ with its transformation $Z$ given by $Z = sin \Big(\frac{W}{W+1}+2\Big)$ when estimating different nuisance models. \\

\pagebreak
\section*{Additional plots and tables}

\begin{figure}[h]
\begin{center}
\includegraphics[scale=0.4]{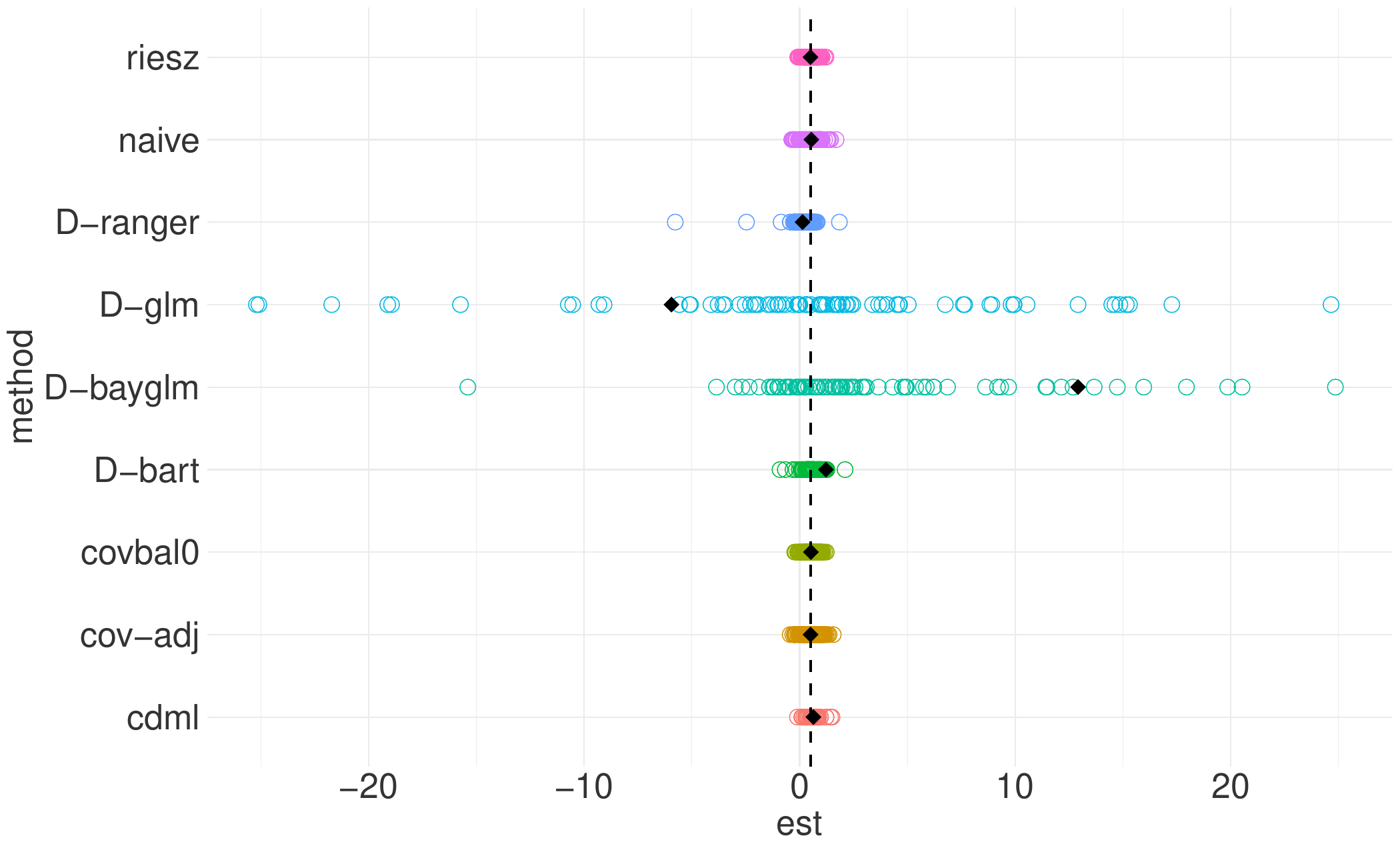}
\end{center}
\caption{The colored dots represent the estimates of $\satett(v=1,1)$ by method under practical positivity violations. The black diamond in each row represents the mean of the estimates. The dashed line indicates the true effect in $V=1$, in this case the true subgroup specific treatment effect is $0.5$.\label{fig:ppv1}}
\end{figure}

\begin{figure}[h]
\begin{center}
\includegraphics[scale=0.4]{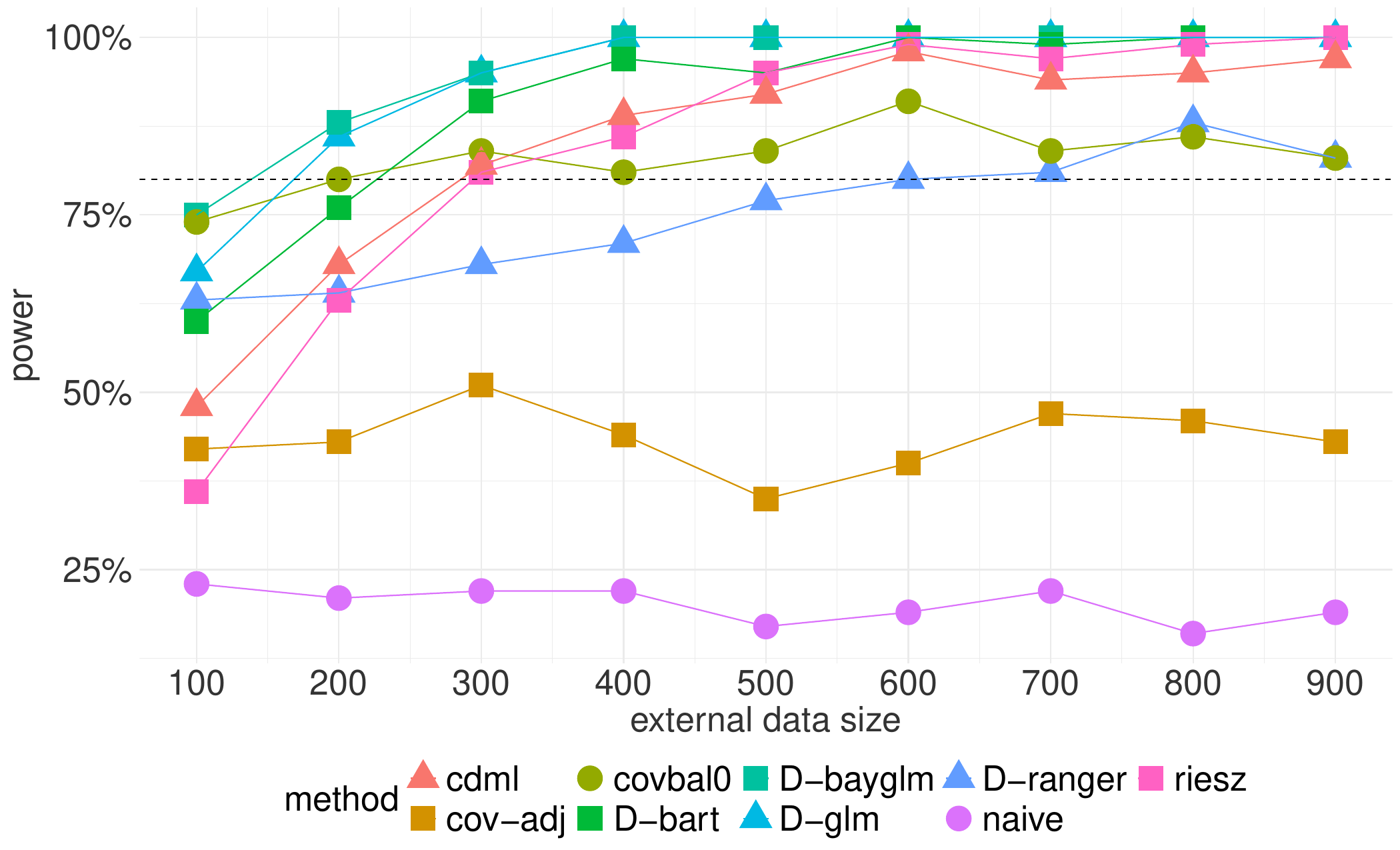}
\end{center}
\caption{Power of the estimators in table \ref{table_methods} as a function of external data size to detect a subgroup specific effect in the subgroup defined by $V=0$.\label{fig:powv0} }
\end{figure}

\begin{figure}[]
\begin{center}
\includegraphics[scale=0.4]{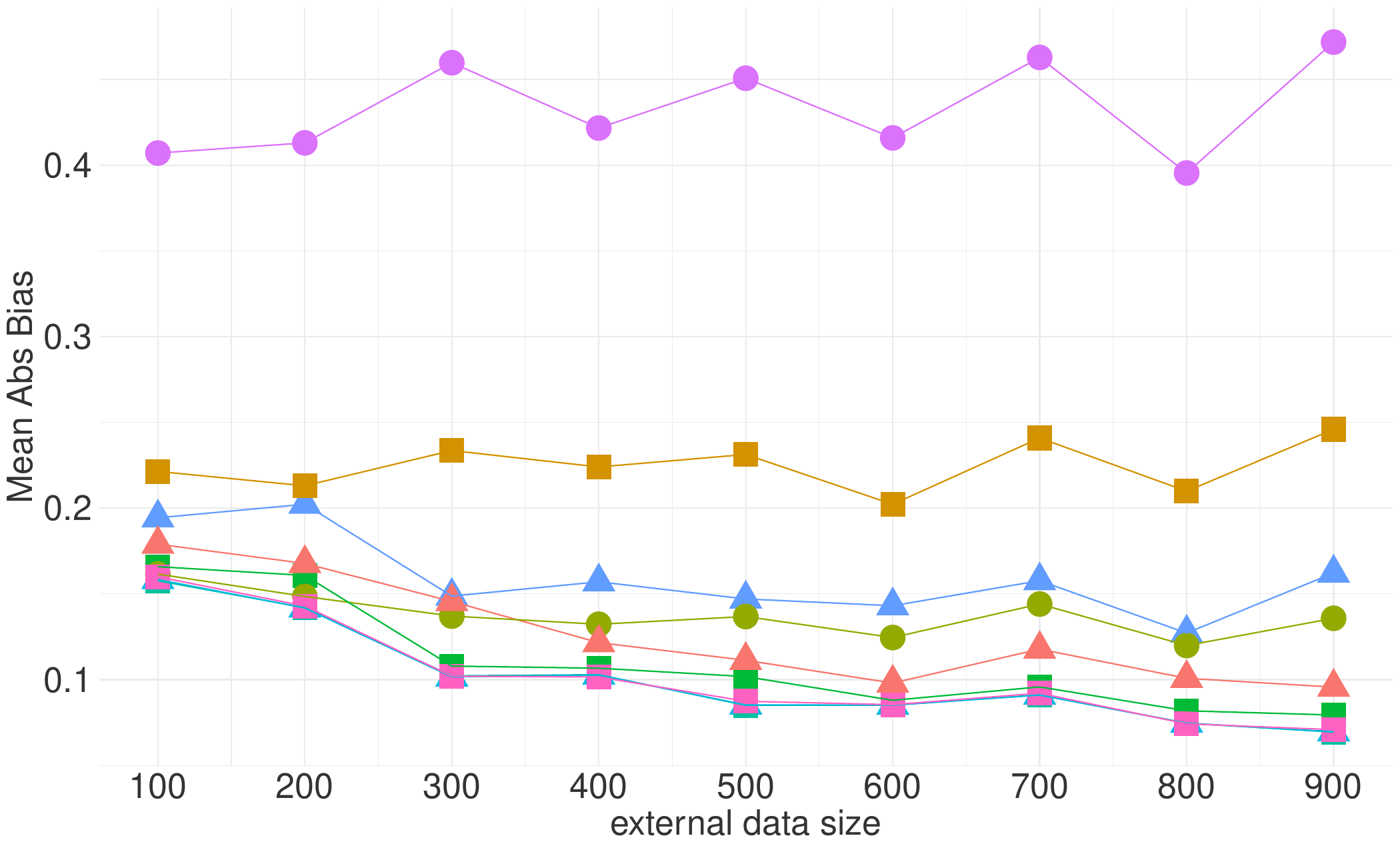}
\includegraphics[scale=0.4]{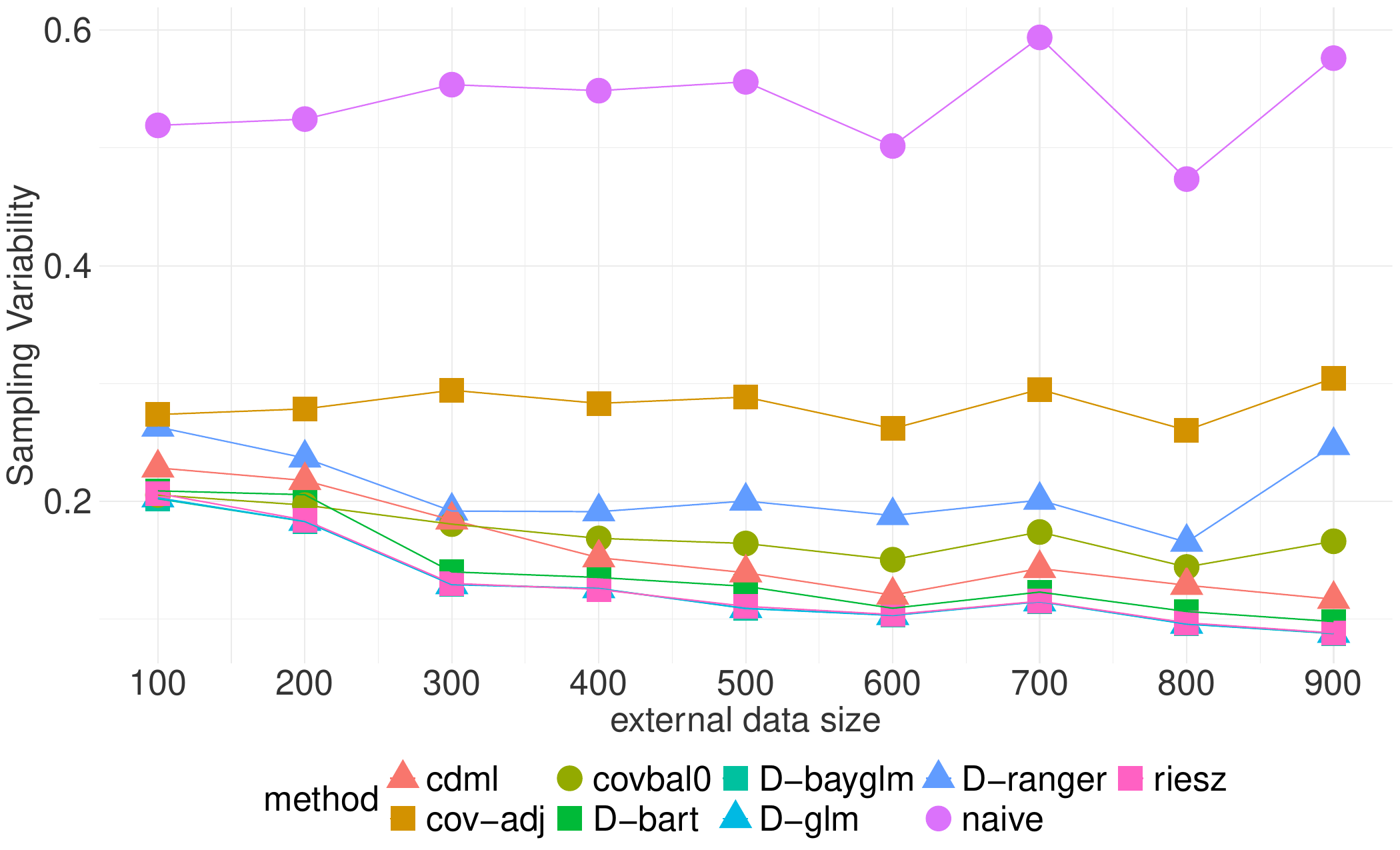}
\end{center}
\caption{Mean absolute bias of the estimators in table \ref{table_methods} as a function of external data size (top). The sampling variability of the same estimators
as a function of external data size (bottom) for subgroup $V=0$.\label{fig:biasvarv0}}
\end{figure}

\begin{figure}[]
\begin{center}
\includegraphics[scale=0.4]{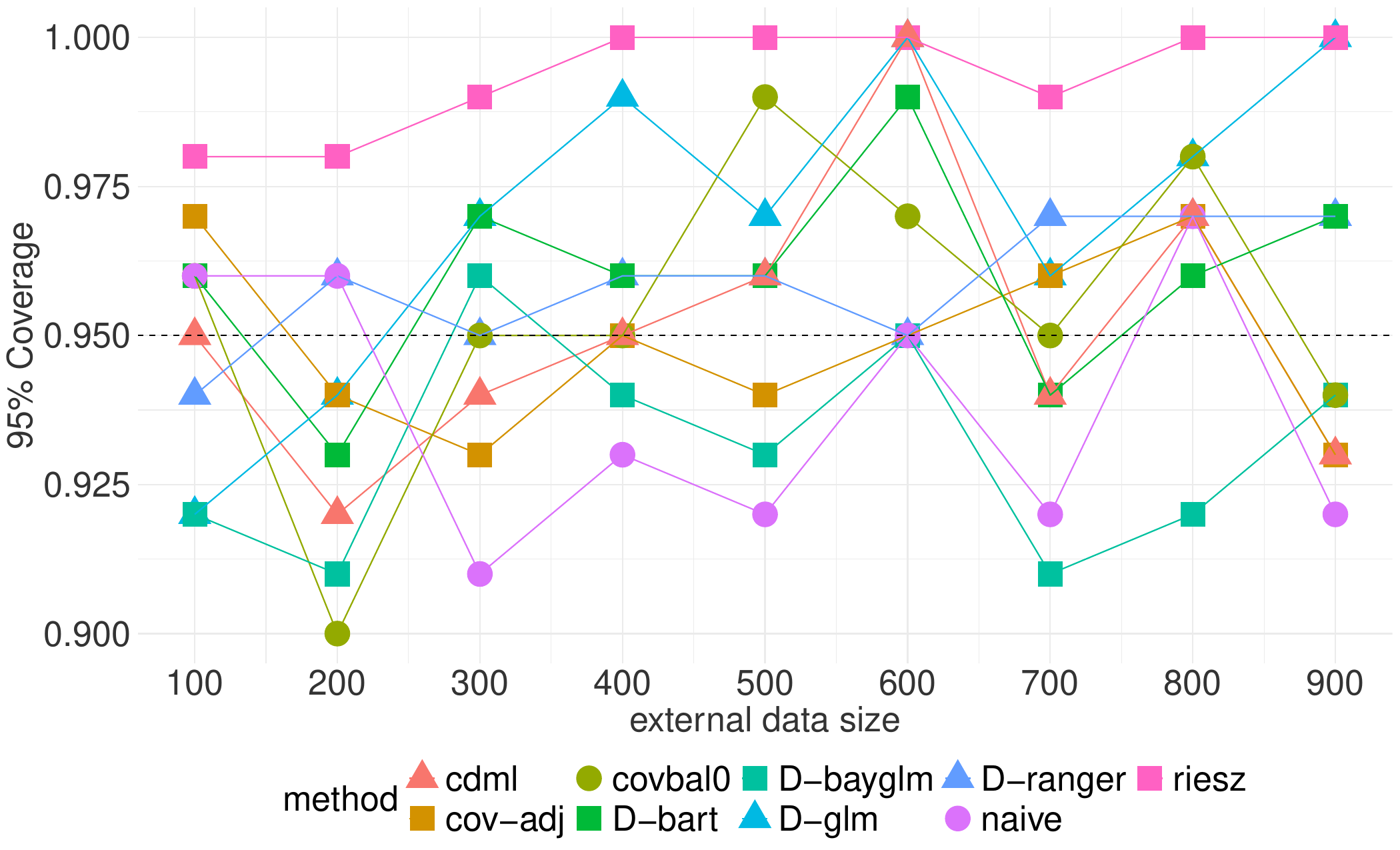}
\end{center}
\caption{Coverage of the $95\%$ confidence interval of the estimators in table \ref{table_methods} as a function of external data size for the subgroup $V=0$.\label{fig:covv0}}
\end{figure}

\begin{table}[]
\centering
\begin{tabular}{lccccccccc}

                                             \multicolumn{10}{c}{\textbf{External data size}} \\  
\hline
 \textbf{Method} & \textbf{100} & \textbf{200} & \textbf{300} & \textbf{400} & \textbf{500} & \textbf{600} & \textbf{700} & \textbf{800} & \textbf{900} \\
 \hline
\texttt{cov-adj}    & 0.98 & 0.96 & 0.92 & 0.97 & 0.94 & 0.94 & 0.94 & 0.95 & 0.94 \\
\texttt{D-glm}     & 0.97 & 0.98 & 0.94 & 0.99 & 0.96 & 0.99 & 0.99 & 1.00 & 0.96 \\ 
\texttt{D-bayglm}  & 0.97 & 0.95 & 0.90 & 0.91 & 0.94 & 0.95 & 0.96 & 0.94 & 0.92 \\
\texttt{D-ranger}  & 0.98 & 0.98 & 0.93 & 0.96 & 0.96 & 0.96 & 0.99 & 0.94 & 0.92 \\
\texttt{D-bart}    & 0.96 & 0.94 & 0.92 & 0.93 & 0.89 & 0.96 & 0.96 & 0.86 & 0.94 \\
\texttt{covbal0}    & 0.95 & 0.99 & 0.93 & 0.97 & 0.97 & 0.96 & 0.97 & 0.95 & 0.96 \\
\texttt{riesz}      & 0.98 & 1.00 & 0.99 & 0.99 & 1.00 & 0.99 & 1.00 & 1.00 & 0.99 \\
\texttt{cdml}       & 0.97 & 0.98 & 0.95 & 0.95 & 0.94 & 0.98 & 0.96 & 0.95 & 0.97 \\ 
\texttt{naive}      & 0.95 & 0.93 & 0.92 & 0.94 & 0.90 & 0.94 & 0.91 & 0.92 & 0.92 \\
\hline
\end{tabular}
\caption{Coverage of the $95\%$ confidence interval of the estimators in table \ref{table_methods} as a function of external data size for the subgroup $V=0$. \label{table_cov0}}
\end{table}

\clearpage

\begin{table*}
\caption{Difference in mean CDSS within the patient subgroup defined by DUP averaged across 100 replications of the case study analysis with a new random seed used on each iteration. The column \textbf{SEn/SE} is the ratio of the standard error of the naive estimator over the standard error of each method, a ratio greater than 1 indicates an increase in precision. When comparing the results displayed in Table 6 corresponding to the average of the real world data application with the RNG seed changing 100 times, to the results of Table 5, where a fixed seed was used, we see that the magnitude and direction of the subgroup specific effect estimates were virtually the same across all the methods listed in Table 1. The estimated standard errors and confidence interevals were also similar across methods.\label{table_applied_avg}}
\label{postable}
\vspace{5mm}
\centering
\begin{tabular}{@{}lcccc|cccc@{}}
\hline
\multicolumn{5}{c|}{ \textbf{DUP $\geq 18$}} & \multicolumn{4}{c}{ \textbf{DUP $< 18$}} \\
\hline
\textbf{Method} & \textbf{Est} & \textbf{SE} & \textbf{CI} & \textbf{SEn/SE} & \textbf{Est} & \textbf{SE} & \textbf{CI} & \textbf{SEn/SE} \\
\hline
\texttt{cov-adj}   & 1.25 & 0.80  &  (-0.32 , 2.82)   &  1.01  & -0.88 & 0.89  & (-2.63 , 0.88)  &  1.07 \\
\texttt{D-glm}    & 1.96 & 0.67  &  (0.65 , 3.26)    &  1.21  & -0.80 & 0.67  & (-2.11 , 0.50)  &  1.43 \\
\texttt{D-bayglm} & 1.96 & 0.82  &  (0.34 , 3.56)    &  0.98  & -0.78 & 0.74  & (-2.25 , 0.68) &  1.29 \\
\texttt{D-ranger} & 1.46 & 0.58  &  (0.33 , 2.61)    &  1.38  & -0.67 & 0.55  & (-1.76 , 0.41) &  1.72 \\
\texttt{D-bart}   & 1.63 & 1.07  & (-0.47 , 3.70)    &  .76  & -0.51 & 0.89  & (-2.27 , 1.20) &  1.07 \\
\texttt{covbal0}   & 1.12 & 0.58  & (-0.02 , 2.26)    &  1.39  & -0.29 & 0.61  & (-1.49 ,  0.91) &  1.58 \\
\texttt{riesz}     & 1.68 & 0.89  & (-0.08  , 3.43)   &  .91  & -0.63 & 0.90  & (-2.40 , 1.14) &  1.07 \\
\texttt{cdml}      & 1.48 & 0.71  &  (0.09 , 2.88)    &  1.13  & -0.59 & 0.79  & (-2.14 , 0.95) &  1.20 \\
\texttt{naive}     & 1.15 & 0.81  & (-0.43  , 2.73)   &  1  & -0.75 & 0.96  & (-2.63 , 1.13) &  1 \\
\hline
\end{tabular}
\end{table*}

\clearpage

\begin{figure}[h]
\begin{center}
\includegraphics[scale=0.4]{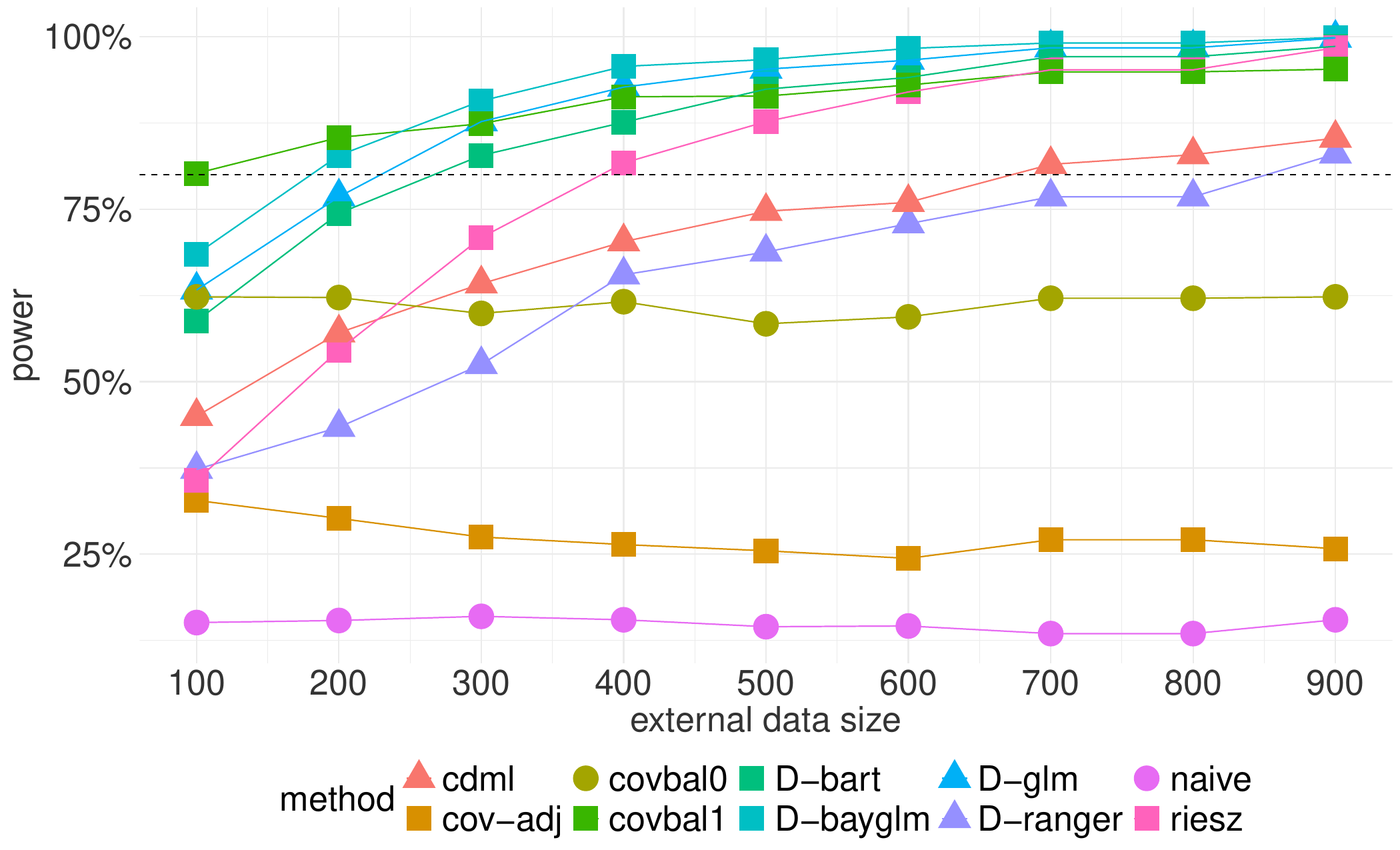}
\end{center}
\caption{Power of the estimators in Table \ref{table_methods} as a function of external data size to detect a subgroup specific effect in the subgroup defined by $V=0$ including covariate balancing with penalty term set to 1.\label{fig:powv0cov1} }
\end{figure}

\begin{figure}[]
\begin{center}
\includegraphics[scale=0.4]{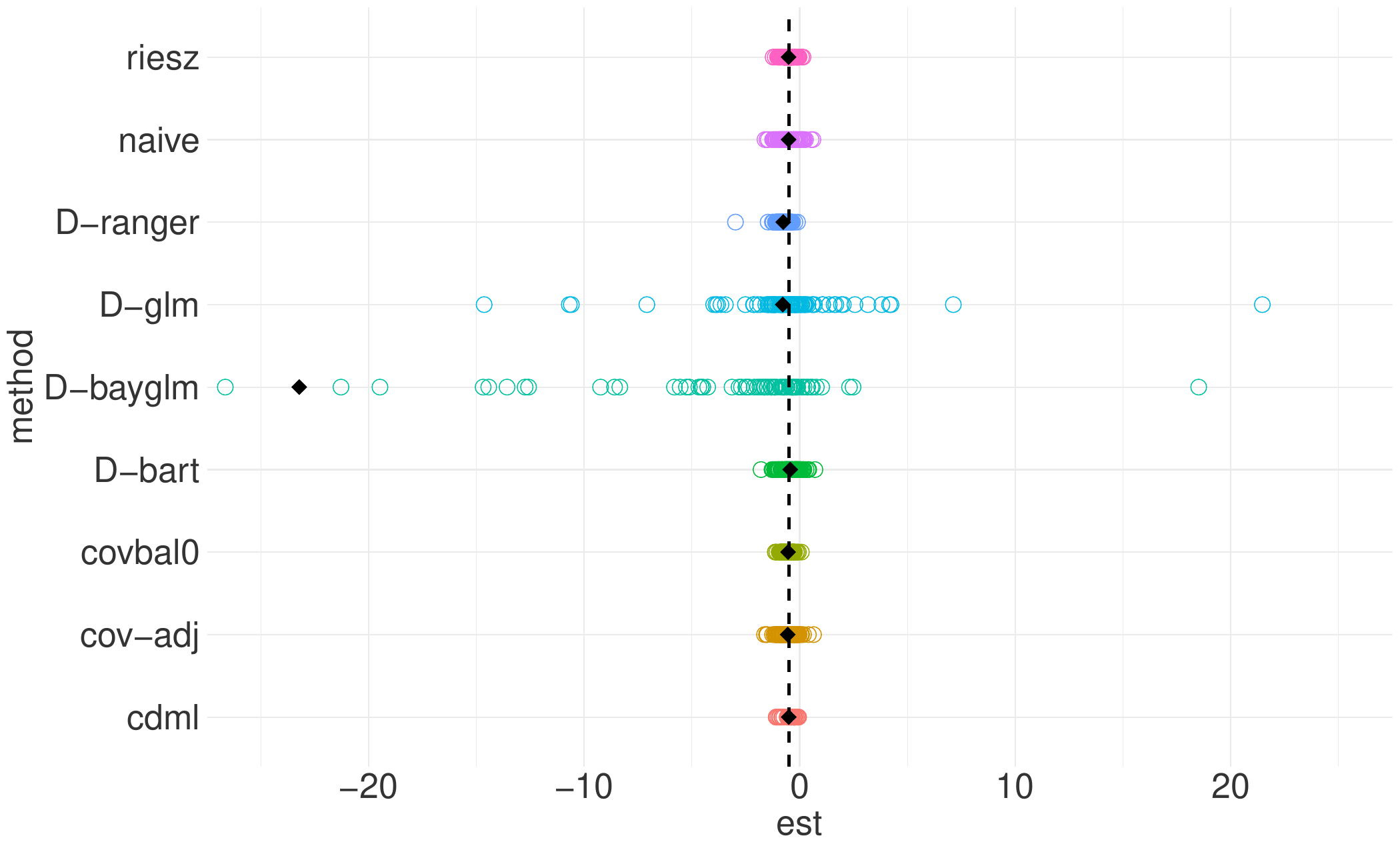}
\end{center}
\caption{The colored dots represent the estimates of $\satett(v=0,1)$ by method under practical positivity violations. The black diamond in each row represents the mean of the estimates. The dashed line indicates the true effect in $V=0$, in this case it is $-0.5$.\label{fig:ppv0}}
\end{figure}

\clearpage

\begin{table*}[]
\caption{Mean absolute bias, variance and coverage of the $95\%$ confidence interval for each method in scenario 2: positivity violations in subgroup $V=0$.\label{table_ppv0}}
\vspace{5mm}
\centering
\begin{tabular}{@{}lccc@{}}
\hline
\multicolumn{1}{c}{\textbf{Method}} & \multicolumn{1}{c}{\textbf{MAB}} & \multicolumn{1}{c}{\textbf{Variance}}  & \multicolumn{1}{c}{\textbf{Coverage}}\\
\hline
\texttt{cov-adj}    & 0.33	& 0.41  & .95    \\
\texttt{D-glm}     & 2.90  & 9.70  & .99    \\
\texttt{D-bayglm}  & 23.50 & 97.35 & .69    \\
\texttt{D-ranger}  & 0.30  & 0.32  &  .92   \\
\texttt{D-bart}    & 0.34  & 0.43	& .95    \\
\texttt{covbal0}    & 0.20  & 0.24  &  0.96  \\
\texttt{riesz}      & 0.20  & 0.26  &  1.00  \\
\texttt{cdml}       &  0.27 & 0.32  &  0.89  \\
\texttt{naive}      & 0.38	& 0.46  & .94    \\
\hline
\end{tabular}
\end{table*}

\clearpage

\begin{figure}[]
\caption{Mean absolute bias of the estimators in table \ref{table_methods} under model misspecification. The sampling variability of the same estimators
under the same misspecification scenarios (bottom) for the subgroup $V=0$.\label{fig:biasvarv0}}

\begin{center}
\includegraphics[scale=0.4]{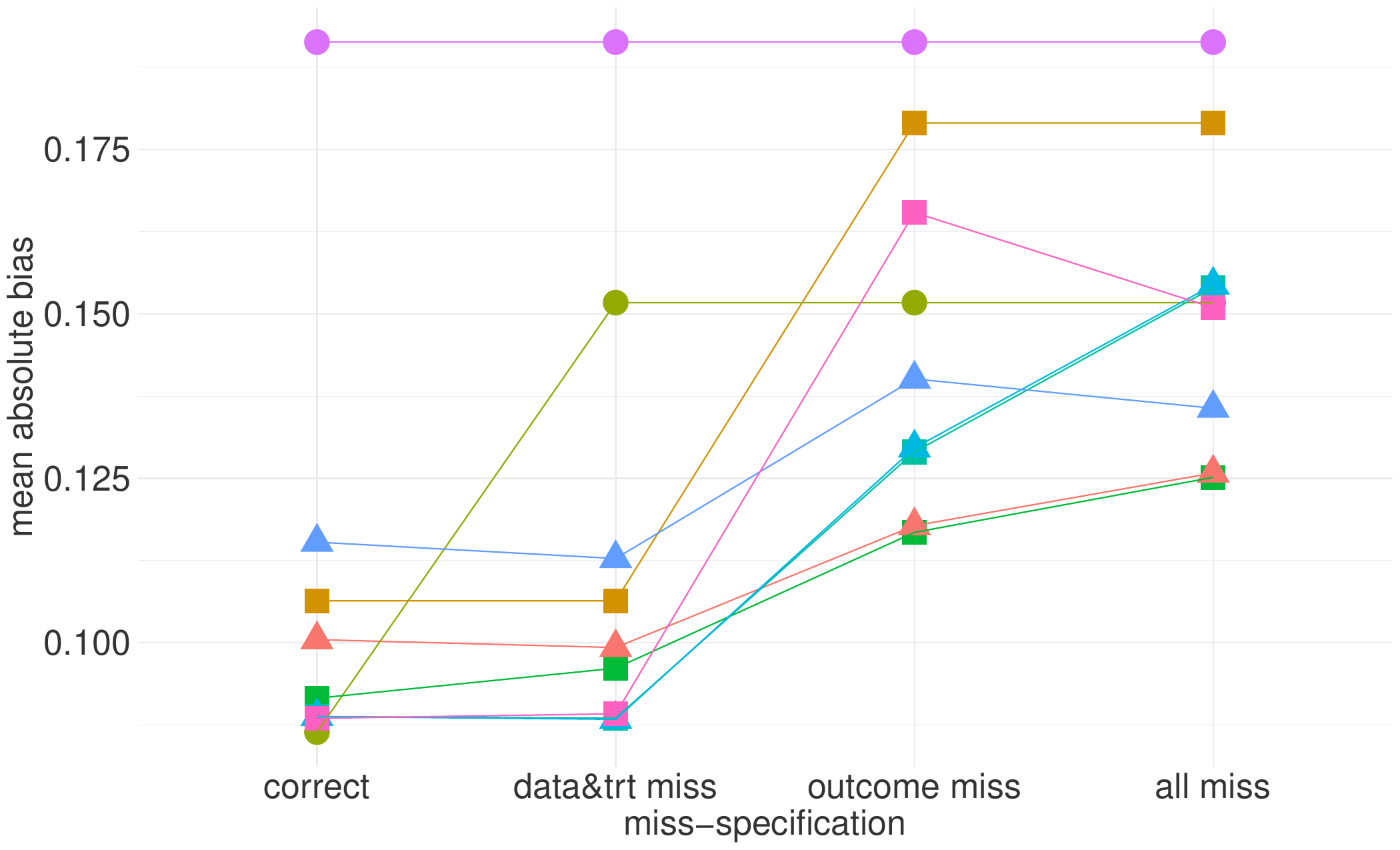}
\includegraphics[scale=0.4]{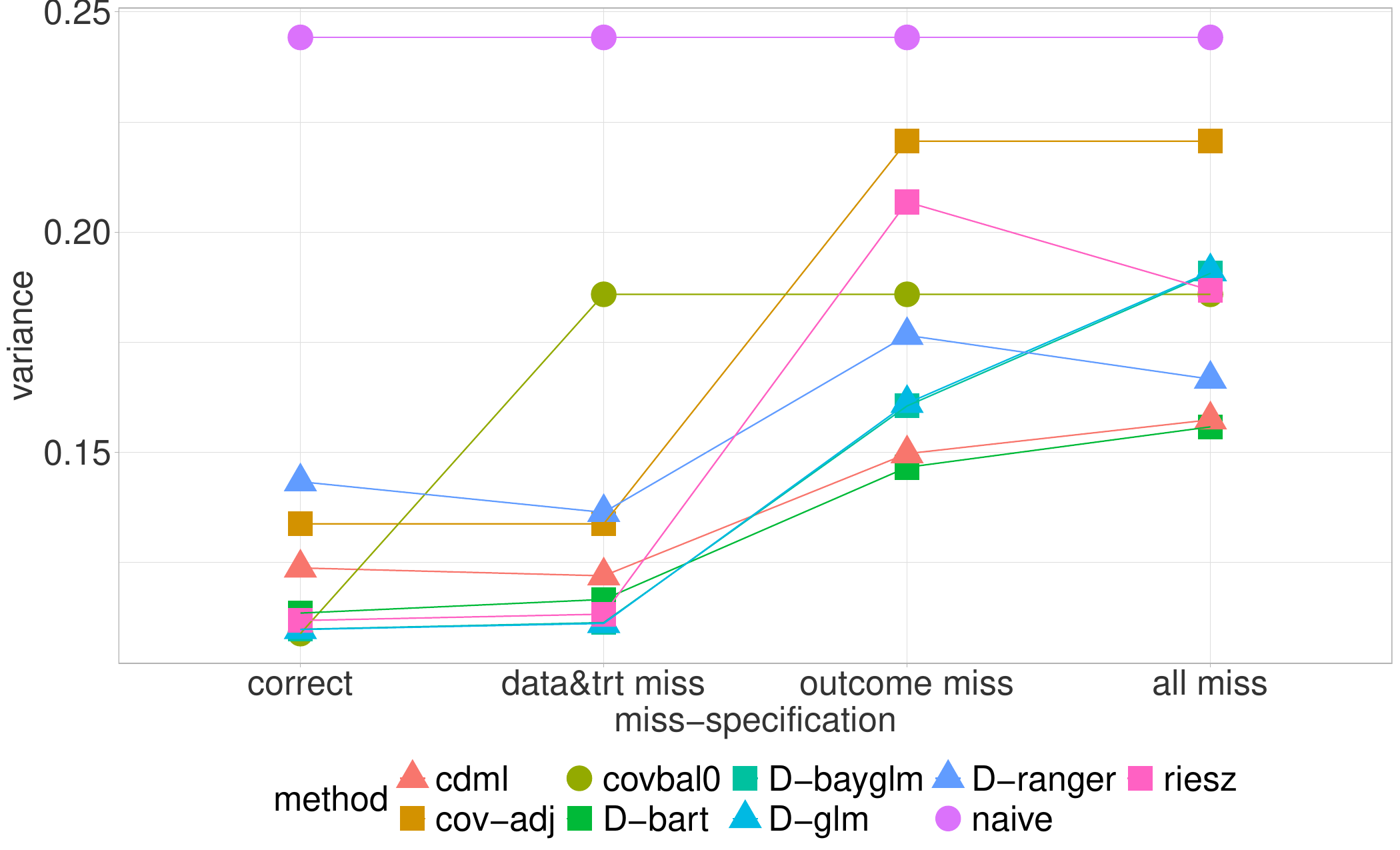}
\end{center}
\end{figure}

\begin{table*}[]
\caption{Coverage of the $95\%$ confidence interval for each method by misspecification scenario in subgroup $V=0$. In the first column all models are correctly specified, in the second column the data and treatment models are misspecified, in the third column only the outcome model is misspecified, the forth column is when all models are misspecified. \label{table_missv0}}
\vspace{5mm}
\centering
\begin{tabular}{@{}ccccc@{}}
\hline
\multicolumn{1}{c}{\textbf{Method}} & \multicolumn{1}{c}{\textbf{all correct}} & \multicolumn{1}{c}{\textbf{data \& treatment miss}} & \multicolumn{1}{c}{\textbf{outcome miss}} & \multicolumn{1}{c}{\textbf{all miss}} \\
\hline
\texttt{cov-adj}      & 0.97 & 0.97 & 0.98 & 0.98 \\
\texttt{D-glm}       & 0.97 & 0.97 & 0.97 & 0.95 \\
\texttt{D-bayglm}    & 0.98 & 0.96 & 0.97 & 0.94 \\
\texttt{D-ranger}    & 0.73 & 0.77 & 0.75 & 0.77 \\
\texttt{D-bart}      & 0.96 & 0.96 & 0.99 & 0.95 \\
\texttt{covbal0}      & 0.95 & 0.97 & 0.97 & 0.97 \\
\texttt{riesz}        & 1.00 & 1.00 & 0.99 & 1.00 \\
\texttt{cdml}         & 0.99 & 0.98 & 0.96 & 0.96 \\
\texttt{naive}        & 0.98 & 0.98 & 0.98 & 0.98 \\
\hline
\end{tabular}
\end{table*}

\ignore{
\begin{table}[h]
\centering
\begin{tabular}{|l|c|c|c|}
\hline
\textbf{Category} & \textbf{DUP $\geq 18$} (N=133) & \textbf{DUP $< 18$} (N=77) & \textbf{Overall (N=210)} \\
\hline
\textbf{Age} & & & \\
Mean (SD) & 24.1 (5.06) & 22.5 (4.26) & 23.6 (4.83) \\
\hline
\textbf{Sex} & & & \\
Female & 35 (26.3\%) & 23 (29.9\%) & 58 (27.6\%) \\
Male & 98 (73.7\%) & 54 (70.1\%) & 152 (72.4\%) \\
\hline
\textbf{Race} & & & \\
American Indian/Alaska Native & 3 (2.3\%) & 4 (5.2\%) & 7 (3.3\%) \\
Asian & 18 (13.5\%) & 18 (23.4\%) & 36 (17.1\%) \\
Black or African American & 45 (33.8\%) & 16 (20.8\%) & 61 (29.0\%) \\
Hawaiian or Pacific Islander & 1 (0.8\%) & 0 (0\%) & 1 (0.5\%) \\
Other & 1 (0.8\%) & 1 (1.3\%) & 2 (1.0\%) \\
White & 65 (48.9\%) & 38 (49.4\%) & 103 (49.0\%) \\
\hline
\textbf{Employment Status} & & & \\
No & 115 (86.5\%) & 64 (83.1\%) & 179 (85.2\%) \\
Yes & 18 (13.5\%) & 13 (16.9\%) & 31 (14.8\%) \\
\hline
\textbf{Education Level} & & & \\
College & 46 (34.6\%) & 34 (44.2\%) & 80 (38.1\%) \\
Complete high school & 41 (30.8\%) & 21 (27.3\%) & 62 (29.5\%) \\
No high school & 46 (34.6\%) & 22 (28.6\%) & 68 (32.4\%) \\
\hline
\textbf{Marital Status} & & & \\
Married & 10 (7.5\%) & 4 (5.2\%) & 14 (6.7\%) \\
Never Married & 115 (86.5\%) & 71 (92.2\%) & 186 (88.6\%) \\
Widowed/divorced/separated & 8 (6.0\%) & 2 (2.6\%) & 10 (4.8\%) \\
\hline
\end{tabular}
\caption{Demographic Information for the combined RAISE plus DECIPHER data set, complete cases.}
\end{table}
}

\begin{table}[ht]
\centering
\resizebox{\textwidth}{!}{  
\begin{tabular}{|l|c|c|c|c|c|c|}
\hline
\textbf{} & \multicolumn{2}{c|}{\textbf{DUP $\geq 18$}} & \multicolumn{2}{c|}{\textbf{DUP $< 18$}} & \multicolumn{2}{c|}{\textbf{DECIPHER Overall}} \\
\hline
 & \textbf{Control} & \textbf{Treatment} & \textbf{Control} & \textbf{Treatment} & \textbf{Control} & \textbf{Treatment} \\
 & (N=10) & (N=13) & (N=16) & (N=12) & (N=26) & (N=25) \\
\hline
\textbf{Age} & & & & & & \\
Mean (SD) & 23.9 (5.17) & 24.5 (4.74) & 22.8 (3.89) & 23.0 (5.56) & 23.2 (4.36) & 23.8 (5.10) \\
Median [Min, Max] & 23.5 [17.0, 34.0] & 25.0 [16.0, 31.0] & 23.0 [17.0, 32.0] & 23.0 [15.0, 35.0] & 23.0 [17.0, 34.0] & 24.0 [15.0, 35.0] \\
\hline
\textbf{Sex} & & & & & & \\
Female & 3 (30.0\%) & 4 (30.8\%) & 6 (37.5\%) & 7 (58.3\%) & 9 (34.6\%) & 11 (44.0\%) \\
Male & 7 (70.0\%) & 9 (69.2\%) & 10 (62.5\%) & 5 (41.7\%) & 17 (65.4\%) & 14 (56.0\%) \\
\hline
\textbf{Race} & & & & & & \\
Asian & 6 (60.0\%) & 7 (53.8\%) & 8 (50.0\%) & 9 (75.0\%) & 14 (53.8\%) & 16 (64.0\%) \\
Black or African American & 1 (10.0\%) & 1 (7.7\%) & 1 (6.3\%) & 0 (0\%) & 2 (7.7\%) & 1 (4.0\%) \\
White & 3 (30.0\%) & 4 (30.8\%) & 6 (37.5\%) & 3 (25.0\%) & 9 (34.6\%) & 7 (28.0\%) \\
Other & 0 (0\%) & 1 (7.7\%) & 1 (6.3\%) & 0 (0\%) & 1 (3.8\%) & 1 (4.0\%) \\
\hline
\textbf{Employment Status} & & & & & & \\
No & 7 (70.0\%) & 12 (92.3\%) & 9 (56.3\%) & 11 (91.7\%) & 16 (61.5\%) & 23 (92.0\%) \\
Yes & 3 (30.0\%) & 1 (7.7\%) & 7 (43.8\%) & 1 (8.3\%) & 10 (38.5\%) & 2 (8.0\%) \\
\hline
\textbf{Education} & & & & & & \\
College & 4 (40.0\%) & 5 (38.5\%) & 6 (37.5\%) & 5 (41.7\%) & 10 (38.5\%) & 10 (40.0\%) \\
Complete high school & 3 (30.0\%) & 6 (46.2\%) & 6 (37.5\%) & 5 (41.7\%) & 9 (34.6\%) & 11 (44.0\%) \\
No high school & 3 (30.0\%) & 2 (15.4\%) & 4 (25.0\%) & 2 (16.7\%) & 7 (26.9\%) & 4 (16.0\%) \\
\hline
\textbf{Marital Status} & & & & & & \\
Married & 1 (10.0\%) & 2 (15.4\%) & 1 (6.3\%) & 3 (25.0\%) & 2 (7.7\%) & 5 (20.0\%) \\
Never Married & 9 (90.0\%) & 11 (84.6\%) & 15 (93.8\%) & 9 (75.0\%) & 24 (92.3\%) & 20 (80.0\%) \\
Widowed/divorced/separated  & 9 (90.0\%) & 11 (84.6\%) & 15 (93.8\%) & 9 (75.0\%) & 24 (92.3\%) & 20 (80.0\%) \\
\hline
\end{tabular}
}
\caption{Demographic data for DECIPHER participants}
\end{table}

\begin{table}[ht]
\centering
\resizebox{\textwidth}{!}{  
\begin{tabular}{|l|c|c|c|c|c|c|}
\hline
\textbf{} & \multicolumn{2}{c|}{\textbf{DUP $\geq 18$}} & \multicolumn{2}{c|}{\textbf{DUP $< 18$}} & \multicolumn{2}{c|}{\textbf{RAISE Overall}} \\
\hline
 & \textbf{Control} & \textbf{Treatment} & \textbf{Control} & \textbf{Treatment} & \textbf{Control} & \textbf{Treatment} \\
 & (N=96) & (N=14) & (N=46) & (N=3) & (N=142) & (N=17) \\
\hline
\textbf{Age} & & & & & & \\
Mean (SD) & 23.8 (4.98) & 26.2 (5.83) & 22.3 (4.13) & 23.3 (4.17) & 23.3 (4.76) & 25.7 (5.58) \\
Median [Min, Max] & 22.8 [15.8, 38.1] & 24.2 [19.3, 38.0] & 21.7 [16.0, 34.3] & 23.1 [19.2, 27.5] & 22.5 [15.8, 38.1] & 23.7 [19.2, 38.0] \\
\hline
\textbf{Sex} & & & & & & \\
Female & 22 (22.9\%) & 6 (42.9\%) & 9 (19.6\%) & 1 (33.3\%) & 31 (21.8\%) & 7 (41.2\%) \\
Male & 74 (77.1\%) & 8 (57.1\%) & 37 (80.4\%) & 2 (66.7\%) & 111 (78.2\%) & 10 (58.8\%) \\
\hline
\textbf{Race} & & & & & & \\
American Indian/Alaska Native & 3 (3.1\%) & 0 (0\%) & 3 (6.5\%) & 1 (33.3\%) & 6 (4.2\%) & 1 (5.9\%) \\
Asian & 5 (5.2\%) & 0 (0\%) & 1 (2.2\%) & 0 (0\%) & 6 (4.2\%) & 0 (0\%) \\
Black or African American & 35 (36.5\%) & 8 (57.1\%) & 15 (32.6\%) & 0 (0\%) & 50 (35.2\%) & 8 (47.1\%) \\
Hawaiian or Pacific Islander & 1 (1.0\%) & 0 (0\%) & 0 (0\%) & 0 (0\%) & 1 (0.7\%) & 0 (0\%) \\
White & 52 (54.2\%) & 6 (42.9\%) & 27 (58.7\%) & 2 (66.7\%) & 79 (55.6\%) & 8 (47.1\%) \\
\hline
\textbf{Employment Status} & & & & & & \\
No & 85 (88.5\%) & 11 (78.6\%) & 41 (89.1\%) & 3 (100\%) & 126 (88.7\%) & 14 (82.4\%) \\
Yes & 11 (11.5\%) & 3 (21.4\%) & 5 (10.9\%) & 0 (0\%) & 16 (11.3\%) & 3 (17.6\%) \\
\hline
\textbf{Education} & & & & & & \\
College & 30 (31.3\%) & 7 (50.0\%) & 23 (50.0\%) & 0 (0\%) & 53 (37.3\%) & 7 (41.2\%) \\
Complete high school & 29 (30.2\%) & 3 (21.4\%) & 8 (17.4\%) & 2 (66.7\%) & 37 (26.1\%) & 5 (29.4\%) \\
No high school & 37 (38.5\%) & 4 (28.6\%) & 15 (32.6\%) & 1 (33.3\%) & 52 (36.6\%) & 5 (29.4\%) \\
\hline
\textbf{Marital Status} & & & & & & \\
Married & 7 (7.3\%) & 0 (0\%) & 0 (0\%) & 0 (0\%) & 7 (4.9\%) & 0 (0\%) \\
Never Married & 83 (86.5\%) & 12 (85.7\%) & 44 (95.7\%) & 3 (100\%) & 127 (89.4\%) & 15 (88.2\%) \\
Widowed/divorced/separated & 6 (6.3\%) & 2 (14.3\%) & 2 (4.3\%) & 0 (0\%) & 8 (5.6\%) & 2 (11.8\%) \\
\hline
\end{tabular}
}
\caption{Demographic data for RAISE participants}
\end{table}

\begin{table}[ht]
\centering
\resizebox{\textwidth}{!}{  
\begin{tabular}{|l|c|c|c|c|c|c|}
\hline
\textbf{} & \multicolumn{2}{c|}{\textbf{DUP $\geq 18$}} & \multicolumn{2}{c|}{\textbf{DUP $< 18$}} & \multicolumn{2}{c|}{\textbf{Combined Overall}} \\
\hline
 & \textbf{Control} & \textbf{Treatment} & \textbf{Control} & \textbf{Treatment} & \textbf{Control} & \textbf{Treatment} \\
 & (N=106) & (N=27) & (N=62) & (N=15) & (N=168) & (N=42) \\
\hline
\textbf{Age} & & & & & & \\
Mean (SD) & 23.8 (4.97) & 25.4 (5.30) & 22.4 (4.04) & 23.1 (5.17) & 23.3 (4.69) & 24.6 (5.31) \\
Median [Min, Max] & 23.0 [15.8, 38.1] & 24.8 [16.0, 38.0] & 22.1 [16.0, 34.3] & 23.1 [15.0, 35.0] & 22.7 [15.8, 38.1] & 24.0 [15.0, 38.0] \\
\hline
\textbf{Sex} & & & & & & \\
Female & 25 (23.6\%) & 10 (37.0\%) & 15 (24.2\%) & 8 (53.3\%) & 40 (23.8\%) & 18 (42.9\%) \\
Male & 81 (76.4\%) & 17 (63.0\%) & 47 (75.8\%) & 7 (46.7\%) & 128 (76.2\%) & 24 (57.1\%) \\
\hline
\textbf{Race} & & & & & & \\
American Indian/Alaska Native & 3 (2.8\%) & 0 (0\%) & 3 (4.8\%) & 1 (6.7\%) & 6 (3.6\%) & 1 (2.4\%) \\
Asian & 11 (10.4\%) & 7 (25.9\%) & 9 (14.5\%) & 9 (60.0\%) & 20 (11.9\%) & 16 (38.1\%) \\
Black or African American & 36 (34.0\%) & 9 (33.3\%) & 16 (25.8\%) & 0 (0\%) & 52 (31.0\%) & 9 (21.4\%) \\
Hawaiian or Pacific Islander & 1 (0.9\%) & 0 (0\%) & 0 (0\%) & 0 (0\%) & 1 (0.6\%) & 0 (0\%) \\
White & 55 (51.9\%) & 10 (37.0\%) & 33 (53.2\%) & 5 (33.3\%) & 88 (52.4\%) & 15 (35.7\%) \\
Other & 0 (0\%) & 1 (3.7\%) & 1 (1.6\%) & 0 (0\%) & 1 (0.6\%) & 1 (2.4\%) \\
\hline
\textbf{Employment Status} & & & & & & \\
No & 92 (86.8\%) & 23 (85.2\%) & 50 (80.6\%) & 14 (93.3\%) & 142 (84.5\%) & 37 (88.1\%) \\
Yes & 14 (13.2\%) & 4 (14.8\%) & 12 (19.4\%) & 1 (6.7\%) & 26 (15.5\%) & 5 (11.9\%) \\
\hline
\textbf{Education} & & & & & & \\
College & 34 (32.1\%) & 12 (44.4\%) & 29 (46.8\%) & 5 (33.3\%) & 63 (37.5\%) & 17 (40.5\%) \\
Complete high school & 32 (30.2\%) & 9 (33.3\%) & 14 (22.6\%) & 7 (46.7\%) & 46 (27.4\%) & 16 (38.1\%) \\
No high school & 40 (37.7\%) & 6 (22.2\%) & 19 (30.6\%) & 3 (20.0\%) & 59 (35.1\%) & 9 (21.4\%) \\
\hline
\textbf{Marital Status} & & & & & & \\
Married & 8 (7.5\%) & 2 (7.4\%) & 1 (1.6\%) & 3 (20.0\%) & 9 (5.4\%) & 5 (11.9\%) \\
Never Married & 92 (86.8\%) & 23 (85.2\%) & 59 (95.2\%) & 12 (80.0\%) & 151 (89.9\%) & 35 (83.3\%) \\
Widowed/divorced/separated & 6 (5.7\%) & 2 (7.4\%) & 2 (3.2\%) & 0 (0\%) & 8 (4.8\%) & 2 (4.8\%) \\
\hline
\end{tabular}
}
\caption{Demographic data for combined DECIPHER + RAISE participants}
\end{table}

\clearpage

\begin{figure}[]
\caption{One thousand confidence intervals generated according to the criteria in Scenario 1 for each method in Table \ref{table_methods}. The horizontal red line denotes the true sub-group average treatment effect for those subjects with V=1, by design this was $0.5$. The highlighted blue lines are confidence intervals constructed when the random seed was fixed across each iteration of the analysis. Each orange line represent a corresponding confidence interval constructed with a new random seed selected at the start of each iteration.    \label{fig:civ1}}

\begin{center}
\includegraphics[scale=0.50]{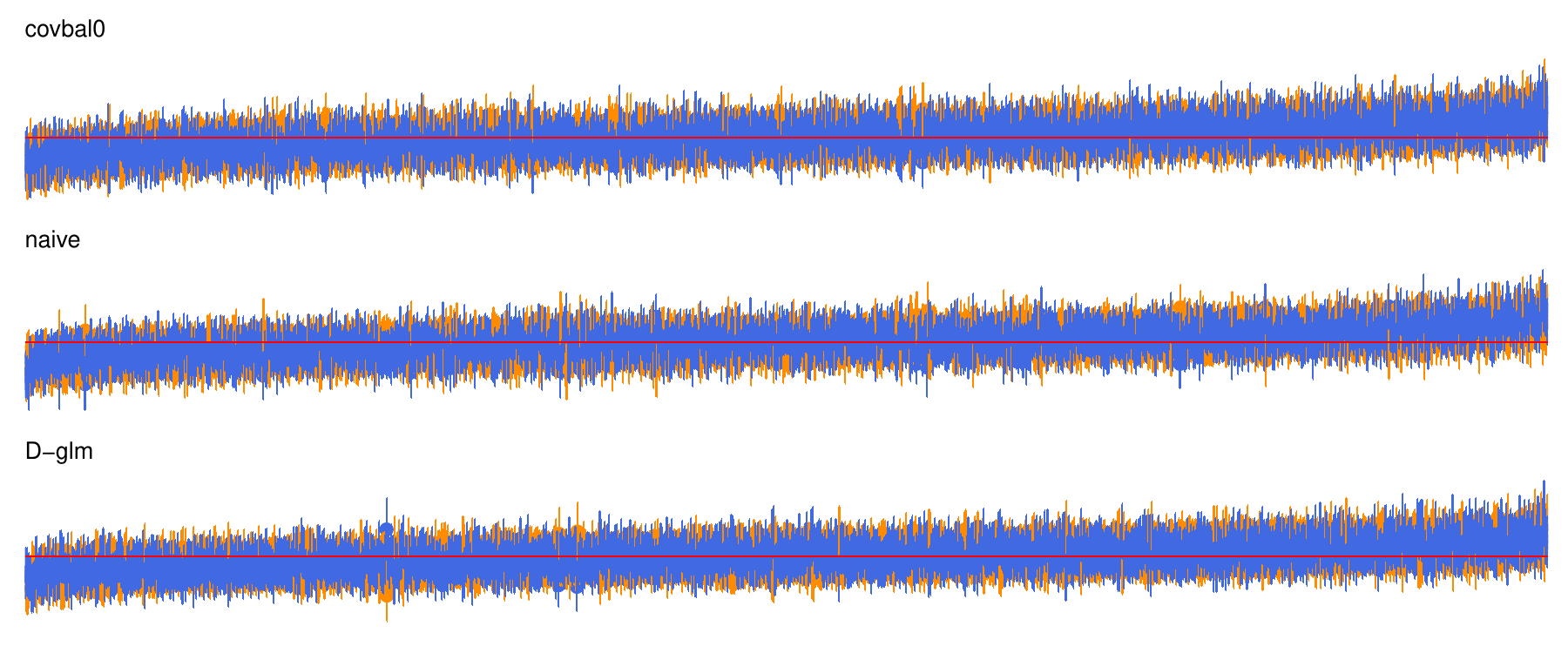}
\includegraphics[scale=0.50]{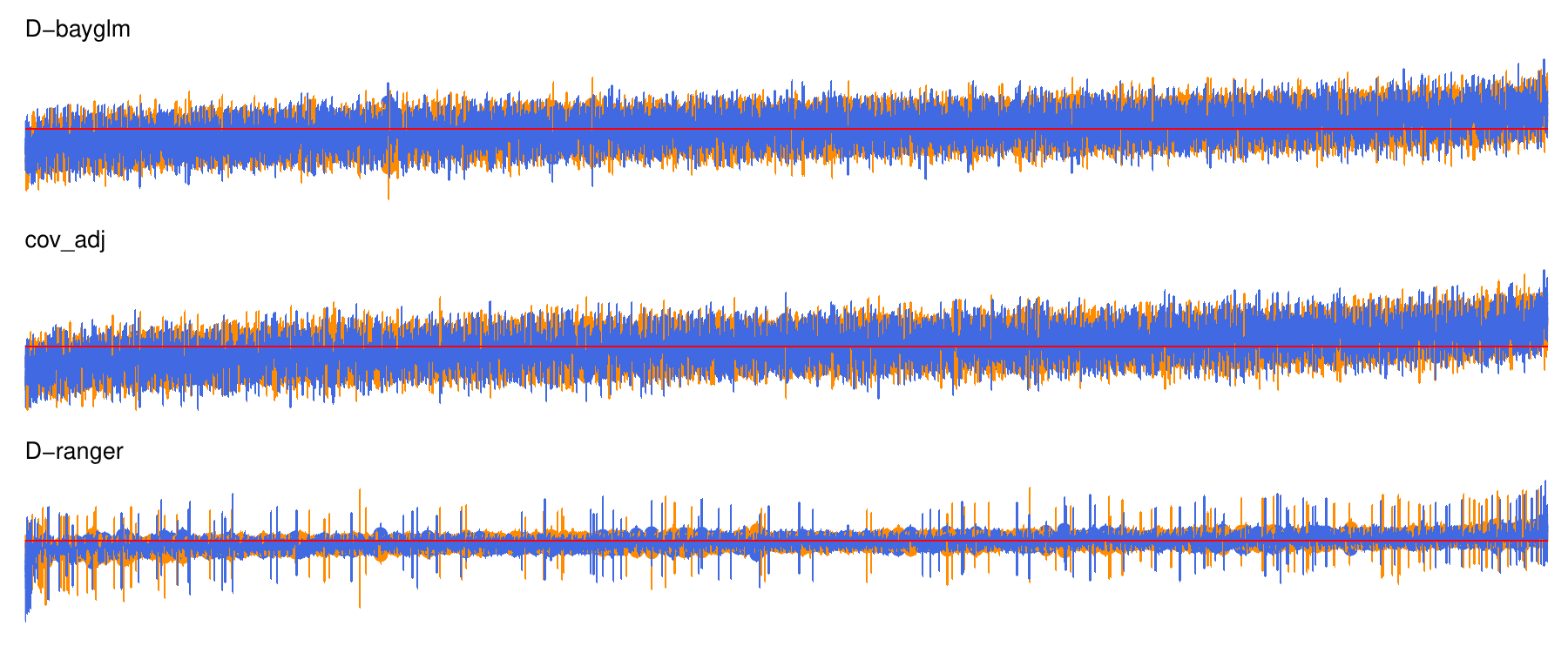}
\includegraphics[scale=0.50]{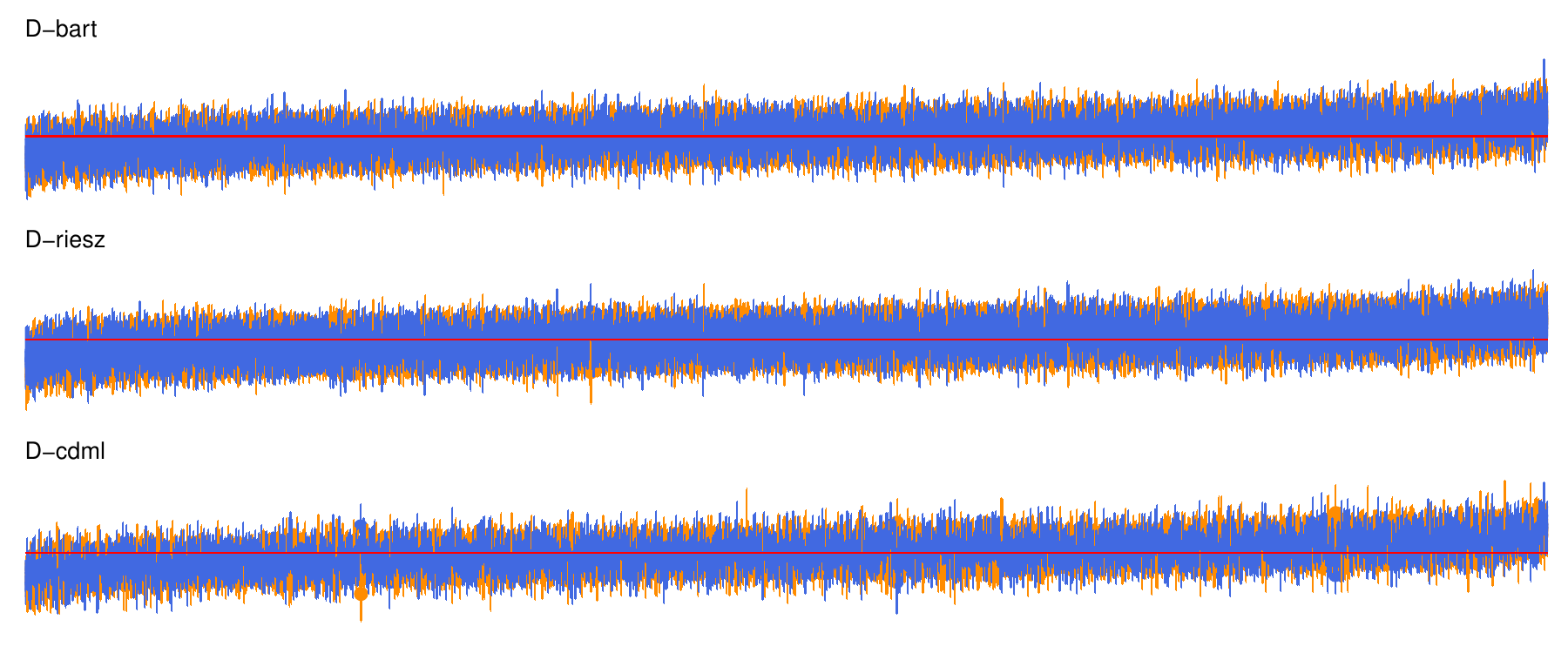}
\end{center}
\end{figure}

\clearpage

\begin{figure}[]
\caption{Sampling distribution of the subgroup specific treatment effect estimates for each method listed in Table \ref{table_methods} for subgroup V=1. The estimated density highlighted red represents those estimates obtained with a fixed seed. The estimated density highlighted in green represents estimates obtained with a new random seed selected at each of the $1000$ iterations. \label{fig:densv1}}
\begin{center}
\includegraphics[scale=0.5]{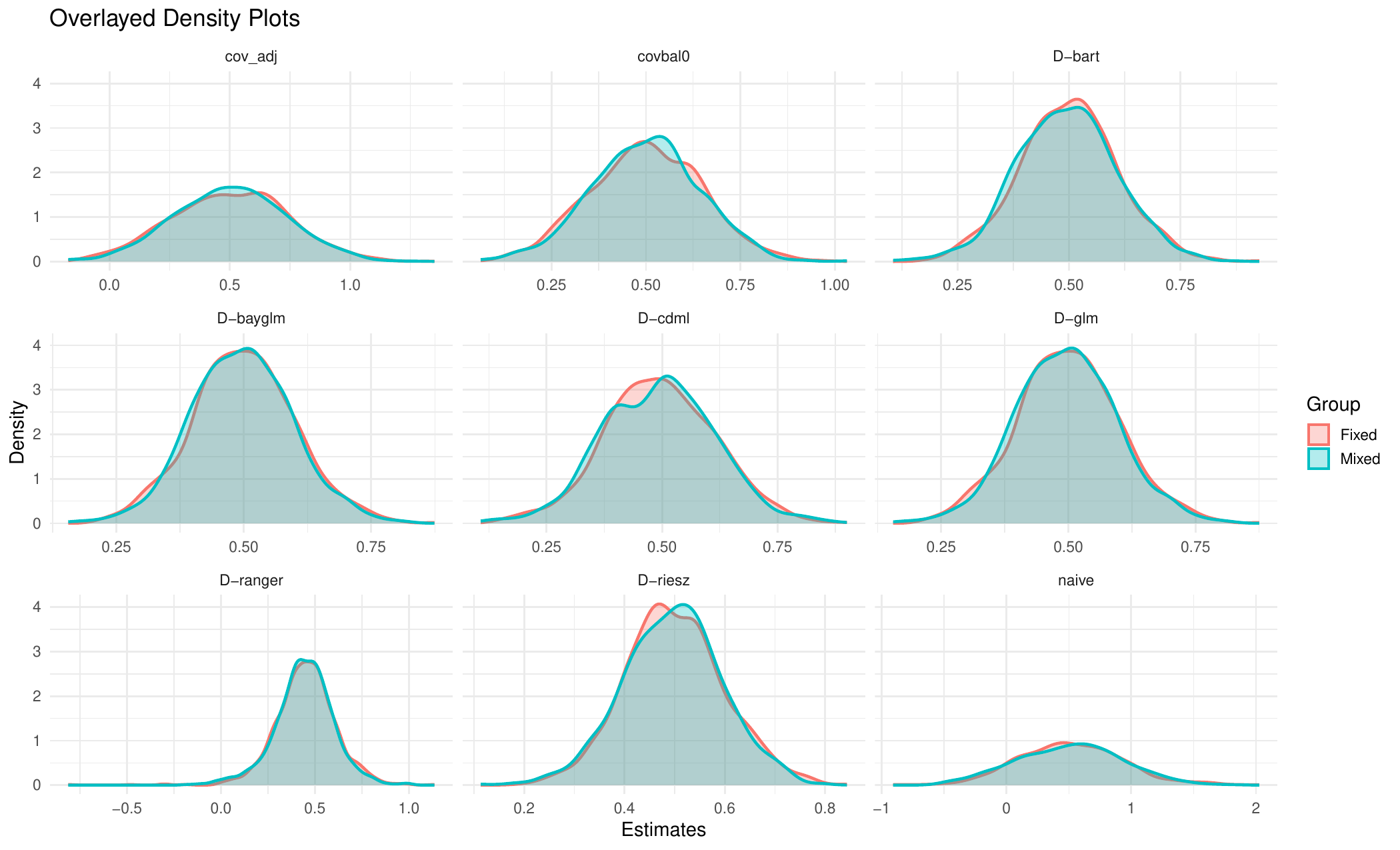}
\end{center}
\end{figure}

\clearpage

\begin{figure}[]
\caption{One thousand confidence intervals generated according to the criteria in Scenario 1 for each method in Table \ref{table_methods}. The horizontal red line denotes the true sub-group average treatment effect for those subjects with V=0, by design this was $-0.5$. The highlighted blue lines are confidence intervals constructed when the random seed was fixed across each iteration of the analysis. Each orange line represent a corresponding confidence interval constructed with a new random seed selected at the start of each iteration. \label{fig:civ0}}

\begin{center}
\includegraphics[scale=0.50]{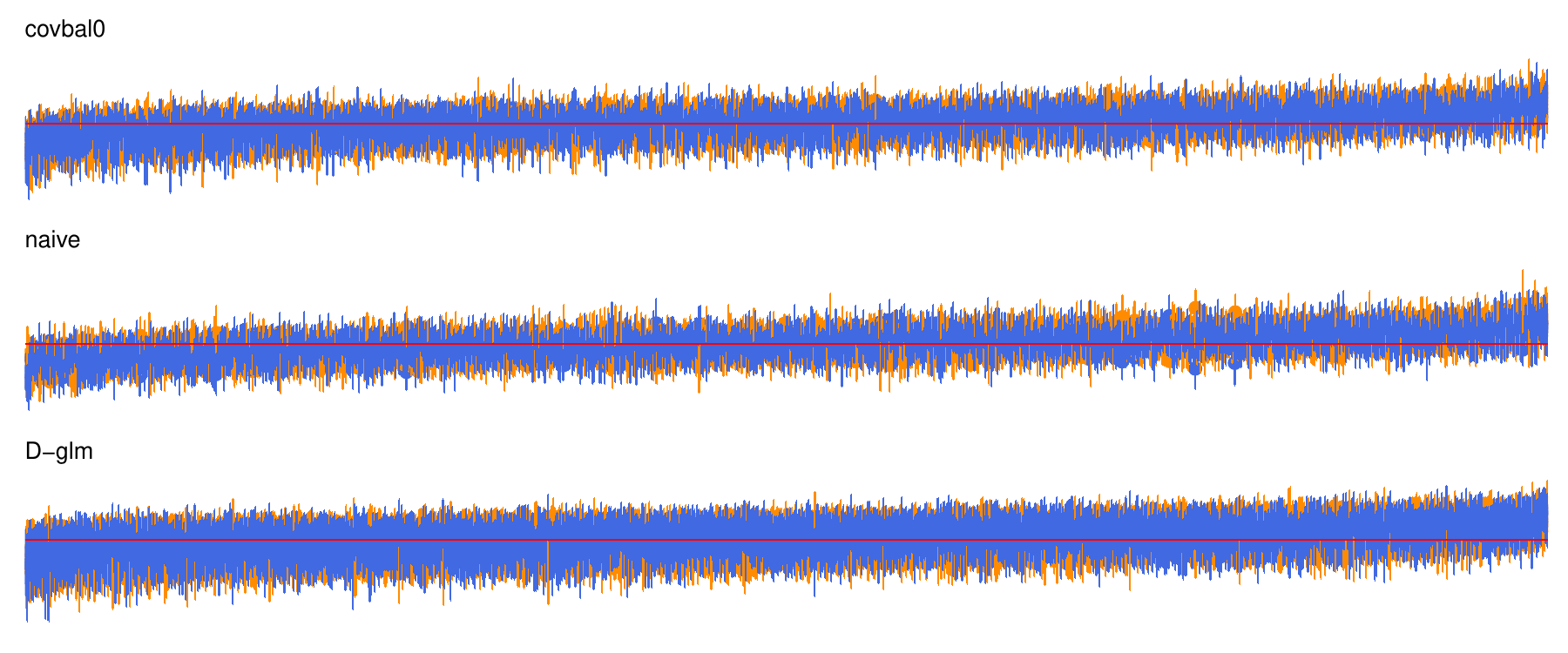}
\includegraphics[scale=0.50]{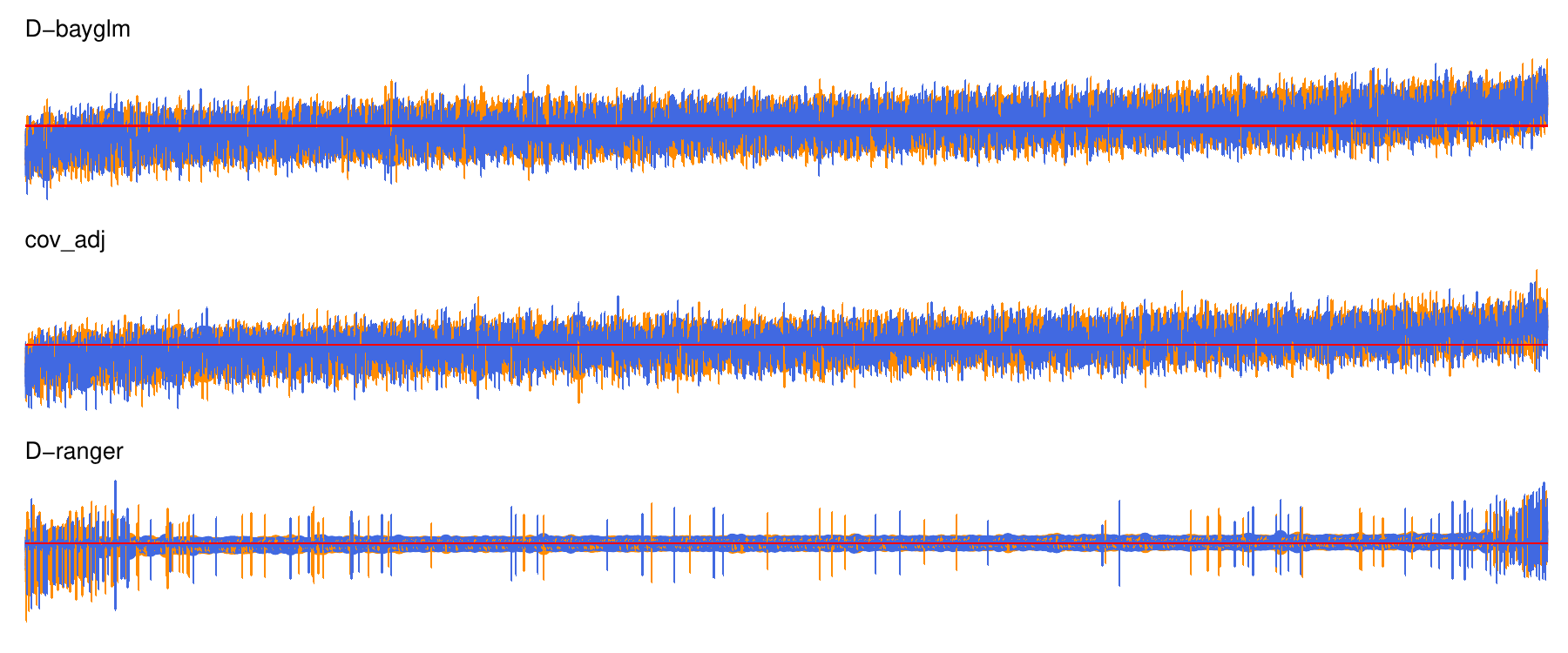}
\includegraphics[scale=0.50]{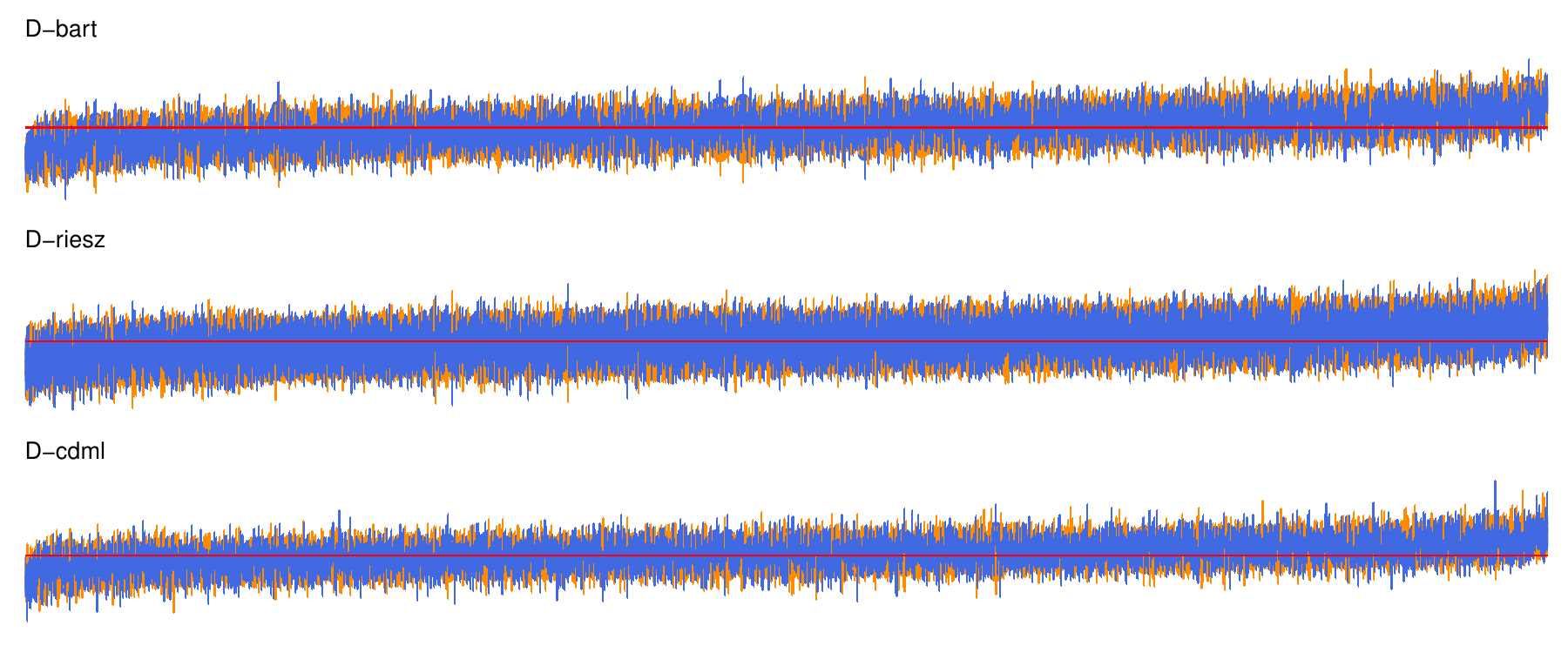}
\end{center}
\end{figure}

\clearpage

\begin{figure}[]
\caption{Sampling distribution of the subgroup specific treatment effect estimates for each method listed in Table \ref{table_methods} for subgroup V=0. The estimated density highlighted red represents those estimates obtained with a fixed seed. The estimated density highlighted in green represents estimates obtained with a new random seed selected at each of the $1000$ iterations.  \label{fig:densv0}}
\begin{center}
\includegraphics[scale=0.5]{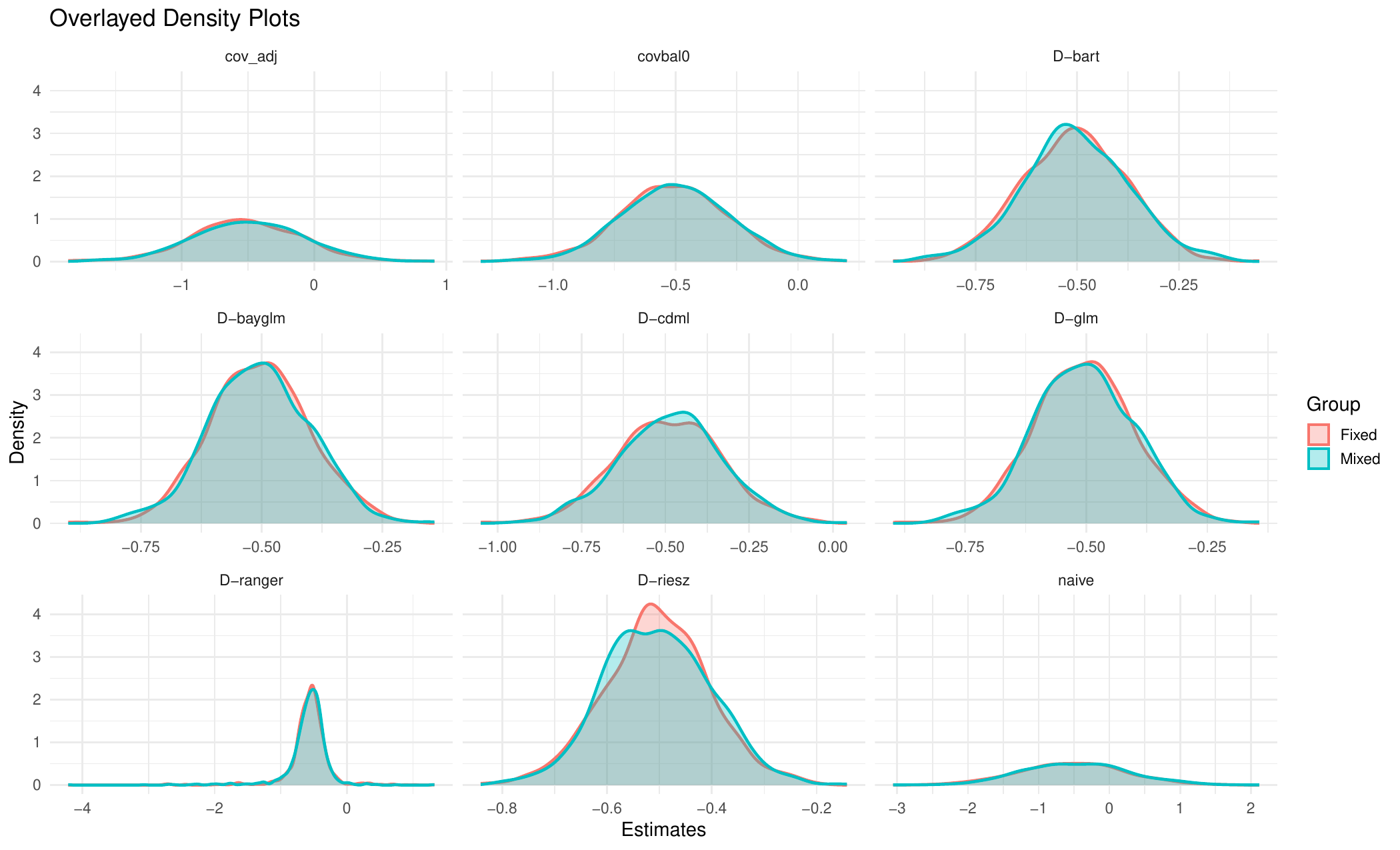}
\end{center}
\end{figure}

\clearpage

\begin{table}[]
\centering
\begin{tabular}{lccccccccc}
 
                                             \multicolumn{10}{c}{\textbf{External data size}} \\  
\hline
 \textbf{Method} & \textbf{100} & \textbf{200} & \textbf{300} & \textbf{400} & \textbf{500} & \textbf{600} & \textbf{700} & \textbf{800} & \textbf{900} \\
\hline
\texttt{D-glm}    & 0.00620 & 0.00620 & 0.00689 & 0.00539 & 0.00689 & 0.00315 & 0.00620 & 0.00539 & 0.00539 \\
\texttt{D-ranger} & 0.01107 & 0.00801 & 0.00708 & 0.00780 & 0.00780 & 0.00768 & 0.00817 & 0.00817 & 0.00843 \\
\texttt{cov-adj}   & 0.00807 & 0.00751 & 0.00620 & 0.00858 & 0.00905 & 0.00539 & 0.00751 & 0.00858 & 0.00807 \\
\texttt{D-bayglm} & 0.00620 & 0.00443 & 0.00689 & 0.00689 & 0.00689 & 0.00443 & 0.00620 & 0.00443 & 0.00620 \\
\texttt{D-bart}   & 0.00539 & 0.00539 & 0.00539 & 0.00539 & 0.00443 & 0.00315 & 0.00443 & 0.00315 & 0.00539 \\
\texttt{covbal0}   & 0.00807 & 0.00689 & 0.00315 & 0.00689 & 0.00807 & 0.00443 & 0.00620 & 0.00315 & 0.00443 \\
\texttt{naive}     & 0.00949 & 0.00807 & 0.00689 & 0.00807 & 0.00751 & 0.00443 & 0.00689 & 0.00807 & 0.01028 \\
\texttt{D-riesz}  & 0.00264 & 0.00223 & 0.00223 & 0.00344 & 0.00223 & 0.00223 & 0.00223 & 0.00223 & 0.00223 \\
\texttt{D-cdml}   & 0.00791 & 0.00715 & 0.00597 & 0.00727 & 0.00662 & 0.00683 & 0.00641 & 0.00649 & 0.00676 \\
\hline
\end{tabular}
\caption{Monte Carlo SE of the estimated coverage based on calculations described in Morris et al. \citep{morris2019}, for each method in Table \ref{table_methods} as a function of external data size for the subgroup $V=1$. \label{table_mc1}}
\end{table}

\begin{table}[]
\centering
\begin{tabular}{lccccccccc}
 
                                             \multicolumn{10}{c}{\textbf{External data size}} \\  
\hline
 \textbf{Method} & \textbf{100} & \textbf{200} & \textbf{300} & \textbf{400} & \textbf{500} & \textbf{600} & \textbf{700} & \textbf{800} & \textbf{900} \\
\hline
\texttt{D-glm}     & 0.00858 & 0.00751 & 0.00539 & 0.00315 & 0.00539 & 0.00000 & 0.00620 & 0.00443 & 0.00000 \\
\texttt{D-ranger}  & 0.00589 & 0.00581 & 0.00513 & 0.00557 & 0.00474 & 0.00581 & 0.00453 & 0.00453 & 0.00474 \\
\texttt{cov-adj}   & 0.00539 & 0.00751 & 0.00807 & 0.00689 & 0.00751 & 0.00689 & 0.00620 & 0.00539 & 0.00807 \\
\texttt{D-bayglm}  & 0.00858 & 0.00905 & 0.00620 & 0.00751 & 0.00807 & 0.00689 & 0.00905 & 0.00858 & 0.00751 \\
\texttt{D-bart}    & 0.00620 & 0.00807 & 0.00539 & 0.00620 & 0.00620 & 0.00315 & 0.00751 & 0.00620 & 0.00539 \\
\texttt{covbal0}   & 0.00620 & 0.00949 & 0.00689 & 0.00689 & 0.00315 & 0.00539 & 0.00689 & 0.00443 & 0.00751 \\
\texttt{naive}     & 0.00620 & 0.00620 & 0.00905 & 0.00807 & 0.00858 & 0.00689 & 0.00858 & 0.00539 & 0.00858 \\
\texttt{D-riesz}   & 0.00244 & 0.00200 & 0.00200 & 0.00223 & 0.00223 & 0.00223 & 0.00344 & 0.00344 & 0.00299 \\
\texttt{D-cdml}    & 0.00634 & 0.00634 & 0.00581 & 0.00605 & 0.00612 & 0.00597 & 0.00494 & 0.00581 & 0.00612 \\
\hline
\end{tabular}
\caption{Monte Carlo SE of the estimated coverage based on calculations described in Morris et al. \citep{morris2019}, for each method in Table \ref{table_methods} as a function of external data size for the subgroup $V=0$. \label{table_mc0}}
\end{table}

\end{document}